\documentstyle[preprint,tighten,aps]{revtex}

\begin{document}
\draft
\preprint{IFUP-TH 11/97}
\title{
The two-point correlation function of
three-dimensional \protect\\ 
$\protect\bbox{{\rm O}(N)}$ models: critical limit and anisotropy.}
\author{Massimo Campostrini, Andrea Pelissetto, Paolo Rossi, and Ettore Vicari}
\address{Dipartimento di Fisica dell'Universit\`a 
and I.N.F.N., I-56126 Pisa, Italy}

\date{\today}

\maketitle

\begin{abstract}
In three-dimensional O($N$) models,
we investigate the low-momentum behavior of the two-point
Green's function $G(x)$ in the critical region of the symmetric phase.
We consider physical systems whose criticality is characterized by a
rotational-invariant fixed point.
Several approaches are exploited, such as strong-coupling expansion of
lattice non-linear ${\rm O}(N)$ $\sigma$ models, $1/N$-expansion,
field-theoretical methods within the $\phi^4$ continuum formulation.

Non-Gaussian corrections to the 
universal low-momentum behavior of $G(x)$ are evaluated,
and found to be very small. 

In non-rotational invariant physical systems with 
${\rm O}(N)$-invariant  interactions, the vanishing of space-anisotropy 
approaching the rotational-invariant fixed point is described 
by a critical exponent $\rho$, which is universal and is related to 
the leading irrelevant operator
breaking rotational invariance. At $N=\infty$ one finds $\rho=2$.
We show that, for all values of $N\geq 0$, $\rho\simeq 2$.

\end{abstract}
\pacs{PACS numbers: 05.70.Jk, 64.60.Fr, 75.10.Hk, 75.40.Cx}


\section{Introduction}
\label{introduction}

Three-dimensional O($N$) models describe many important critical phenomena
in nature. The statistical properties of ferromagnetic materials are described
by the case $N=3$, where the lagrangian field represents the magnetization.
The helium superfluid transition, whose order parameter is the complex quantum
amplitude of helium atoms, corresponds to $N=2$.
The case $N=1$ (i.e. Ising-like systems) describes 
liquid-vapour transitions in classical
fluids or critical binary fluids, where the order parameter is the density.
O($N$) models in the limit $N\rightarrow 0$ describe the statistical properties
of long polymers.

The critical behavior of the
two-point correlation function $G(x)$ of the order parameter is
relevant in the description of critical scattering observed 
in many esperiments,
typically neutron scattering in ferromagnetic materials,
light and X rays in liquid-gas systems, ...
In Born's approximation
the cross section $\Gamma_{fi}$ for incoming particles (i.e. neutrons or
photons) of momentum $p_i$ and final outgoing momentum $p_f$ is proportional to the
component $k=p_f-p_i$ of the Fourier transform of $G(x)$ 
\begin{equation}
\Gamma_{fi}\propto \widetilde{G}(k=p_f-p_i).
\label{gamma}
\end{equation} 
As a consequence of the critical
behavior of the two-point function $G(x)$ at $T_c$,
\begin{equation}
\widetilde{G}(k)\sim {1\over  k^{2-\eta}},
\label{eq1}
\end{equation}
the cross section for $k\rightarrow 0$ (forward scattering) 
diverges as $T\rightarrow T_c$.
When strictly at criticality the relation (\ref{eq1}) holds at all momentum
scales. In the vicinity of the critical point 
where the relevant correlation length $\xi$
is large but finite, the behavior (\ref{eq1}) occurs for
$\Lambda \gg k\gg 1/\xi$, where $\Lambda$ is a generic cut-off related
to the microscopic structure of the statistical system, for example
the inverse lattice spacing in the case of lattice models.
At low momentum, $k\ll 1/\xi$,
experiments show that $G(x)$  is well approximated by a Gaussian
behavior, 
\begin{equation}
{\widetilde{G}(0)\over\widetilde{G}(k)}
\simeq 1 + {k^2\over M_G^2},
\label{gaubeh}
\end{equation}
where $M_G\sim 1/\xi$ is a mass scale defined at zero momentum
(for a general discussion see e.g. Ref.~\cite{Fi-Bu}).

We will specifically consider systems with an ${\rm O}(N)$-invariant
Hamiltonian in the symmetric phase, where the ${\rm O}(N)$
symmetry is unbroken. Furthermore, we will only consider systems with
a rotationally-symmetric fixed point. Interesting members of this
class are systems defined on highly symmetric lattices, i.e.
Bravais or two-point basis lattices with a tetrahedral or larger
discrete rotational symmetry.

In this paper we focus on the low-momentum behavior of the
two-point correlation function of the order parameter, which coincides with
the lagrangian field, in three-dimensional ${\rm O}(N)$ models.
We want to estimate the deviations from Eq.~(\ref{gaubeh}) 
in the critical region of the symmetric phase, 
i.e. for $0<T/T_c - 1\ll 1$, and in the low-momentum regime,
i.e. $k^2\lesssim M_G^2$.
We focus on two quite different sources of deviations:

(i) Scaling corrections to Eq.~(\ref{gaubeh}), depending
on the ratio $k^2/M_G^2$, and reflecting the non-Gaussian
nature of the fixed point. 

(ii) Non-rotationally invariant scaling violations,
reflecting a microscopic anisotropy in the space distribution
of the spins. This phenomenon may be 
relevant, for example, in the study of ferromagnetic materials, 
where the atoms lie on the sites of a lattice giving rise to a space anisotropy
which may be observed in neutron-scattering experiments. 
In these systems the anisotropy vanishes 
in the critical limit, and $G(x)$ approaches a rotationally invariant
form. It should be noticed that this phenomenon is 
different from the breakdown of the ${\rm O}(N)$ symmetry
in the interaction, which has been widely
considered in the literature~\cite{Zinn}.

In our study of 
the critical behavior of the two-point function of the order parameter 
$G(x)$ we will consider several approaches.
We analyze the strong-coupling expansion of $G(x)$,
\begin{equation}
G(x)\equiv\langle \vec{s}(x) \cdot \vec{s}(0)\rangle,
\label{gx}
\end{equation}
for the lattice ${\rm O}(N)$ non-linear $\sigma$ model  with nearest-neighbor
interaction
\begin{equation}
S_L = -N\beta \sum_{{\rm links}\ \langle xy\rangle}
\vec{s}_{x}\cdot \vec{s}_{y},
\label{actionlatt}
\end{equation}
which we have calculated
up to 15th order on the simple cubic lattice and 21st order on the 
diamond lattice.
We also perform a detailed study using the $1/N$-expansion,
whose results, beside clarifying physical mechanisms,
are also useful as benchmarks for the strong-coupling analysis. 
Moreover we compute the first few non-trivial terms of 
the $\epsilon$-expansion and of the $g$-expansion (i.e. 
expansion in the four-point renormalized 
coupling at fixed dimension $d=3$) of
the two-point function for the corresponding $\phi^4$ 
continuum formulation of ${\rm O}(N)$ models:
\begin{equation}
S_{\phi^4}= \int d^d x \;\;{1\over 2}\partial_\mu \phi(x)\partial_\mu \phi(x)
+ {1\over 2}\mu_0^2\phi^2 + {1\over 4!}g_0(\phi^2)^2.
\label{phi4}
\end{equation}
We recall
that non-linear $\sigma$ models and $\phi^4$ models possessing
the same internal symmetry ${\rm O}(N)$
describe the same critical behavior.
By universality our study provides information on the behavior 
of the physical systems mentioned above in the critical region of the
high-temperature phase. 
A short report of our study can be found in Ref.~\cite{d3letter}.

The first systematic study of the critical behavior of $G(x)$ is due
to Fisher and Aharony~\cite{Fi-Bu,Fisher,Aharony}. They computed 
$G(x)$ in the $\epsilon$-expansion up to terms $O(\epsilon^2)$
\cite{Fisher} and in the large-$N$ expansion to order $1/N$~\cite{Aharony};
moreover some estimates of the non-Gaussian corrections
for $N=1$ and $N=3$ were derived from strong-coupling series for various
lattices~\cite{Fisher,Ta-Fi}.
Their calculations confirmed experimental observations that 
non-Gaussian corrections are small in the low-momentum region. 

In this paper we reconsider the problem of determining the
non-Gaussian correction to $G(x)$ in the low-momentum regime using the 
different approaches we mentioned above.
Our strong-coupling analysis of the $N$-vector model on the simple cubic and
diamond lattice leads to rather accurate results
which considerably improve earlier calculations.
This is achieved essentially for two reasons:
longer strong-coupling series are available, and, more importantly, 
we consider improved estimators which allow  more stable
extrapolations to the critical limit.
Our strong-coupling analysis is integrated and supported 
by results obtained from the $O(1/N)$,
$\epsilon$- and $g$-expansion. We 
compute the expansion of $G(x)$ up to four-loops, $O(g^4)$, in
fixed dimension $d=3$ and we extend the results
of~\cite{Fisher} by calculating the next three-loop term $O(\epsilon^3)$.
The results of the various approaches are reasonably
consistent among each other: the $g$-expansion and the analysis of the 
strong-coupling series provide in general the most precise estimates,
together with the $1/N$ expansion for $N\gtrsim 16$. The $\epsilon$-expansion
is somewhat worse but still consistent, perhaps because of the limited 
number of terms (one term less than in the $g$-expansion).

We also discuss the space anisotropy in $G(x)$ induced by the lattice 
structure. 
For the class of systems we consider,  $G(x)$ becomes 
rotationally-invariant at criticality: when $\beta\to\beta_c$, 
so that $M_G\to 0$, the anisotropic deviations vanish as 
$M_G^\rho$, where $\rho$ is a universal critical exponent.
From a field-theoretical point of view, 
space anisotropy is due to non-rotationally invariant 
${\rm O}(N)$-symmetric irrelevant
operators in the effective Hamiltonian, whose presence depends essentially on
the symmetries of the physical system or of the lattice formulation.
The exponent $\rho$ is  related to the critical effective dimension
of the leading irrelevant operator breaking rotational invariance.
On $d$-dimensional lattices with cubic symmetry the leading operator
has canonical dimension $d+2$.
In the large-$N$ limit, where the canonical dimensions determine the
scaling properties, one finds $\rho=2$ with very small $O(1/N)$ corrections.
A strong-coupling analysis supported by a two-loop 
$\epsilon$-expansion and three-loop $g$-expansion computation indicate
that $\rho$ remains close 
to its canonical value for all $N\ge 0$, with deviations
of approximately 1\% for small values of $N$. 
It should be noted that the exponent $\rho$ which controls 
the recovery of rotational invariance is different from $\omega$, the leading
subleading exponent, since they are related to different irrelevant
operators. This means --- and this may be of relevance for numerical 
calculations --- that the recovery of rotational invariance 
is unrelated to the disappearance of the subleading corrections controlled 
by $\omega$: in practice, as $\rho\approx 2$ while $0.5\lesssim \omega
\lesssim 1$~\cite{Zinn,Ma},  
rotational invariance is recovered long before the scaling region.

We also investigated the recovery of rotational invariance
in two-dimensional models. On the square lattice,
for $N=1$ (Ising model) and $N\ge 3$, we show that $\rho=2$.
This leads us to conjecture that 
$\rho=2$ holds exactly for all two-dimensional models  on the 
square lattice. Similarly we conjecture that $\rho=4$ (resp. $\rho = 3$)
are the exact values of the exponents 
for the triangular (resp. honeycomb) lattice.
A Monte Carlo and exact-enumeration study~\cite{CGPS} for $N=0$ on the 
square lattice is consistent with this conjecture. 

We should mention that our results on space anisotropy 
are also relevant in the discussion of the linear response of the 
system in presence of an external (anisotropic) field.

The paper is organized as follows:

In Sec.~\ref{sec1} we fix the notation and introduce a general
parametrization of $G(x)$ which includes
the off-critical and non-spherical dependence.

In Sec.~\ref{sec2} we analyze the
critical behavior of $G(x)$ at low momentum.
We present calculations based on various approaches:
$1/N$-expansion (up to $O(1/N)$), 
$g$-expansion (up to $O(g^4)$), $\epsilon$-expansion (up to $O(\epsilon^3)$), 
and an analysis
of the strong-coupling expansion of $G(x)$ on the cubic and diamond lattice.

In Sec.~\ref{sec3} the anisotropy of $G(x)$ is studied
in the critical region.
We present large-$N$ and $O(1/N)$ calculations on various lattices,
and a strong-coupling analysis
of some non-spherical moments of $G(x)$ on cubic and diamond lattice. 
Again, the analysis of the first non-trivial terms of 
the $g$-expansion and the $\epsilon$-expansion is presented.
Anisotropy in $G(x)$ is also studied in
two-dimensional ${\rm O}(N)$ models.

In App.~\ref{appexlargeN} we collect explicit 
formulae for the large-$N$ limit
of $G(x)$ for the nearest-neighbor $N$-vector model
on the cubic, f.c.c. and diamond lattices.

In Apps.~\ref{appscfun} and \ref{app1oncalc} 
 we present some details of our 
$O(1/N)$ calculations.

In App.~\ref{appcube} we present the 
15th-order strong-coupling expansion
of the two-point function on the cubic lattice.

In App.~\ref{appdia} we report the 21st-order
strong-coupling series
of the magnetic susceptibility  and of the second moment of
$G(x)$ on the diamond lattice for $N=1,2,3$.

\section{The two-point Green's function}
\label{sec1}

\subsection{Hypercubic lattices}
\label{sec1sub1}

In this Section we discuss the general behaviour of the 
two-point spin-spin correlation function in lattice
${\rm O}(N)$ non-linear $\sigma$ models. 
We consider a generic Hamiltonian defined on a hypercubic lattice
\begin{equation}
{\cal H} = -  \sum_{x,y} J(x-y) \vec{s}_x \cdot \vec{s}_y 
\label{hamiltonian-hypercubic}
\end{equation}
where the sum runs over all lattice sites. 
We will later extend our analysis to other lattices. 
Let us define
\begin{equation}
\overline{k}^2(k) = {2} \left[ \widetilde{J}(k) - \widetilde{J}(0) \right]
\label{overlinek2}
\end{equation}
where $\widetilde{J}(k)$ denotes the Fourier transform of $J(x)$. 
In spite of the notation, we are not assuming 
that $\overline{k}^2(k)$ is a sum of the type $\sum_\mu f(k_\mu)$. 
We consider models for which,
by a suitable normalization of the inverse temperature $\beta$,
one finds
\begin{equation}
\overline{k}^2(k) = k^2 + O(k^4),
\end{equation}
so that the critical limit is rotationally invariant.
Moreover we make the following assumptions:

(1) The interaction $J(x)$ is short-ranged so that $\overline{k}^2$ is 
continuous;

(2) The function $J(x)$ (and thus also $\overline{k}^2$) is invariant 
under all the symmetries of the lattice;

(3) The interaction is ferromagnetic, so that $\overline{k}^2= 0$ 
only for $k=0$ in the Brillouin zone.

Beside the leading (universal) rotationally-invariant critical behaviour, we  
are interested in understanding the effects of the lattice structure 
on the two-point function and the recovery of rotational invariance.
For this reason, our analysis must take into account the
irrelevant operators which break rotational invariance.
It is natural to expand $\overline{k}^2(k)$ in multipoles 
by writing 
\begin{equation}
\overline{k}^2 (k) = \sum_{l=0}^\infty \sum_{p=1}^{p_l} 
          e_{2l}^{(p)}(k^2)\, Q_{2l}^{(p)}(k). 
\label{expansion-tildeJ}
\end{equation}
Here the functions $Q_{2l}^{(p)}(k)$ are multipole combinations which are 
invariant under the symmetries of the lattice. Their expressions
can be obtained from the fully
symmetric traceless tensors of rank $2l$, 
$T_{2l}^{\alpha_1\ldots\alpha_{2l}}(k)$~\cite{RB,CMPS},  
by considering all the cubic-invariant combinations, which can be 
obtained by setting equal an even number of indices larger than or equal to 
four and then summing over them. 
Odd-rank  terms are absent in the expansion (\ref{expansion-tildeJ}) 
because of the parity symmetry $x\to -x$. Moreover, there is no
rank-two term,  i.e. $Q_2(k) = 0$, due to the discrete
rotational symmetry of the lattice.
The summation over $p$ in Eq.~(\ref{expansion-tildeJ})
is due to the fact that, for given $l$, there are 
in general many multipole combinations which  are
cubic invariant. The number $p_l$ depends
on the dimensionality $d$ of the space; more precisely it is 
a non-decreasing function of $d$. It can be computed from the 
following generating function (for the 
derivation see Appendix \ref{appcube})
\begin{equation}
\sum_{l=0}^\infty p_lt^l = \prod_{m=2}^d {1\over 1-t^m}.
\end{equation}
For notational simplicity, 
we will suppress the explicit dependence on $p$
in all the following 
formulae, but the reader 
should remember that it is understood in the notation. 

Let us give the explicit expressions of 
$Q_{2l}^{(p)}(k)$ for the first few values of $l$. 
We set $Q_{0}(k) \equiv 1$. 
For $l=2$ there is only one invariant combination, 
i.e. $p_2 =1$, which can be derived from
\begin{eqnarray}
T_4^{\alpha\beta\gamma\delta} (k) =&&
   k^\alpha k^\beta k^\gamma k^\delta - 
{k^2\over (d+4)} {\rm Perm}_{\alpha\beta\gamma\delta}
\left(\delta^{\alpha\beta}k^\gamma k^\delta\right)  
\nonumber \\
  && + {(k^2)^2 \over (d+2) (d+4)} {\rm Perm}_{\alpha\beta\gamma\delta}
    \left(\delta^{\alpha\beta} \delta^{\gamma\delta}\right),
\end{eqnarray}
where ${\rm Perm}_{\alpha_1...\alpha_n}(...)$ represents the sum of
the non-trivial
permutations of its arguments. One then defines
\begin{equation}
Q_{4}(k) = \sum_\mu T_4^{\mu\mu\mu\mu} = 
   k^4 - {3\over d+2} (k^2)^2,  
\label{spinfour}
\end{equation}
where the notation $k^n \equiv \sum_\mu k^n_\mu$ is used.
For $l=3$, $p_3=1$ for all $d>2$.  From
\begin{eqnarray}
&&T_6^{\mu\nu\alpha\beta\gamma\delta} (k) = 
     k^\mu k^\nu k^\alpha k^\beta k^\gamma k^\delta 
- {k^2 \over (d+8)} 
    {\rm Perm}_{\alpha\beta\gamma\delta\mu\nu}
\left(\delta^{\mu\nu} k^\alpha k^\beta k^\gamma k^\delta\right) + \\
 && + {(k^2)^2\over (d+6)(d+8)}
    {\rm Perm}_{\alpha\beta\gamma\delta\mu\nu}
\left(\delta^{\mu\nu} \delta^{\alpha\beta} k^\gamma
     k^\delta\right)  - {(k^2)^3 \over (d+4)(d+6)(d+8)} 
    {\rm Perm}_{\alpha\beta\gamma\delta\mu\nu}
    \left(\delta^{\mu\nu} \delta^{\alpha\beta}
     \delta^{\gamma\delta}\right),
\nonumber 
\end{eqnarray}
one finds
\begin{equation}
Q_{6}(k) = \sum_\mu T^{\mu\mu\mu\mu\mu\mu}_6 (k) = 
           k^6 - {15 k^2 k^4\over d+8} + 
         {30 (k^2)^3 \over (d+4) (d+8)}.
\label{spin6cubic}
\end{equation}
In $d=2$ it is easy to verify that $Q_6(k)=0$ so that $p_3 = 0$.
For $l=4$ and $d>3$ two different $Q_{8}^{(p)}(k)$ can be extracted 
from the corresponding tensor $T_8^{\alpha_1...\alpha_8}$:
\begin{eqnarray}
Q_{8}^{(1)}(k) &=& \sum_\mu T_8^{\mu\mu\mu\mu\mu\mu\mu\mu} ,\\
Q_{8}^{(2)}(k) &=& \sum_{\mu\nu} T_8^{\mu\mu\mu\mu\nu\nu\nu\nu}  .
\end{eqnarray}
When $d=2,3$ the two combinations are not independent. Indeed
$Q_8^{(2)}=2Q_8^{(1)}$ so that $p_4 = 1$. Higher values of $l$ can be dealt 
with similarly.

In order to study the formal continuum limit of the Hamiltonian
defined in (\ref{hamiltonian-hypercubic}),
we expand $e_{2l}(k^2)$ in powers of $k^2$. 
We write (the sum over different multipoles 
with the same value of $l$ being understood in the notation)
\begin{equation}
 \overline{k}^2(k) = \sum_{l=0}^{\infty}\sum_{m=0}^\infty
    e_{2l,m} \,(k^2)^m Q_{2l}(k),
\label{aj}
\end{equation}
where $e_{0,0}\equiv 0$ and $e_{0,1}\equiv 1$.
Inserting back in Eq.~(\ref{hamiltonian-hypercubic}) 
one sees that Eq.~(\ref{aj})
represents an expansion in terms of the irrelevant operators
\begin{equation}
O_{2l,m}(x)=\vec{s}(x)\cdot \Box^m Q_{2l}(\partial) \vec{s}(x),
\label{irrop}
\end{equation}
 where $\Box \equiv \sum_\mu \partial_\mu^2$. 
The leading operator that breaks  rotational
invariance is the four-derivative term 
\begin{equation}
O_4(x)\equiv O_{4,0}(x)=\vec{s}(x)\cdot Q_{4}(\partial)\vec{s}(x),
\label{dp2term}
\end{equation}
which has  canonical dimensions $d+2$.

Let us now consider the Green's function 
\begin{equation}
G(x;\beta) \equiv \langle\vec{s}_0\cdot\vec{s}_x\rangle,
\label{glatt}
\end{equation}
 and its Fourier 
transform
$\widetilde{G}(k;\beta)$. We define a 
zero-momentum mass-scale $M_G(\beta)$ by 
\begin{equation}
M_G (\beta) \equiv {1\over \xi_G(\beta)}
\end{equation}
where $\xi_G(\beta)$ is the second-moment correlation length 
\begin{equation}
\xi_G^2(\beta) = {1\over 2d} 
   {\sum_x |x|^2 G(x;\beta)\over \sum_x  G(x;\beta)}.
\end{equation}
Since there is a 
one-to-one correspondence between $M_G(\beta)$ and $\beta$, one may
consider $\widetilde{G}(k;\beta)$ as a function of $M_G(\beta)$ 
instead of $\beta$. Indeed, for the purpose of studying the critical limit, 
it is natural to consider $\widetilde{G}(k;\beta)$ as a function
of $k$ and $M_G$. In complete
analogy to our discussion of $\widetilde{J}(k)$, we analyze the behaviour of 
$\widetilde{G}(k,M_G)$ in terms of multipoles
(again a sum over different multipole combinations with the same value of
$l$ is understood, see Eq.~(\ref{expansion-tildeJ})):
\begin{equation}
\widetilde{G}^{-1}(k,M_G) = 
   \sum_{l=0}^\infty g_{2l}(y,M_G) Q_{2l}(k), 
\label{gm1}
\end{equation}
where $y=k^2/M_G^2$. Notice that 
\begin{equation}
Q_{2l}(k) = Q_{2l}\left( k/ M_G\right) M_G^{2l}.
\end{equation}

For the purpose of studying the universal properties of the critical
limit of $G(x)$, in which $M_G\to 0$ keeping $k/M_G$ fixed, 
it is important to understand the behavior of the functions 
$g_{2l}(y,M_G)$ when $M_G\rightarrow 0 $. The naive limit 
does not exist. However, as long as the contributions
to $\widetilde{G}^{-1}(k,M_G)$ are originated by the insertion of
individual (irrelevant) operators without any mixing among different
operators with the same symmetry properties, one can apply standard
results in renormalization theory.
In this case one can establish some universal properties.
For a generic choice of
$\widetilde{J}(k)$ this holds only for the functions $g_0(y,M_G)$
and $g_4(y,M_G)$.
Indeed for higher values of $l$ there are mixings among different operators
which make the renormalization of the functions $g_{2l}(y,M_G)$ more 
complicated. Consider, for instance, the case $l=3$ in the 
large-$N$ limit, where the operators have canonical dimensions.
In this case terms proportional to $Q_4(k)$ are depressed as $M_G^2$,
while terms proportional to $Q_6(k)$ are depressed as $M_G^4$. However
it is easy to see that the multipole decomposition of $Q_4(k)^2$,
which is also depressed as $M_G^4$, contains a term of the form
$k^2 Q_6(k)$. This means that there are two operators contributing to 
$g_6(k,M_G)$, $O_{4,0}(x)^2$ and $O_{6,0}(x)$. An analogous
argument applies to higher values of $l$. Notice that for the particular case
of $l=3$ the mixing should disappear in the limit $y\to 0$: thus for 
$M_G\to 0$ $g_6(0,M_G)$ can be directly related to the renormalization
properties of the operator $O_{6,0}(x)$.
 
For $l=0$ and $l=2$ standard results of renormalization theory show that,
if $Z_{2l}(M_G)\equiv g_{2l}(0,M_G)$, the following limit exists:
\begin{equation}
\lim_{M_G\to 0} {g_{2l}(y,M_G)\over Z_{2l}(M_G)} =
   \widehat{g}_{2l} (y),
\label{definitionghat}
\end{equation}
where $\widehat{g}_{2l} (y)$ is a smooth function which is normalized
so that $\widehat{g}_{2l} (0) = 1$.
The function $\widehat{g}_{2l}(y)$ is universal,
in the sense that it is independent of the specific Hamiltonian.

The function $\widehat{g}_{4} (y)$ can also be obtained by considering 
the linear response of the system to an external field possessing the 
corresponding symmetry properties. One considers the one-particle
irreducible two-point function with an insertion 
of a $O_{2l,0}(x)$ operator at zero momentum, i.e.
\begin{equation}
\Gamma_{O_{2l}}(x,M_G) \equiv \int dz \ 
     \langle O_{2l,0}(z)\, \vec{s}(0)\cdot \vec{s}(x)\rangle^{irr} 
\label{gamma2ldef}
\end{equation}
and the corresponding Fourier transform 
$\widetilde{\Gamma}_{O_{2l}}(k,M_G)$. Setting
\begin{equation}
\overline{Z}_{2l}(M_G)\equiv \lim_{k\to 0}
   {\widetilde{\Gamma}_{O_{2l}}(k,M_G)\over Q_{2l}(k)}, 
\end{equation}
the following limit exists
\begin{equation}
\lim_{M_G\to 0} 
   {\widetilde{\Gamma}_{O_{2l}}(k,M_G)\over \overline{Z}_{2l}(M_G)} =
   \widehat{g}_{2l} (y) Q_{2l}(k) .
\label{limiteGamma}
\end{equation}
For $l=2$ the function defined by the previous equation coincides 
with that defined in (\ref{definitionghat}); moreover for $M_G\to 0$,
${Z_4(M_G) /\overline{Z}_{4}(M_G) }$
is a finite (non-universal) constant,
meaning that both quantities have the same singular behavior for 
$M_G\to 0$. For higher values of 
$l$, formula (\ref{limiteGamma}) still holds, but there is no 
easy relation between $\widehat{g}_{2l} (y)$ and $g_{2l}(y,M_G)$
as defined in Eq. (\ref{gm1}), at least for generic Hamiltonians.
Indeed, at least in principle,
one may consider specific forms of $\widetilde{J}(k)$
enjoying the property that all contributions 
$g_{2n}(k,M_G)$ with $0<n<\overline{l}$ 
vanish in the critical limit, for a given
value of $\overline{l}$. 
In lattice quantum field theory this is essentially the spirit 
of Symanzik's improvement program~\cite{Symanzik}. In this case
formula (\ref{definitionghat}) is valid for $l= \overline{l}$ and the 
corresponding function $\widehat{g}_{2l} (y)$ coincides with that 
defined by Eq. (\ref{limiteGamma}).

The functions 
$\widehat{g}_{2l} (y)$ defined in Eq. (\ref{limiteGamma}) 
have a regular expansion in $y$ around $y=0$:
\begin{equation}
\widehat{g}_{2l} (y) = 1 + c_{2l,1} y + c_{2l,2} y^2 + \ldots.
\label{expansionghat}
\end{equation}
$c_{0,1} = 1$ due to the definition of the second-moment correlation 
length.

The renormalization constant $\overline{Z}_{2l}(M_G)$ is instead non-universal.
For $M_G\to 0$ it behaves as 
\begin{equation} 
  \overline{Z}_{2l} (M_G) \approx \overline{z}_{2l} M_G^{-\eta_{2l}},
\label{etal}
\end{equation}
where $\eta_{2l}$ is a critical exponent which depends 
only on the spin of the representation (i.e. it does not depend on 
the additional index $p$ which has always been understood in the notation, 
see Eq.~(\ref{expansion-tildeJ})), and $\overline{z}_{2l}$ is a 
non-universal constant which depends on the lattice 
and on the Hamiltonian (and the additional index $p$). An analogous expression
is valid for $Z_4(M_G)$ (and for $Z_{2\overline{l}}$ for the special
Hamiltonians we have discussed before): for $M_G\to 0$ we have 
$Z_4(M_G) \approx z_4 M_G^{-\eta_{4}}$. For $l=0$, as a consequence of 
our definitions,
\begin{equation}
Z_0(M_G)\sim M_G^{2-\eta}, 
\end{equation}
where $\eta$ is the standard anomalous
dimension of the field.
More generally $\sigma_{2l} \equiv \eta-\eta_{2l}$ 
is the anomalous dimension of the irrelevant operator $O_{2l,0}(x)$.

In two dimensions the renormalization constants diverge only logarithmically 
and thus we write for $l\ne 0$ 
\begin{equation}
   \overline{Z}_{2l} (M_G) \approx 
   \overline{z}_{2l} (\ln M_G)^{\gamma_{2l}}\left[1 + O\left(
   \frac{1}{\ln M_G}\right)\right].
\end{equation}
The anomalous dimensions $\gamma_{2l}$ are universal
while the prefactor $\overline{z}_{2l}$ depends on the details of 
the interaction.

We can now discuss the critical limit of Eq. (\ref{gm1}). 
Using the previous formulae we can write for $M_G\to 0$
\begin{equation}
{\tilde{G}^{-1}(k,M_G)\over Z_0(M_G)} \approx
\hat{g}_0(y) + \hbox{\rm ``rot. inv. sublead."} +\, 
 {z_4\over z_0} M_G^{2 + \eta - \eta_4} \hat{g}_4(y)  Q_4(k/M_G) +\, \ldots
\label{Gm1expansion}
\end{equation}
where ``rot. inv. sublead." indicates rotationally-invariant 
subleading corrections
and the dots stand for terms which vanish faster as $M_G\to 0$. From 
Eq. (\ref{Gm1expansion}) one immediately convinces oneself that the 
anisotropic effects in $G(x)$ vanish for $M_G\to 0$ as $M_G^\rho$ where
$\rho$ is a universal critical exponent given by
\begin{equation}
\rho = 2 + \eta - \eta_4 .
\label{rhoexpr}
\end{equation}
We must notice that the exponent $\rho$ is not related to the exponent 
$\omega$ which characterizes the critical behaviour of the 
``subleading" terms which vanish as $M_G^\omega$, as they are connected to 
different (rotationally-invariant) 
irrelevant operators. Finally notice that the leading
term breaking rotational invariance is universal apart from 
a multiplicative constant, the factor $z_4/z_0$. 

Let us now consider  the small-momentum limit
in which $y\rightarrow 0$ keeping $M_G$ fixed. In this case
one can write for $l=0,2$ (or in the special case we have discussed above
for $l=0,\overline{l}$)
\begin{equation}
g_{2l}(y,M_G) =\, \sum_{m=0}^\infty 
   u_{2l,m}(M_G) y^m.  
\end{equation}
By comparing this expansion with Eq.~(\ref{expansionghat}) and using 
Eq.~(\ref{definitionghat}), one recognizes that
\begin{equation}
Z_{2l}(M_G) = u_{2l,0}(M_G),
\label{z2l}
\end{equation}
and
\begin{equation}
c_{2l,m} = \lim_{M_G\to 0} {u_{2l,m} (M_G) \over u_{2l,0} (M_G)}.
\label{c2lm}
\end{equation}
In the following sections 
we will use this formula to derive estimates for $c_{2l,m}$. Indeed 
the functions $u_{2l,m} (M_G)$ can be determined by computing 
dimensionless invariant ratios of moments of $G(x;\beta)$:
\begin{equation}
q_{2l,m} (\beta) = \sum_x (x^2)^m Q_{2l}(x) G(x;\beta)  .
\label{llm}
\end{equation}

It is interesting to notice that the expansion 
(\ref{Gm1expansion}) implies some universality properties for some ratios 
of $q_{2l,m}$. It is easy to verify that 
\begin{eqnarray}
R_{4,m,n} (\beta) = {q_{0,n}(\beta) q_{4,m}(\beta) \over 
                     q_{0,m}(\beta) q_{4,n}(\beta)}
\label{Rratios}
\end{eqnarray}
is universal for $T\to T_c$; indeed the constant $z_4/z_0$ drops out in the 
ratio. Notice that this means that not only $R_{4,m,n}$ 
does not depend on the particular Hamiltonian, but also that
it is independent of the lattice structure as long as 
$O_{4,0}(x)$ is the leading operator breaking rotational invariance.

\subsection{Other regular lattices}
\label{sec1sub2}

All the considerations of the previous subsection 
can be extended without changes 
to other lattices with cubic symmetry, such as the b.c.c. and the f.c.c. 
lattices.
For other Bravais lattices the same general formulae hold, but 
different multipole combinations will appear in the 
expansion, according to the symmetry of the lattice.
In general a larger number of multipole combinations with given spin
appears when considering lattices with a lower symmetry. 
It is important to notice that in order to have a rotationally-invariant 
critical
limit no multipole $Q_{l}(k)$ with $l=2$ should appear in the expansion of 
the Hamiltonian. 
Thus our considerations apply only to highly symmetric lattices 
with a tetrahedral or larger discrete rotational group.

As an example of a non-cubic-like
lattice  let us consider the two-dimensional triangular lattice. 
It is invariant under rotation of $\pi/3$. The relevant 
multipoles are 
\begin{equation}
T_{6l} (k) = (-k^2)^{3l} \cos (6l \theta) = 
  \sum_{m=0}^{3l} {6l\choose{2m}} k_y^{2m} (ik_x)^{6l-2m},
\label{t6l}
\end{equation} 
where we have set $k_x = |k| \cos \theta$, $k_y = |k| \sin\theta$ 
and we have assumed one of the generators of the lattice to be parallel
to the $x$-axis. Thus in this case we write 
\begin{equation}
\overline{k}^2(k) = \sum_{l=0}^\infty 
          T_{6l}(k) e_{6l} (k^2),
\label{kbartriangolo}
\end{equation}
and a similar expression  for the expansion of the two-point function. 
For the triangular lattice the first operator which breaks 
rotational invariance has dimension $d+4$.
This is a consequence 
of the fact that the triangular lattice has a larger symmetry group 
with respect to the square lattice. 
We define moments corresponding to $T_{6l}(k)$  by
\begin{equation}
t_{6l,m} (\beta) = \sum_x (x^2)^m T_{6l}(x) G(x;\beta)  .
\label{tlm}
\end{equation}
The arguments given in the previous subsection can be generalized to 
the triangular lattice in a straightforward way. One derives 
an expansion of the form (\ref{Gm1expansion}) with $\rho = 4 + \eta - \eta_6$,
$T_6(k/M_G)$ and $\hat{g}_6(y)$ substituting 
$Q_4(k/M_G)$ and $\hat{g}_4(y)$.

\subsection{Non-Bravais lattices}
\label{sec1sub3}

Up to now we have considered regular (Bravais) lattices.
However other important lattice structures are represented
by lattices with basis. Particular examples are the honeycomb lattice
in two dimensions and the diamond lattice in three dimensions.
These lattices are generically defined by the set of points $\vec{x}$ such that
\begin{equation}
\vec{x}=\vec{x}\,' + p\vec{\eta}_p,\quad\quad 
\vec{x}\,'=\sum_i l_i\vec{\eta}_i,
\label{posx}
\end{equation}
where $\vec{\eta}_p$ is the so-called basis vector joining
the two points of the  basis, and $\vec{\eta}_i$ are the generators
of the underlying regular lattice. Here $p=0,1$ and
$l_i\in Z$. For the honeycomb lattice $\vec{\eta}_i$ are the generators
of a triangular lattice while for the diamond lattice 
$\vec{\eta}_i$ are the generators
of a f.c.c. lattice. Due to the breaking of translational
invariance one distinguishes between correlations
between points with the same value of $p$
(i.e. points belonging to the same regular lattice) and
points with different $p$. In general the  components $G_{pp'}$
of the two-point correlation function can be written in the form
\begin{equation}
G_{00}(x-y)=G_{11}(x-y) = \int {dk\over V_B}
 e^{ik(x-y)} {1\over \Delta(k,M_G)}
\label{g00}
\end{equation}
and
\begin{equation}
G_{01}(x-y)=G_{10}(y-x) = \int {dk\over V_B}
 e^{ik(x-y)} { H(k,M_G)\over \Delta(k,M_G)},
\label{g10}
\end{equation}
where the integrals are performed over the Brillouin zone of the
corresponding underlying regular lattice, $V_B$ being its volume. 
$G_{11}(x)$ 
and therefore $\Delta(k,M_G)$ have the
symmetries of the underlying regular lattice and thus can be expanded 
as in the first subsection.
On the other hand, $H(k,M_G)$ does not have the symmetry of
the regular lattice, but only the reduced symmetry of 
the full lattice. 
For the Gaussian  model with nearest-neighbor interactions defined 
on the honeycomb and diamond lattices (and also on their 
$d$-dimensional generalization), 
it is easy to realize that, when $M_G\rightarrow 0$,
\begin{equation}
\Delta(k,M_G)\longrightarrow d\left[ 1 - |H(k,0)|^2 \right]+M_G^2,
\label{ddd}
\end{equation}
and $\Delta(k,M_G)$ turns out to be the inverse propagator
for the Gaussian theory defined on the corresponding regular lattice.
In App.~\ref{diamondNi} we present  a more detailed analysis of the Gaussian
theory with nearest-neighbor interactions on the diamond lattice.
 
Because of the reduced symmetry, additional multipoles which are not 
parity-invariant appear in the expansion of $H(k,M_G)$.
In the case of the honeycomb lattice the symmetry 
of the triangular lattice is reduced to 
$\theta\rightarrow
\theta+\case{2\pi}{3}$. Assuming that one of the links leaving a site 
is parallel to the $x$-axis, one can write
\begin{equation}
H(k,M_G)=\sum_{l=0}^\infty T_{3l}(k) h_{3l}(y,M_G),
\label{BBh}
\end{equation}
where we have extended the definition (\ref{t6l}) to include odd multipoles:
\begin{equation}
T_{3l}(k)=(-k^2)^{3l/2}\cos(3l\theta)=
\sum_{m=0}^{3l/2} {3l\choose{2m}} k_y^{2m} (ik_x)^{3l-2m}.
\label{BH}
\end{equation}
The factor $i$ in this equation 
insures that the functions $h_{3l}(y,M_G)$ are real for all $l$.

For the diamond lattice one can write
\begin{equation}
H(k,M_G)=\sum_{l=0}^\infty \sum_{p=1}^{p_l} Q_{l}^{(p)}(k) 
h_{l}^{(p)}(y,M_G),
\label{BBd}
\end{equation}
where $Q_{l}^{(p)}(k)$ are multipoles constructed from
$T_l^{\alpha_1...\alpha_l}$ as in the case of the cubic lattices.  
The only difference is that now odd-spin operators are allowed,
belonging to the  class
\begin{equation}
Q_{2l+3}(k)=ik_1k_2k_3Q_{2l}(k),
\label{q2l3}
\end{equation}
where we have assumed the natural orientation of the underlying f.c.c.
lattice (see App.~\ref{diamondNi}).

For these lattices,
it is not straightforward to make contact with the field-theoretical 
approach. The problem is writing down operators in the effective 
Hamiltonian that break the parity symmetry. These operators must have 
an odd number of derivatives, but, if they are
bilinear in a real field $\phi$, they give after
integration only boundary terms. The solution to this apparent puzzle
comes from the fact that the effective Hamiltonian for models on 
lattices with basis is naturally written down in terms of 
two fields, defined on the two regular sublattices (for the 
diamond lattice see App.~\ref{diamondNi}). 

As in the regular lattice case, 
we can associate to the breaking of the parity symmetry
a universal exponent $\rho_p$. In principle it can be derived 
from the critical dimension of the lower-dimensional operator
breaking this symmetry. From a practical point of view it is simpler
to consider moments of $G(x)$. For the diamond lattice one defines
$\rho_p$ from the behavior, for $M_G\to 0$, of the odd moments
$q_{3,m}(\beta)$, i.e.
\begin{equation}
{q_{3,m}(\beta)\over q_{0,0}(\beta)} \sim M_G^{-3 - 2m + \rho_p} .
\end{equation}
The same formula applies to the honeycomb lattice with the 
obvious substitutions, $q_{0,0}\to t_{0,0}$, $q_{3,m} \to t_{3,m}$.

\section{Critical behavior of 
$\protect\bbox{G(\lowercase{x})}$ at low momentum}
\label{sec2}

\subsection{Parametrization of the spherical limit of
$\protect\bbox{G(x)}$ at low momentum}
\label{sec2sub1}

According to the discussion presented in the previous section, in the critical
limit multipole contributions are depressed by powers of $M_G$,
hence for $\beta\rightarrow\beta_c$
\begin{equation}
{\widetilde{G}(0;\beta)\over\widetilde{G}(k;\beta)}
\longrightarrow \widehat{g}_0\left(y\right).
\end{equation}
where, again, $y=k^2/M^2_G$.
As already stated by Eq.~(\ref{expansionghat}),
$\widehat{g}_0(y)$ can be expanded in powers of $y$ around $y=0$:
\begin{equation}
\widehat{g}_0(y)=1 + y + \sum_{i=2}^\infty c_i y^i,
\label{lexp}
\end{equation}
where $c_i\equiv c_{0,i}$.
For generalized Gaussian theories $c_i=0$.
As discussed in Sec. \ref{sec1sub1}
the coefficients $c_i$ of the low-momentum expansion of
$\widehat{g}_0(y)$ can be related to the critical limit of 
appropriate dimensionless ratios of spherical moments 
\begin{equation}
m_{2j}\equiv q_{0,j}=\sum_x |x|^{2j} G(x),
\label{m2j}
\end{equation}
or of the corresponding weighted moments
\begin{equation}
\overline{m}_{2j}\equiv {m_{2j}\over m_0}.
\label{barm2j}
\end{equation}
It is easy to compute the behavior of $\overline{m}_{2j}$ for $M_G\to 0$:
\begin{eqnarray}
&&\overline{m}_2 \approx 2d M_G^{-2},\label{en8}  \\
&&\overline{m}_4 \approx 8d(d+2)(1-c_2)M_G^{-4},\nonumber \\
&&\overline{m}_6 \approx 48d(d+2)(d+4)\left(1-2c_2+c_3\right)M_G^{-6},
\nonumber 
\end{eqnarray}
etc.., where $d$ is the lattice dimension.
Then, introducing the quantities
\begin{equation}
v_{2j}={1\over 2^{j} j! \prod_{i=0}^{j-1}(d+2i)}\overline{m}_{2j}M_G^{2j},
\label{vj}
\end{equation}
one may compute $\hat{u}_i\equiv u_{0,i}/u_{0,0}$ from
the following combinations of $v_{2j}$ 
\begin{eqnarray}
\hat{u}_2&=& 1-v_4,\nonumber \\
\hat{u}_3&=& 1-2v_4+v_6,
\label{cvj}
\end{eqnarray}
etc... By definition,
see Eqs. (\ref{c2lm}) and (\ref{lexp}), 
in the critical limit $\hat{u}_i\rightarrow c_i$.

Another important
quantity which characterizes the low-momentum behavior of
$\widehat{g}_0(y)$ is the critical limit of the ratio 
$M^2/M_G^2$,
\begin{equation}
S_M\equiv \lim_{\beta\rightarrow\beta_c} {M^2\over M_G^2},
\label{smdef}
\end{equation}
where $M$ is the mass-gap of the
theory, that is the mass determining the long-distance exponential
behavior of $G(x)$.  The value of $S_M$
is related to the negative zero 
$y_0$ of $\widehat{g}_0(y)$ which is closest to the origin by
\begin{equation}
y_0=-S_M.
\label{y0sm}
\end{equation}
The constant $S_M$ is one in
Gaussian models (i.e., when $\widehat{g}_0(y)=1+y$), like the large-$N$
limit of ${\rm O}(N)$ models.

Let us now consider the relation between the renormalization constant
$Z_G$ defined at zero momentum, 
\begin{equation}
Z_G \equiv \chi M^2_G = Z_0^{-1}M_G^2,
\end{equation}
where $Z_0$ has been introduced in Eq.~(\ref{definitionghat}),
and the on-shell renormalization constant $Z$, which is defined by
\begin{equation}
\widetilde{G}(k) \longrightarrow  {Z\over M^2+k^2}
\end{equation}
when $k\rightarrow iM$.
The mass gap $M$ and the constant $Z$ determine the 
large-distance behavior of $G(x)$; indeed for $|x|\rightarrow \infty$
\begin{equation}
G(x)\longrightarrow {Z\over 2 M} \left({M\over 2\pi |x|} \right)^{{d-1\over2}}
e^{-M|x|}.
\label{largexbehavior}
\end{equation}
The critical limit $S_Z$ of the ratio $Z_G/Z$ is a universal quantity given by
\begin{equation}
S_Z=\lim_{\beta\rightarrow \beta_c}
{Z_G\over Z} = {\partial \over \partial y}\widehat{g}_0(y)|_{y=y_0}.
\label{SZg}
\end{equation}
In a Gaussian theory $Z_G=Z$.

\subsection{$\protect\bbox{1/N}$-expansion} 
\label{sec2sub2}

In the large-$N$ limit the difference 
\begin{equation}
\widehat{g}_0(y)-(1+y)
\label{ly}
\end{equation}
is depressed by a factor $1/N$. 
It can be derived from the $1/N$ expansion of the self-energy
in the continuum formulation.
One finds~\cite{Aharony,Abeetal} 
\begin{equation}
\widehat{g}_0(y) = 1 + y + {1\over N} \phi_1(y) +
O\left( {1\over N^2}\right).
\label{g01oN}
\end{equation}
where, for $d=3$, 
\begin{equation}
\phi_1(y)  = 
{2\over \pi}\int_0^\infty dz 
{z\over {\rm arctan}\left( \case{1}{2} \sqrt{z}\right)}
\left[ 
{1\over 4\sqrt{yz}}\ln\left( {y+z+2\sqrt{yz}+1\over
y+z-2\sqrt{yz}+1}\right) - {1\over z+1} + {y(3-z)\over
(z+1)^3}\right].
\label{g01oN2}
\end{equation}
A general discussion of the $O(1/N)$ correction to $\widehat{g}_0(y)$
in $d$-dimension is presented in App.~\ref{appscfun}.
In particular Eq.~(\ref{g01oN2}) can be derived from
Eqs.~(\ref{x1}),  (\ref{x2}) and (\ref{t3}).
The coefficients $c_i$ of the low-momentum expansion of
 $\widehat{g}_0(y)$ turn out  to be very small.
Writing them as 
\begin{equation}
c_i={c_i^{(1)}\over N}+O\left( {1\over N^2}\right),
\label{ciNN}
\end{equation}
one obtains
\begin{eqnarray}
c_2^{(1)}&=&-0.00444860...,\label{ciN}\\
c_{3}^{(1)}&=&0.000134410...,\nonumber \\
c_{4}^{(1)}&=&-0.0000065805...,\nonumber \\
c_{5}^{(1)}&=&0.0000004003...,\nonumber 
\end{eqnarray}
etc.. 
For sufficiently large $N$ we then expect
\begin{equation}
c_i\ll c_2\ll 1 \;\;\;\;\;\;\;\;\;{\rm for}\;\;\;\; i\geq 3.
\label{crel}
\end{equation}  
As we shall see from the analysis of the strong-coupling expansion of $G(x)$,
the pattern (\ref{crel}) is verified also at low values of $N$.

The ratio $S_M\equiv M^2/M_G^2$ is obtained 
by evaluating the negative zero $y_0$ of $\widehat{g}_0(y)$ closest
to the origin:
\begin{equation}
S_M=-y_0= 1 + {1\over N} \phi_1(-1) + O\left( {1\over N^2}\right), 
\label{sm1N}
\end{equation}
where 
\begin{equation}
\phi_1(-1) = -0.00459002....
\label{nsm1N}
\end{equation}
Moreover using Eq.~(\ref{SZg}), one finds 
\begin{equation}
S_Z= 1 + {1\over N} \phi_1'(-1) + 
O\left( {1\over N^2}\right)
\label{ratioz1oN}
\end{equation}
where 
\begin{equation}
\phi_1'(-1)= 0.00932894....
\end{equation}

As expected from the relations (\ref{crel}) among the coefficients $c_i$,
a comparison with Eqs.~(\ref{ciN}) shows that 
the non-Gaussian corrections to
$S_M$ and $S_Z$ are essentially determined  by the term proportional to
$(k^2)^2$ in $\tilde{G}^{-1}(k)$, through the approximate relations 
\begin{eqnarray}
S_M&\simeq& 1 + c_2, 
\label{sc2}\\
S_Z&\simeq& 1 - 2c_2,
\label{sc3}
\end{eqnarray}
with corrections of $O(c_3)$.

\subsection{$\protect\bbox{g}$-expansion in three dimensions}
\label{sec2sub2c}

Another approach to the study of the critical behavior
in the symmetric phase of $O(N)$ models is based on 
the so-called $g$-expansion,
the perturbative expansion at fixed dimension $d=3$
for the corresponding $\phi^4$ continuum formulation~\cite{Parisig}.
The perturbative series which are obtained in this way 
are asymptotic; nonetheless accurate results can be obtained 
using a Borel transformation and a conformal mapping which
take into account their large-order behavior.
As general references on this method see for instance
Refs.~\cite{Zinn} and \cite{Parisi}.
This technique has led to very precise estimates
of the critical exponents.

Starting from the continuum action (\ref{phi4}), one renormalizes the
theory at zero momentum using the following renormalization
conditions for
the irreducible two- and four-point correlation functions of the
field $\phi$:
\begin{eqnarray}
\Gamma^{(2)}(p)_{\alpha\beta} &=&
Z_G^{-1} \Gamma^{(2)}_R(p) \,\delta_{\alpha\beta}, 
\label{s2e7a}  \\
\Gamma^{(4)}(0,0,0,0)_{\alpha\beta\gamma\delta} &=&
-Z_G^{-2} {g\over 3}M_G\, \delta_{\alpha\beta\gamma\delta} ,
\label{s2e7b}
\end{eqnarray}
where 
\begin{equation}
\Gamma^{(2)}_R(p) = 
M_G^2+p^2+O(p^4),
\end{equation} 
and $\delta_{\alpha\beta\gamma\delta} \equiv
\delta_{\alpha\beta}\delta_{\gamma\delta}
+\delta_{\alpha\gamma}\delta_{\beta\delta}
+\delta_{\alpha\delta}\delta_{\beta\gamma}$.
When $M_G\rightarrow 0$ the renormalized coupling constant is driven toward
an infrared stable zero $g^*$ of the $\beta$-function
\begin{equation}
\beta(g)\equiv M_G{\partial g\over \partial M_G}|_{g_0,\Lambda}\;.
\end{equation}

The universal function $\widehat{g}_0(y)$ is related to the 
renormalized function 
\begin{equation}
f(g,y)\equiv M_G^{-2}\Gamma^{(2)}_R(k)
\end{equation}
by
\begin{equation}
\widehat{g}_0(y) = \lim_{g\rightarrow g^*} f(g,y).
\end{equation}
We computed the first three non-trivial orders
of the non-Gaussian corrections to $\widehat{g}_0(y)$.
A calculation up to four loops gave
\begin{eqnarray}
&&f(g,y)= 1+y+\bar{g}^{2} Z_g^2 Z_G {N+2\over (N+8)^2}
\varphi_2(y) +\bar{g}^{3}Z_g^3 Z_G^{3/2}{N+2\over (N+8)^2}
\varphi_3(y)+\\
&&\bar{g}^4Z_g^4 Z_G^2 {N+2\over (N+8)^2}\left[ 
{(N+2)\over (N+8)^2} \varphi_{4,1}(y)+ 
{(N^2+6N+20)\over (N+8)^2} \varphi_{4,2}(y)+
{(5N+22)\over (N+8)^2} \varphi_{4,3}(y) 
\right] + O(\bar{g}^{5}),
\label{g0gexp}\nonumber
\end{eqnarray}
where $\bar{g}$ is the rescaled coupling~\cite{Zinn}
\begin{equation}
\bar{g}= {(N+8)\over 48\pi}g,
\label{rescou}
\end{equation}
$Z_g$ is the renormalization constant of the coupling (defined by
$g_0=M_G g Z_g$)
\begin{equation}
Z_g = 1 + \bar{g} + \left[ 1 - {2(41N+190)\over 27 (N+8)^2} \right]\bar{g}^2
+ O(\bar{g}^3),
\label{zg}
\end{equation}
and $Z_G$ is the zero-momentum renormalization of the field
\begin{equation}
Z_G = 1 - {4(N+2)\over 27 (N+8)^2}\bar{g}^2 + O(\bar{g}^3).
\label{zgg}
\end{equation}
A simple derivation of the
two and three-loop functions $\varphi_2(y)$ and $\varphi_3(y)$
is presented in App.~\ref{appscfun} (cfr. Eqs.~(\ref{wwphi})).
In particular using the results of Refs.~\cite{Nickel3,Rajantie} one finds
\begin{equation}
\varphi_2(y)= 
4\ln (1+\case{1}{9}y) + 24{{\rm arctan}
(\sqrt{y}/3)\over \sqrt{y}}-8 - {4\over 27}y. 
\label{varphi2}
\end{equation}
We shall not report the expressions of the four-loop functions
$\varphi_{4,j}(y)$ because 
they are not very illuminating.

The coefficients of the low-momentum expansion 
can be easily obtained
from Eq.~(\ref{g0gexp}) by calculating
the zero-momentum derivatives of the functions
$\varphi_{n,j}(y)$. We write
\begin{eqnarray}
c_i = && {N+2\over (N+8)^2}h_i^{(2)}\bar{g}^2  
+ {N+2\over (N+8)^2}\left(2h_i^{(2)}+ h_i^{(3)}\right)\bar{g}^3+
{N+2\over (N+8)^2}\Bigg\{ 
2h_i^{(2)} \left[ 1 - {8(7N+32)\over 3 (N+8)^2}\right]
\nonumber \\
&&\;\;+ 3h_i^{(3)} 
+ h_i^{(4,1)} {(N+2)\over (N+8)^2}
+ h_i^{(4,2)} {(N^2+6N+20)\over (N+8)^2}
+ h_i^{(4,3)} {(5N+22)\over (N+8)^2}\Bigg\}\bar{g}^4
+O(\bar{g}^5),
\label{cigexp}
\end{eqnarray}
where we have introduced the coefficients
\begin{equation}
h_i^{(n,j)} = {1\over i!} {d^i\over dy^i}\varphi_{n,j}(y)|_{y=0}.
\label{kkk}
\end{equation}
In Table~\ref{hcoeff} we report the numerical values of 
$h_i^{(k,j)}$ for $i\leq 5$.

By evaluating the zero of $\widehat{g}_0(y)$ closest to the origin,
one obtains
\begin{equation}
S_M = 1 +\bar{g}^{2}{N+2\over (N+8)^2} 
\varphi_2(-1)+\bar{g}^{3} {N+2\over (N+8)^2} 
\left[ 2\varphi_2(-1)+\varphi_3(-1)\right] + O(\bar{g}^{4}),
\label{sgexp}
\end{equation}
which numerically leads to
\begin{equation}
S_M-1 = -\bar{g}^{2}{N+2\over (N+8)^2} 
0.00521783 \left[ 1 + \bar{g}\times  0.054182 + O(\bar{g}^2)\right].
\label{sgexp2}
\end{equation}
Moreover
\begin{equation}
S_Z = 1 + \bar{g}^{2}{N+2\over (N+8)^2} 
\varphi_2'(-1)+\bar{g}^{3} {N+2\over (N+8)^2} 
\left[ 2 \varphi_2'(-1)+\varphi_3'(-1)\right] + O(\bar{g}^{4}),
\label{ratioZgexp}
\end{equation}
which numerically leads to
\begin{equation}
S_Z-1 = \bar{g}^{2}{N+2\over (N+8)^2} 
0.0107349 \left[ 1 + \bar{g} \times 0.041829 + O(\bar{g}^{2})\right].
\label{ratioZgexp2}
\end{equation}
A comparison of the  $g$-expansions of $c_i$, $S_M$ and $S_Z$ shows
that the approximate relations (\ref{sc2}) and (\ref{sc3})  are  valid
for all values of $N$ and not only for $N\to\infty$ as shown in the 
previous subsection.

In order to get quantitative estimates, 
one must perform a  resummation of  the series and  then evaluate it 
at the fixed-point value of the coupling $\bar{g}^*$.
Although the terms of the $g$-expansion we have calculated are only three
for $c_i$ and two for $S_M$ and $S_Z$,
we have tried to extract quantitative
estimates taking into account also the following facts:

(i) The $g$-expansion is Borel-summable~\cite{bores}
(see also e.g. Refs.~\cite{Zinn}
and \cite{Parisi} for a discussion of this issue),
and the singularity closest to the origin
of the Borel transform (corresponding
to the rescaled coupling $\bar{g}$) is known~\cite{B-L-Z}:
$b_s=-0.75189774\times (N+8)$. 

(ii) The fixed point value $\bar{g}^*$ of $\bar{g}$ has been 
accurately determined
by analyzing  a much longer expansion (to $O(g^7)$) of the corresponding 
$\beta$-function~\cite{Nickel,Zinng,Nickel2,Antonenko}. 
Indipendent and consistent estimates of 
$\bar{g}^*$  have been obtained by other approaches, such as  
strong-coupling expansion of 
lattice non-linear ${\rm O}(N)$ $\sigma$ 
models~\cite{ONgr,Reisz} (for $N=1$ see also 
Refs.~\cite{Bakerold,Bakergr2,Z-L-F,B-C-g}), and
Monte Carlo lattice simulations 
(only data for $N=1$ are available~\cite{K-P,Tsypin,BakerKawa,Kim}).

We have followed the procedure described in Ref.~\cite{LeG-Zinn} (see
also Ref.~\cite{Zinn}), where the perturbative expansion
in powers of $\bar{g}$ is summed using a Borel transformation
and a conformal mapping which takes into account its large-order behavior.
We transform the series 
\begin{equation}
R(g)=\sum_{k=0}R_k g^k
\end{equation}
into
\begin{equation}
R(g) = \sum_{k=0} B_k \int_0^\infty e^{-t} u(gt)^k  dt
\label{rgu}
\end{equation}
where
\begin{equation}
u(x) = {\sqrt{x-b_s} - \sqrt{-b_s}\over \sqrt{x-b_s} + \sqrt{-b_s}}. 
\end{equation}
The coefficients $B_k$ are determined by comparing the expansion in powers
of $g$ of the r.h.s. of Eq.~(\ref{rgu}) with the original
expansion. 
Since the $g$-series of $c_i$, $S_M-1$ and $S_Z-1$,
have the form $R(\bar{g})=\bar{g}^2\sum_{k=0} a_i \bar{g}^i$,
one may apply the resummation method to either
$R(\bar{g})$ or $R(\bar{g})/\bar{g}^2$. In Table~\ref{summary} 
we present results for both choices. 
Following the suggestions of Ref.~\cite{LeG-Zinn} we
also tried more refined resummations, changing formula
(\ref{rgu}) to weaken the singularity of the Borel transform.
We did not find any significant difference.
In our calculations we used the estimates of $\bar{g}^*$ 
obtained from the analysis of the 
$\beta$-function by \cite{Nickel,Zinng,Antonenko}. They are reported 
in Table~\ref{betac}. For small values of $N$ slightly lower values 
of $\bar{g}^*$
were computed in Ref. \cite{Murray-Nickel} taking into account the 
non-analiticity of the $\beta$-function at the critical point \cite{Nickel2}. 
This difference is however too small to be quantitatively relevant in our 
calculations.

It is difficult to estimate the uncertainty of the results:
the fluctuations of the results with respect to the method 
we used to resum the perturbative series 
indicate an error of $\lesssim 20\%$ on $c_i$ and
$S_M$ for small values of $N$. As $N$ increases the estimates 
become more stable. The final results are in good agreement
with the estimates by other methods.

\subsection{$\protect\bbox{\epsilon}$-expansion}
\label{sec2sub2b}

The universal function $\widehat{g}_0(y)$ can be computed perturbatively
in $\epsilon=4-d$ using the continuum $\phi^4$ theory~\cite{Wilson}. 
The leading order is simply $\widehat{g}_0(y) = 1 + y$. The first 
correction appears at order $\epsilon^2$ and was computed
by Fisher and Aharony~\cite{Fisher}. We have extended 
the series, calculating the $O(\epsilon^3)$ term, obtaining
\begin{equation}
\widehat{g}_0(y)=1+y + \epsilon^2 {N+2\over (N+8)^2}
\left\{ 1 + \epsilon\left[ 1+{6(3N+14)\over (N+8)^2}\right] \right\}
\psi_2(y)+\epsilon^3 {N+2\over (N+8)^2}
\psi_3(y) +  O(\epsilon^4),
\label{epsg0}
\end{equation}
where
\begin{eqnarray}
\psi_2(y) &=& 2\int_0^\infty \sqrt{z(1+\case{1}{4}z)}
\ln \left( \sqrt{1+\case{1}{4}z} +\case{1}{2}\sqrt{z}\right)
h(y,z),\label{psi2}\\
h(y,z) &=& -{1\over 1+z} + {y\over (1+z)^3}
+{1\over 2yz}\left( 1 + y + z - 
\sqrt{1 + 2y+2z+y^2-2yz+z^2}\right).
\nonumber 
\end{eqnarray}
We do not report the explicit expression of $\psi_3(y)$ because 
it is not very illuminating.
It can however be obtained from Eqs. (\ref{x11}), (\ref{x12}) and (\ref{x14})
of App.~\ref{appscfun}, where we show how to derive 
the functions $\psi_2(y)$  and $\psi_3(y)$ 
from the $O(1/N)$ calculation of $\widehat{g}_0(y)$
in $d$ dimensions. 

Setting 
\begin{equation}
c_i = \epsilon^2 {N+2\over (N+8)^2} \widehat{c}_i,  
\label{epsexp}
\end{equation}
one finds
\begin{eqnarray}
\widehat{c}_2 &=&-0.00752024...
\times \left[ 1 + \epsilon  \left( 
{6(3N+14)\over (N+8)^2} - 0.249301...\right)+ O(\epsilon^2)\right],
\label{cprime}\\
\widehat{c}_3 &=& 0.000191931... 
\times \left[ 1 + \epsilon  \left( 
{6(3N+14)\over (N+8)^2} - 0.130607...\right)+ O(\epsilon^2)\right],
 \nonumber \\
\widehat{c}_4 &=&-0.0000081420... 
\times \left[ 1 + \epsilon  \left( 
{6(3N+14)\over (N+8)^2} - 0.003053...\right)+ O(\epsilon^2)\right],
\nonumber \\
\widehat{c}_5 &=& 0.0000004391... 
\times \left[ 1 + \epsilon  \left( 
{6(3N+14)\over (N+8)^2} + 0.117278...\right)+ O(\epsilon^2)\right],
 \nonumber 
\end{eqnarray}
etc.... 

One also obtains
\begin{equation}
S_M= 1 + \epsilon^2 {N+2\over (N+8)^2} 
\left[ 1 + {6(3N+14)\over (N+8)^2}\epsilon \right]\psi_2(-1) + 
\epsilon^3 {N+2\over (N+8)^2}
\psi_3(-1) +O(\epsilon^4),
\label{epss}
\end{equation}
where 
\begin{equation}
\psi_2(-1) = -0.00772078...,\qquad 
\psi_3(-1) = 0.00189984...,
\end{equation}
and
\begin{equation}
S_Z = 1 + \epsilon^2 {N+2\over (N+8)^2} 
\left[ 1 + {6(3N+14)\over (N+8)^2}\epsilon \right]\psi_2'(-1)+ 
\epsilon^3 {N+2\over (N+8)^2}\psi_3'(-1) +O(\epsilon^4),
\label{ZGZo}
\end{equation}
where 
\begin{equation}
\psi_2'(-1)=0.0156512...,\qquad
\psi_3'(-1)=-0.0038246....
\end{equation}

In order to get quantitative estimates from the 
perturbative $\epsilon$-expansion, one should first resum the series and  
then evaluate the resulting expression at $\epsilon=1$.
Usually resummations are performed assuming the Borel summability of the
$\epsilon$-series.
As in the case of the $g$-expansion, a considerable improvement is 
obtained if one uses the knowledge of
the singularity of the Borel transform~\cite{B-L-Z},
$b_s=-(N+8)/3$. 
We have used the procedure described in the previous subsection.
Again, since the $\epsilon$-series of $c_i$, $S_M-1$ and $S_Z-1$,
have the form $R(\epsilon)=\epsilon^2\sum_{k=0} a_i \epsilon^i$,
we applied the resummation method
to $R(\epsilon)$ and to
$R(\epsilon)/\epsilon^2$. In Table~\ref{summary} we present results
for  both choices. Since we use a series with only two terms
the estimates are not very precise as the large difference between
the results obtained with the two methods indicates. 

One can also try to get estimates 
for two-dimensional $O(N)$ models, i.e. for $\epsilon=2$.
By resumming the series of $c_2(\epsilon)$ and $S_M(\epsilon)$,
we find:
$c_2=-0.0010$ and $S_M=0.9989$
for $N=1$, which must be compared with the exact results~\cite{ON-d2-b}
$c_2=-0.000793...$ and $S_M=0.999196...$;
$c_2=-0.0013$ and $S_M=0.9987$ for $N=3$, to be compared with
the strong-coupling results~\cite{d2letter} $c_2=-0.0012(2)$ and
$S_M=0.9987(2)$. In both cases the agreement is satisfactory.
Instead, when resumming the series divided by $\epsilon^2$
the agreement is poorer. We find $c_2=-0.0026$ and $S_M=0.9973$
for $N=1$ and $c_2=-0.0028$ and $S_M=0.9971$ for $N=3$.
A posteriori, it thus appears that the estimates obtained from the 
resummation of  the complete series $R(\epsilon)$ are more reliable. 
This is confirmed by the three-dimensional analysis where 
the estimates obtained by considering $R(\epsilon)$
are those which are in better agreement with
the strong-coupling and $g$-expansion estimates.

For quantities which are exactly known in two dimensions, one can modify the 
$\epsilon$-series to obtain a new expansion which gives the exact value
for $\epsilon = 2$. This can achieved~\cite{L-Z} by defining for a generic 
observable $R$, with $\epsilon$-series $R(\epsilon)$,
\begin{equation}
\bar{R}(\epsilon) =\, {R(\epsilon) - R(\epsilon=2)\over 2 - \epsilon}
\end{equation}
and then rewriting 
\begin{equation}
\widehat{R}(\epsilon) = R^{exact} + (2-\epsilon)\bar{R}(\epsilon),
\label{constrained1}
\end{equation}
where $R^{exact}$ is the exactly known value of $R$ in two dimensions. 
In other words one subtracts to the original series its value for 
$\epsilon = 2$ and then sums the exact two-dimensional result.
Estimates for $\epsilon = 1$ are obtained by resumming the perturbative
expansion of $\bar{R}(\epsilon)$ and computing $\widehat{R}(1)$.
We applied the method to the Ising model, i.e. $N=1$, since in this 
case, the coefficients $c_i$ and $S_M$ are exactly known
\cite{ON-d2-b}. As before, one can also apply the same procedure to
$R(\epsilon)/\epsilon^2$, defining
\begin{equation}
\bar{R}_2(\epsilon) =\, {1\over 2-\epsilon} 
   \left[ {R(\epsilon)\over \epsilon^2} - 
          {R(\epsilon=2)\over 4} \right]
\end{equation}
and then writing
\begin{equation}
\widehat{R}_2(\epsilon) = {\epsilon^2\over 4} R^{exact} + 
          \epsilon^2 (2-\epsilon)\bar{R}_2 (\epsilon).
\label{constrained2}
\end{equation}
From a conceptual point of view, Eq. (\ref{constrained2}) appears
preferable to Eq. (\ref{constrained1}). Indeed Eq. (\ref{constrained1})
gives the exact value for $d=2$, but it does not reproduce the correct value
for $d=4$. Eq. (\ref{constrained2}) instead 
predicts correctly $\widehat{R}_2(\epsilon) \sim O(\epsilon^2)$ for 
$\epsilon\to 0$. In any case we report the results obtained with both 
methods in Table~\ref{summary}. They are referred to as 
``improved"-$\epsilon$ expansion.
The estimates are in good agreement with the other results.
Notice also that the large discrepancy between the two different 
resummations of the unconstrained $\epsilon$-expansion is here 
significantly reduced.

\subsection{Strong-coupling analysis}
\label{sec2sub3}

In this subsection we evaluate some of the quantities introduced
in Sec.~\ref{sec2sub1} by analyzing the strong-coupling expansion of
the two-point function $G(x)$
in lattice ${\rm O}(N)$ non-linear $\sigma$ models. We recall
that non-linear $\sigma$ models and $\phi^4$ models possessing
the same internal symmetry ${\rm O}(N)$
describe the same critical behavior, thus giving rise to 
the same universal two-point function $\widehat{g}_0(y)$
in the critical limit $M_G\to 0$.

By employing a character-like approach~\cite{noiMelbourne}, we have 
calculated the strong-coupling expansion of $G(x)$
up to 15th order on the cubic lattice and 21st order on the diamond lattice
for the corresponding nearest-neighbor formulations.
In App.~\ref{appcube} we present the 
15th order strong-coupling expansion
of $G(x)$ on the cubic lattice.
In App.~\ref{appdia} we report the 21st order
strong-coupling series
of the magnetic susceptibility  and of the second moment of
$G(x)$ on the diamond lattice for $N=1,2,3$.

We mention that longer strong-coupling
series, up to 21st order, of the lowest moments of $G(x)$
on the cubic and b.c.c. lattices have been recently calculated
by a linked-cluster expansion technique,
and an updated analysis of the critical exponents $\gamma$ and $\nu$
has been presented~\cite{Bu-Co}.
For $N=0$ even longer series have been calculated for various 
lattices~\cite{GuttmannO0,Guttmann_di,MacDonald}.

In our strong-coupling analysis,
we took special care in devising improved estimators
for the physical quantities $c_i$ and $S_M$, because 
better estimators can greatly improve the stability of the
extrapolation to the critical point. Our search for optimal estimators
was guided by the large-$N$ limit of lattice ${\rm O}(N)$ $\sigma$ models.

In the large-$N$ limit of ${\rm O}(N)$ $\sigma$ models on  the cubic lattice 
the following exact relations hold in the high-temperature phase,
i.e. for $\beta<\beta_c$,
\begin{eqnarray}
\hat{u}_2^{\infty}(M_G)\equiv \hat{u}_2&=& -{1\over 20}M_G^2, \nonumber \\
\hat{u}_3^{\infty}(M_G)\equiv \hat{u}_3&=& {1\over 840} M_G^4,
\label{cci}
\end{eqnarray}
etc..., which vanish for $T\rightarrow T_c$,
i.e. for $M_G^2\rightarrow 0$, leading to the expected result $c_i=0$.
Similarly on the diamond lattice one obtains
\begin{eqnarray}
\hat{u}_2^{\infty}(M_G)\equiv \hat{u}_2&=&  -{1\over 20}M_G^2, \nonumber \\
\hat{u}_3^{\infty}(M_G)\equiv \hat{u}_3&=&  {1\over 7560} M_G^4 
{ 1 + \case{3}{4}M_G^2\over 1 + \case{1}{12}M_G^2},
\label{dci}
\end{eqnarray}
etc...
The above formulae can be  derived from 
the large-$N$ limit of the two-point function
on the cubic and diamond lattice 
given in App.~\ref{appexlargeN}.

We  introduce the following quantities
\begin{equation}
\bar{u}_i \equiv \hat{u}_i - \hat{u}_i^{\infty}(M_G),
\label{baru}
\end{equation}
whose limits for $T\rightarrow T_c$ are still $c_i$.
At $N=\infty$ $\bar{u}_i$ are optimal  estimators of $c_i$, 
since
\begin{equation}
\bar{u}_i(\beta)=\bar{u}^*_i=c_i=0
\label{baru2}
\end{equation}
for $\beta<\beta_c$, i.e. off-critical corrections are absent.
It turns out that the use of $\bar{u}_i$, beside improving the estimates 
for large values of $N$, leads also to more precise
estimates of $c_i$ at low values of $N$.
Strong-coupling series of $\bar{u}_i$ can be easily obtained
from the strong-coupling expansion of $G(x)$.

On the lattice, in the absence of a strict rotational invariance, 
one may actually
define different estimators of the mass-gap having the same critical
limit.  On the cubic lattice one may consider $\mu$ obtained by the
long-distance behavior of the side wall-wall correlation constructed
with $G(x)$, or equivalently the solution of the equation
\begin{equation}
\widetilde{G}^{-1}(i\mu,0,0)=0.
\end{equation}
In view of a strong-coupling
analysis, it is convenient to use another estimator of the mass-gap
derived from $\mu$~\cite{Fe-Wo,Ta-Fi}:
\begin{equation}
M_{\rm c}^2= 2\left( {\rm cosh} \mu - 1\right),
\label{MM}
\end{equation}
which has an ordinary strong-coupling expansion
($\mu$ has a singular strong-coupling expansion, starting with
$-\ln\beta$).  One can easily check that $M_{\rm c}/\mu\rightarrow 1$
in the critical limit.  A similar quantity $M^2_{\rm d}$ 
can be defined on the diamond lattice, as shown in
App.~\ref{diamondNi} (cfr. Eq.~(\ref{Md})).  One may then
consider the dimensionless ratios $M_{\rm c}^2/M_G^2$ and 
$M_{\rm d}^2/M_G^2$ respectively on the cubic
and diamond lattices, and evaluate their fixed-point
limit $S_M$, which by universality must be the same.

In order to determine the coefficients $c_2$ and $c_3$
of the low-momentum expansion of $\widehat{g}_0(y)$ and the mass-ratio $S_M$, 
we analyzed the strong-coupling series of $\bar{u}_2$ and $\bar{u}_3$ 
(defined in Eq.~(\ref{baru})),
and those of the ratios $M_{\rm c}^2/M_G^2$ and
$M_{\rm d}^2/M_G^2$ respectively on the cubic and diamond lattice.

On the cubic lattice
the available series of $\bar{u}_2$, $\bar{u}_3$ and $M_{\rm c}^2/M_G^2-1$ 
are respectively 
of the form $\beta^4\sum_{i=0}^9 a_i\beta^i$, 
$\beta^3\sum_{i=0}^9 a_i\beta^i$,
and $\beta^6\sum_{i=0}^5 a_i\beta^i$; 
except for $N=1$ where they are of the form 
$\beta^6\sum_{i=0}^7a_i\beta^i$,
$\beta^5\sum_{i=0}^7a_i\beta^i$, and 
$\beta^8\sum_{i=0}^3 a_i\beta^i$.
These series can be derived from the strong-coupling
expansion of $G(x)$ presented in App.~\ref{appcube}.
On the diamond lattice
the available series of $\bar{u}_2$, $\bar{u}_3$ 
and $M_{\rm d}^2/M_G^2-1$ 
are respectively 
of the form $\beta^6\sum_{i=0}^{13} a_i\beta^i$, 
$\beta^5\sum_{i=0}^{13} a_i\beta^i$,
and $\beta^6\sum_{i=0}^9 a_i\beta^i$; 
except for $N=1$ where they are of the form 
$\beta^8\sum_{i=0}^{11}a_i\beta^i$,
$\beta^7\sum_{i=0}^{11}a_i\beta^i$, 
and $\beta^8\sum_{i=0}^7 a_i\beta^i$

We constructed various types of approximants to the above series,
such as Pad\'e approximants (PA's), Dlog-Pad\'e approximants (DPA's) 
and first-order inhomogeneous 
integral approximants (IA's)~\cite{Gutt}. We then 
evaluated them at the critical point $\beta_c$
in order to obtain an estimate of the corresponding 
fixed point value. 
For the cubic lattice and most values of $N$, $\beta_c$ 
is available in the literature from
strong-coupling and numerical Monte Carlo studies (see for example
Refs.~\cite{ONgr,Bu-Co,Guttmann_di,Grassberger,Blote,Hasenbush,Chen-Holm}).
When $\beta_c$ was not known (as in the case of
diamond lattice models for $N>0$),
we estimated it by performing a IA analysis
of the strong-coupling series of the magnetic susceptibility
$\chi=\sum_x G(x)$.
In our analysis errors due to
the uncertainty on the value of $\beta_c$ turned out negligible.
The values of $\beta_c$ used in our calculations are reported in 
Table~\ref{betac}.

In the analysis of a series such as 
$A=\beta^m\sum_{i=0}^n a_i\beta^i$, 
we constructed approximants to the $n$th order series 
$\beta^{-m}A=\sum_{i=0}^n a_i\beta^i$,
and then derived the original quantity from them. 
Given a $n$th order series,
we considered the following quasi-diagonal approximants:

(i) $[l/m]$ PA's with
\begin{eqnarray}
&&l+m \geq n-2,\nonumber \\
&&l,m \geq  {n\over 2} - 2;
\label{paap}
\end{eqnarray}

(ii) $[l/m]$ DPA's  with
\begin{eqnarray}
&&l+m+1\geq n-2,\nonumber \\
&&l,m\geq  {n\over 2} - 2;
\label{dpaap}
\end{eqnarray}

(iii) $[m/l/k]$ IA's  with
\begin{eqnarray}
&&m+l+k+2=n,\nonumber \\
&&\lfloor (n-2)/3 \rfloor -1\leq m,l,k \leq \lceil (n-2)/3\rceil +1. 
\label{iaap}
\end{eqnarray}

In the case (i) and (ii), $l,m$ are the orders of 
the polynomials respectively in the numerator and denominator of the PA
of the series at hand, or of the series of
the logarithmic  derivative in the case of DPA.
In the case of integral approximants,
$m,l,k$ are the orders of the polynomial $Q_m$, $P_l$
and $R_k$ defined by the first-order linear differential equation
\begin{equation}
Q_m(x)f^\prime (x)+ P_l(x)f(x)+R_k(x)= O\left( x^{k+l+m+2}\right),
\label{intapprx}
\end{equation}
whose solution provides an approximant of the series at hand.

In Table~\ref{ancub} we show some details of the 
above-described analysis for the 
series on the cubic lattice and for selected values of $N$,
including those physically interesting.
We report there various estimates as obtained from the corresponding
plain series, and the resummations by PA's, DPA's and IA's.
On the cubic lattice, for $N\leq 8$,
the plain series of $\bar{u}_2$ already provides a good estimate 
of $c_2$, indeed
the values at $\beta_c$ of the series using $i$ terms appear to oscillate
around an approximately constant value. In these cases as an estimate
of $c_2$  we can take the average of the values of $\bar{u}_2$ at
$\beta_c$ obtained using $n$ and $n-1$ terms 
(the errors reported in Table~\ref{ancub} are related to their difference). 
For the $c_2$ corresponding to other values of $N$ and for 
$c_3$ and $S_M$,
such oscillations are not observed but the results from the
plain series are close to their resummations. In these cases
in Table~\ref{ancub} we report just the value 
at $\beta_c$ using all available terms of the series.
From the PA, DPA, IA analysis we take the average of the 
values at $\beta_c$ of 
the non-defective approximants using all available
terms of the series. As an indicative error from each analysis we consider
the square root of the variance around the estimate of the results 
from all non-defective approximants specified above.
PA's, DPA's and IA's are considered defective when
they present spurious singularities close to the real axis for ${\rm
Re} \beta \lesssim \beta_c$. More precisely we considered defective 
the approximants with spurious singularities located 
in the region $0\leq {\rm Re}\beta \leq 1.1 \times \beta_c$
(sometimes $0\leq {\rm Re}\beta \leq 1.2 \times \beta_c$)
 and $|{\rm Im}\beta| \leq 0.5\times \beta_c$. Anyway, 
our final results turned out to be quite insensitive to this choice. 
Most of the approximants
we considered turned out non-defective.
Similarly in Table~\ref{andia} we report results of the analysis
of the series on the diamond lattice. In this case we do not 
report estimates from the plain series because 
they appear not to be reliable and very far from their resummations.

Table~\ref{summary} summarizes our calculations.
The final estimates of $c_2$, $c_3$ and $S_M$
reported in Table~\ref{summary} synthetize
the results from all the analyses we performed, and
the reported errors are a rough estimate of the uncertainty.
Final results are rather accurate taking into account
the smallness of the effect we are looking at.
Universality among the cubic and diamond lattices is in
all cases well verified and gives further support to our final
estimates. 
Our results are in good agreement with the estimates obtained from the 
$g$- and $1/N$ expansion. Only for $c_2$ and small values of $N$ 
one notices a small discrepancy, probably due to 
confluent singularities,
which are not properly handled by the approximants we considered.
We also tried different resummation methods \cite{Roskies} 
which take into account
the subleading corrections. However, in this case, most of the approximants 
turned out to be defective and no reliable estimate could be obtained.

Our strong-coupling analysis 
represents a substantial improvement with respect to 
earlier results reported in Ref.~\cite{Ta-Fi}
for the Ising model,
and obtained by an analysis
of the strong-coupling series calculated in Refs.~\cite{Fi-Bu,Ri-Fi}:
$c_2=-5.5(1.5)\times 10^{-4}$,
$c_3=0.05(2)\times 10^{-4}$  on the cubic lattice,
and $c_2=-7.1(1.5)\times 10^{-4}$ and 
$c_3=0.09(3)\times 10^{-4}$  on the b.c.c. lattice.
Other strong-coupling results can be found  in Ref.~\cite{Fe-Wo}.
Our analysis achieves a considerable improvement with respect to
such earlier works
essentially for two reasons:
we use longer series and improved estimators,
see Eq.~(\ref{baru}), which allow a more stable
extrapolation to the critical limit. 
Estimates from the analysis of the strong-coupling
series of the standard variables $\hat{u}_i$, defined 
in Eq.~(\ref{cvj}), are much less precise, although consistent
with those obtained from $\bar{u}_i$.

\subsection{Summary}
\label{sec2last}

We have studied the low-momentum behavior of the two-point function
in the critical limit by considering several approaches:
$1/N$-expansion, $g$-expansion, $\epsilon$-expansion and
strong-coupling expansion. A summary of our results 
can be found in Table~\ref{summary}. 

From the analysis of 
our strong-coupling series we have obtained quite accurate estimates of
the coefficients $c_2$, $c_3$ of the low momentum expansion 
(\ref{lexp}). 
Asymptotic large-$N$ formulae (\ref{ciN}) and (\ref{sm1N})
are clearly approached by our strong-coupling results, but only at rather
large values of $N$. The same behavior was already 
observed for other quantities
like critical exponents~\cite{Zinn} and the zero-momentum 
renormalized four-point coupling~\cite{ONgr}. 
We have also computed the universal function $\hat{g}_0(y)$ in 
the $g$-expansion in fixed dimension to order $O(g^4)$ and in
$\epsilon$-expansion to order $O(\epsilon^3)$.
The corresponding estimates of $c_2$, $c_3$ and $S_M$ are in 
reasonable agreement with the strong-coupling results.

For all values of $N$ the coefficients $c_2$ and $c_3$ turn out
very small and the pattern (\ref{crel}) is verified. 
Furthermore relation
(\ref{sc2}) is satisfied within the precision of our analysis.

The few existing Monte Carlo results 
for the low-momentum behavior of the
two-point Green's function  are consistent with our determinations but
are by far less precise.
Using Refs.~\cite{n0MC1,n0MC2,n0MC3} one estimates $c_2=- 13(17)\times
10^{-4}$ for self-avoiding walks, which correspond to $N=0$.
In Ref.~\cite{Caselle} the authors give a bound
on $\sqrt{S_M}$ for the Ising model ($N=1$), from which 
$-1.2\times 10^{-3}<S_M-1<0$, which must be  compared with our
estimate $S_M-1=-2.5(5)\times 10^{-4}$. 
Monte Carlo simulations of the $XY$ model ($N=2$) shows that 
$S_M\simeq 1$ within  0.1\%~\cite{Hasenbush}, which is consistent 
with our strong-coupling result 
$S_M-1\simeq -3.5(5)\times 10^{-4}$.

We can conclude that in the critical
region of the symmetric phase
the two-point Green's function for all $N$ from zero to infinity 
is almost Gaussian
in a large region around $k^2=0$, i.e., $|k^2/M_G^2|\lesssim 1$. The
small corrections to Gaussian behavior are dominated
by the $(k^2)^2$ term in the expansion of the inverse propagator.
Via the relation (\ref{gamma}) such low-momentum behavior could be probed 
by scattering esperiments by observing the low-angle variation
of intensity.
A similar low-momentum behavior of the two-point correlation function 
has been found in two-dimensional ${\rm O}(N)$ 
models~\cite{ON-d2-b,d2letter,ON-d2-a}.
Substantial differences from Gaussian behavior appear at
sufficiently large momenta, where $\widetilde{G}(k)$ behaves as
$1/k^{2-\eta}$ with $\eta\neq 0$ (although $\eta$ is rather small:
$\eta\simeq 0.03$ for $0\leq N\leq 3$).

\section{Anisotropy of $\protect\bbox{G(\lowercase{x})}$ 
at low momentum and in the critical regime}
\label{sec3}

In this Section we will study anisotropic effects on the 
two-point function due to the lattice structure. We will 
mainly consider three-dimensional lattices with cubic symmetry.
However, whenever
possible, we will give expressions for general $d$-dimensional
lattices with hypercubic symmetry, so that one can recover the 
results for the square lattice and compare with perturbative 
series in $d=4-\epsilon$. We will also comment briefly and 
present some results for the triangular, honeycomb and diamond lattices.

\subsection{Notations}
\label{sec3sub0}

In the following subsections we will compute the exponent 
$\rho = 2+\eta-\eta_4 $ defined in Eq. (\ref{rhoexpr}). It can be derived 
directly from Eq. (\ref{etal}) or Eq. (\ref{Gm1expansion}) or by studying the 
weighted moments  $\overline{q}_{4,j} = q_{4,j}/m_0$ where
$q_{4,j}$ is defined in Eq.~(\ref{llm}) and $m_0\equiv \chi$.
Indeed for $M_G\to 0$,
\begin{equation}
\overline{q}_{4,j} \sim M_G^{-4 - 2m + \rho}.
\end{equation}
We will also compute the universal
function $\hat{g}_4(y)$. In particular we will be interested in the 
first terms of its expansion in powers of $y$ around $y=0$:
\begin{equation}
\widehat{g}_4(y)= 1 + \sum_{i=1} d_i y^i,
\label{g2def}
\end{equation}
where $d_i\equiv c_{4,i}$ (cfr. Eq.~(\ref{expansionghat})). The coefficients 
$d_i$ can be easily obtained from the expressions of the moments
$q_{4,m}$. For $M_G\to 0$, we find
\begin{eqnarray}
&&{\overline{q}_{4,1}\over \overline{q}_{4,0}} \longrightarrow
    4 (d+8) \left(1-\case{1}{2}d_1\right) M_G^{-2} ,\nonumber \\
&& {\overline{q}_{4,2}\over \overline{q}_{4,0}} \longrightarrow
    24 (d+8)(d+10) 
    \left(1-\case{2}{3}d_1-\case{2}{3}c_2+\case{1}{3}d_2\right) M_G^{-4},
\label{en9}
\end{eqnarray}
and so on. From (\ref{en9}) it is easy to derive expressions for 
$r_i\equiv u_{4,i}/u_{4,0}$ whose critical limit is $d_i$. In particular
\begin{equation}
r_1 = 2 - {M^2_G\over 2(d+8)} {\overline{q}_{4,1} \over \overline{q}_{4,0}}.
\label{definizioner}
\end{equation}

\subsection{Breaking of rotational invariance in the 
large-$\protect\bbox{N}$ limit}
\label{sec3sub1}

In the large-$N$ limit lattice ${\rm O}(N)$ models become massive
Gaussian theories which can be solved exactly. If one considers 
theories defined on Bravais lattices one has in the large-$N$ limit
\begin{equation}
  \tilde{G}^{-1}(k) = c\beta\, (\overline{k}^2 + M^2_G),
\end{equation}
where $\overline{k}^2$  is defined by Eq. (\ref{overlinek2}). The relation
between $M^2_G$ and $\beta$ is given by the gap equation.
The constant $c$ is lattice-dependent and will not play any role in
the discussion.
Specific examples are the nearest-neighbor Hamiltonians of the 
form (\ref{actionlatt}) for which we have collected in 
Appendix~\ref{appexlargeN} general expressions for various three-dimensional
lattices, the cubic, diamond and f.c.c. lattice. Analogous formulae
for some two-dimensional lattices, the square, triangular,
and honeycomb lattice can be found in Ref.~\cite{ON-d2-a}.
The function  $\overline{k}^2$ has the properties 
mentioned at the beginning of Sect. \ref{sec1sub1} and a multipole
expansion of the type (\ref{expansion-tildeJ}) for lattices 
with cubic symmetry. For other Bravais lattices the only difference is
the presence of different multipole combinations. 
Considering first lattices with (hyper)-cubic symmetry, from 
Eqs. (\ref{expansion-tildeJ}) and (\ref{aj}), we find for $M_G\to 0$
\begin{equation}
\widetilde{G}^{-1}(k) = \, 
  c\beta M^2_G \left[ 1+y +\ M^2_G (e_{2,0} y^2 +
                                     e_{4,0} Q_4(k/M_G))\ldots\right] .
\end{equation}
%
%
Comparing with Eqs. (\ref{gm1}) and (\ref{Gm1expansion}) we get immediately
$\rho=2$ and $\hat{g}_4(y)=1$, i.e. $d_i=0$ for all $i\not=0$.

In the large-$N$ limit one can easily verify the universality properties 
of the ratios defined in Eq. (\ref{Rratios}). Indeed
for generic Hamiltonians in the critical limit $M_G\to 0$
(keeping the dimension of the lattice $d$ generic) we have
\begin{eqnarray}
\overline{m}_{2m} &\longrightarrow& 
   2^m m! \left(\prod_{i=0}^{m-1} (d+2i)\right)\, M_G^{-2m} ,\\
{\overline{q}_{4,m}\over \overline{q}_{4,0}}&\longrightarrow &
2^m (m+1)! \left(\prod_{i=0}^{m-1}(d+8+2i)\right)\, M_G^{-2m} , 
\label{q4suq0}
\end{eqnarray}
and 
\begin{equation}
\overline{q}_{4,0} \longrightarrow - e_{4,0} {24 d(d-1)\over d+2} M_G^{-2} .
\end{equation}
Notice that 
the only dependence on the specific Hamiltonian is in the expression 
of $\overline{q}_{4,0}$. 
(Exact expressions for 
some of these quantities are reported for the theory with nearest-neighbor
interactions on the cubic, diamond and f.c.c. lattice in Table~\ref{momd3}
and on the square lattice in Table~\ref{momd2}).
Universality is then a straightforward consequence of the 
independence of the ratio (\ref{q4suq0}) 
from $e_{4,0}$. It should also be noticed
that 
\begin{equation}
A_{4,m}\equiv { \overline{q}_{4,m}\over \overline{m}_{4+2m} }\sim M_G^{2}.
\label{iNlm}
\end{equation}
This shows that, as expected,
anisotropic moments are suppressed by two powers of $M_G$ 
in agreement with the prediction $\rho = 2$. 
We stress that the universality of $R_{4,m,n}$ is due the
fact that there is only one leading irrelevant operator
breaking rotational invariance.

It is interesting to notice that such a universality does not hold for 
moments $\overline{q}_{6,m}$ 
(or for $\overline{q}_{2l,m}$ for higher values of $l$) 
because of the mixings we have mentioned in Sec. \ref{sec1sub1}. 
For $\overline{q}_{6,m}$ we have for $T\to T_c$
\begin{equation}{\overline{q}_{6,m}\over \overline{q}_{6,0}}\longrightarrow 
 2^m (m+1)! \left(1 + {e_{4,0}^2\over e_{6,0}} {8m\over d+12}\right)
             \left(\prod_{i=0}^{m-1}(d+12+2i)\right)\, M_G^{-2m},
\end{equation}
which
depends on $e_{6,0}$ and $e_{4,0}^2$, a consequence of the fact 
that $Q_4(k)^2$ contains a term of the form $k^2 Q_6(k)$. Thus 
ratios of the form (\ref{Rratios}) built with 
$\overline{q}_{6,m}$ are not universal.

Let us now consider the diamond lattice. In this case not only 
rotational invariance is broken, but also the parity symmetry.
As the leading anisotropic operator is $O_{4,0}(x)$ the behaviour of the 
leading anisotropic corrections is identical to that we have discussed above.
Therefore $\rho = 2$ also in this case. Moreover the invariant ratios 
$R_{4,m,n}$ are identical for the diamond lattice and for
the other Bravais lattices with cubic symmetry. Eq. (\ref{q4suq0}) is 
exact for the diamond lattice as well.

To discuss parity-breaking effects we must consider odd moments of $G(x)$. 
In particular one finds that, for $M_G\to 0$,
\begin{equation}
\overline{q}_{3,0} \equiv {q_{3,0}\over m_0}\longrightarrow {1\over 6\sqrt{3}},
\label{l3chi}
\end{equation}
where
\begin{equation}
q_{3,0}\equiv \sum xyz \, G(x,y,z).
\label{q30}
\end{equation}
Thus parity-breaking effects vanish as $M_G^3$,, i.e. $\rho_p=3$, faster than 
the anisotropic effects we have considered previously.

Finally let us consider lattices which do not have cubic invariance,
such as the triangular and the honeycomb one.
In Table~\ref{momd2} we report the large-$N$ limit of some of the
lowest spherical and non-spherical moments of $G(x)$ for the models
with nearest-neighbor interactions.

For the triangular lattice one should consider the multipole expansion
(\ref{kbartriangolo}). In this case the leading term breaking
rotational invariance is proportional to $T_6(k)$ and thus we have
$\rho = 4$. This is indeed confirmed by the fact that, for $M_G\to 0$,
\begin{equation}
B_{6,m} \equiv {\overline{t}_{6,m} \over \overline{m}_{6 + 2m}} \sim M^4_G ,
\end{equation}
where $\overline{t}_{6,m} = t_{6,m}/m_0$ and $t_{6,m}$ is defined in
Eq. (\ref{tlm}).
As in the cubic case, it is immediate to verify the universality of ratios of 
the form given in Eq. (\ref{Rratios}) 
with $t_{6,m}$ instead of $q_{4,m}$, which
is a consequence of the uniqueness 
of the leading operator breaking rotational invariance. 
Universality follows from the fact that, for $T\to T_c$,
\begin{equation}
{\overline{t}_{6,m}\over \overline{t}_{6,0}} \longrightarrow
    {2^{2m} (m+1)! (m+5)! \over 5!} M_G^{-2m},
\end{equation}
independently of the specific Hamiltonian.

For the honeycomb lattice one must also consider the breaking of 
parity. Considering the odd moment
\begin{equation}
t_{3,0}=\sum (x^3-3y^2x)G(x,y) ,
\end{equation}
(cfr. Eq. (\ref{BH})) one finds
\begin{equation}
\overline{t}_{3,0}\equiv {t_{3,0}\over m_0}\longrightarrow {1\over2}.
\end{equation}
Thus, as in the diamond case, parity breaking effects vanish as $M_G^3$,
i.e. $\rho_p=3$.

\subsection{Analysis to order $\protect\bbox{1/N}$}
\label{sec3sub2}

In the previous subsection we computed the exponent $\rho$ 
for $N\to \infty$ for lattices with cubic symmetry, finding $\rho = 2$.
Now we want to compute the $1/N$ corrections, i.e. the 
value of $\sigma \equiv \sigma_4 = \eta - \eta_4$
(cfr. Eq. (\ref{rhoexpr})), which is the anomalous dimension of the 
operator $O_{4,0}(x)$. More generally we can compute the exponents 
$\eta_{2l}$ defined in Eq. (\ref{etal}) for arbitrary $l$. Notice that in this 
way we will also obtain the $1/N$ correction to $\rho$ for the 
triangular lattice which depends on $\eta_6$.

In $d$ dimensions, we consider
the following representation of the
inverse two-point function where the $O(1/N)$ correction has been
included 
\begin{equation}
\beta^{-1}\widetilde{G}^{-1}(k)=
\overline{k}^2+ \beta^{-1} Z_G^{-1}M_G^2 + {1\over N}
\int {d^dp\over (2\pi)^d} \Delta(p)
\left( {1\over {\overline{p+k}}^2+M_G^2} 
                 - {1\over \overline{p}^2+M_G^2} \right) .
\label{en1}
\end{equation}
Here
$\overline{k}^2$ is the inverse lattice propagator defined in 
Eq. (\ref{overlinek2}),
\begin{equation}
Z_G^{-1}=\, \left.
   {1\over 2} {\partial^2 \widetilde{G}(k)^{-1}\over \partial k_\mu^2}
\right |_{k_\mu=0}=\, {Z_0\over M_G^2} ,
\label{en2}
\end{equation}
and
\begin{equation}
\Delta^{-1}(p)={1\over 2} 
\int {d^dq\over (2\pi)^d} 
{1\over \left(\overline{q+p}^2+M_G^2 \right)\left(\overline{q}^2+M_G^2\right)}.
\label{en4}
\end{equation}

The following statements can be checked explicitly
in Eq.~(\ref{en1}) and hold to all orders of the $1/N$ expansion:
(i) in the limit $M_G\rightarrow 0$ the function $\widetilde{G}^{-1}(k,M_G)$
is spherically symmetric (i.e. it depends only on $y\equiv k^2/M_G^2$,
apart from an overall factor);
(ii) the only non-spherically symmetric contribution that may appear
in $\widetilde{G}^{-1}(k,M_G)$ to $O(M_G^4)$ can be reduced to a spherically
symmetric function multiplied by $Q_4(k)$.
These statements are simply a consequence  of applying the discrete
and continuous symmetry properties to all integrals appearing in the
asymptotic expansion in $M_G$ of the relevant Feynman integrals. 
They prove to all orders in $1/N$ the validity of the expansion 
(\ref{Gm1expansion}).

To compute the anomalous dimension $\eta_{2l}$  to order $1/N$ 
we will use the trick we explained in Sec. \ref{sec1sub1}. If one considers
a particular Hamiltonian such that $g_{2l}(y,M_G)=0$ for 
$0\le l \le \overline{l}$, then $\widetilde{G}^{-1}(k)$ has an 
expansion of the form (\ref{Gm1expansion}) with 
$\eta_4\to \eta_{2\overline{l}}$
and $\widehat{g}_4(y)\to\widehat{g}_{2\overline{l}}(y)$.
In the $1/N$-expansion, to order $1/N$ this can be achieved 
by considering Hamiltonians such that, for $k\to 0$,
(to simplify the notation
from now on we write $l$ instead of $\overline{l}$)
\begin{equation}
\overline{k}^2 = k^2 + r k^{2l} + O(k^{2l+2})
\end{equation}
where $k^{2l}\equiv \sum_\mu k_\mu^{2l}$.
The limit $M_G\to 0$ can then be easily obtained 
by evaluating massless continuum integrals,
and taking the contribution proportional to $r$,
which is the only term relevant to our computation.
In this limit we obtain 
\begin{eqnarray}
\Delta^{-1}(p)\longrightarrow &&
{1\over 2} \int {d^dq\over (2\pi)^d} 
{1\over \left[ (q+p)^2+r(q+p)^{2l}\right]\left[ q^2+rq^{2l}\right]}\nonumber \\
&& \approx 
{1\over 2} \int {d^dq\over (2\pi)^d} 
{1\over q^2 (q+p)^2} -
r\int {d^dq\over (2\pi)^d} 
{q^{2l}\over (q^2)^2(p+q)^2}\nonumber \\
&& \approx \Delta_0^{-1}(p) 
\left( 1 - r B_l{p^{2l}\over p^2}\right),
\label{en13}
\end{eqnarray}
where
\begin{equation}
\Delta_0^{-1}(p) = {1\over 2} (p^2)^{\case{d}{2}-2}
{\Gamma(2-\case{d}{2})\Gamma(\case{d}{2}-1)^2\over
(4\pi)^{\case{d}{2}}\Gamma(d-2)},
\label{en14}
\end{equation}
\begin{equation}
B_l = (4-d){\Gamma\left(\case{d}{2}+2l-2\right)
\Gamma\left(d-2\right)\over 
\Gamma\left(d+2l-3\right)
\Gamma\left(\case{d}{2}-1\right)},
\label{bl}
\end{equation}
and we have discarded rotationally invariant terms proportional to $r$,
since they will not contribute to the
final result.

We must now identify the singular contribution in the limiting form of
Eq.(\ref{en1}):
\begin{eqnarray}
\hskip -20pt
&&\beta^{-1}\widetilde{G}^{-1}(k)\longrightarrow
k^2+rk^{2l}+
{1\over N} \int {d^dp\over (2\pi)^d} 
\Delta_0(p) \left[ 
1 - rB_l\frac{p^{2l}}{p^2}\right]^{-1}
\left[ (p+k)^2 + r (p+k)^{2l}\right]^{-1}\nonumber \\
\hskip -20pt
&&\quad\approx k^2 + 
{1\over N} \int {d^dp\over (2\pi)^d} {\Delta_0(p)\over
(p+k)^2}  + r\Biggl\{
k^{2l} + {1\over N} \int {d^dp\over (2\pi)^d} {\Delta_0(p)\over
(p+k)^2} \Biggl[ B_l{p^{2l}\over p^2}
- {(p+k)^{2l}\over (p+k)^2}\Biggr] \Biggr\}\nonumber \\
\hskip -20pt
&&\quad\approx 
k^2\left( 1 - {1\over N}\eta_1 \ln k\right) 
+ rk^{2l}
\left( 1 - {1\over N}\eta_{2l,1}\ln k\right).
\label{en15}
\end{eqnarray}
The coefficients $\eta_1$ and $\eta_{2l,1}$ are related 
to the $1/N$ expansion of the exponents $\eta$ and $\eta_{2l}$:
\begin{equation}
\eta= {\eta_1\over N}+O\left({1\over N^2}\right),
\label{etaex}
\end{equation}
and
\begin{equation}
\eta_{2l}= {\eta_{2l,1}\over N}+O\left({1\over N^2}\right).
\label{sigex}
\end{equation}
By simple manipulations one obtains
\begin{equation}
\eta_1 = -{4\Gamma(d-2)\over 
\Gamma(2-\case{d}{2})\Gamma(\case{d}{2}-2)
\Gamma(\case{d}{2}-1)\Gamma(\case{d}{2}+1)},
\label{en16}
\end{equation}
and
\begin{equation}
\eta_{2l,1}= {d(d-2)\over
(d-2+4l)(d-4+4l)}\left[
1 + 2{\Gamma(2l+1)\Gamma(d-2)\over \Gamma(2l+d-3)}
\right]\eta_1.
\label{en17}
\end{equation}
Notice the following properties:
\begin{eqnarray}
\eta_{2,1}&=&\eta_1\qquad{\rm for}\quad{\rm  all}\quad d,\\
\eta_{2l,1}&
\rightarrow&\eta_1\qquad{\rm for} \quad d\rightarrow 2,\label{etad2}\\
\eta_{2l,1}&=&{3\over 4l-1}\eta_1\qquad{\rm for} \quad d=3,\\
\eta_{2l,1}&\rightarrow&{3\over l(2l+1)}\eta_1\qquad{\rm for} \quad
d\rightarrow 4. \label{etad4}
\end{eqnarray}
Therefore for $d=3$ we find
\begin{equation}
\eta_1={8\over 3\pi^2},
\label{en18}
\end{equation}
which agrees with the known result, and
\begin{equation}
\sigma=\eta-\eta_4={32\over 21\pi^2N}
          +O\left({1\over N^2}\right).
\label{en19}
\end{equation}
For $\eta$ also the $O(1/N^2)$ and $O(1/N^3)$ corrections are 
known~\cite{V-P-K}.
The coefficient $\sigma_{1}$ of the $1/N$ expansion is very small.
Thus, at least for $N$ sufficiently
large, say  $N\gtrsim 8$, where the $1/N$ expansion is known to work 
reasonably well,
corrections to the Gaussian value of $\rho$ are very small.
In two dimensions, and to $O(1/N)$, there are no
corrections to the Gaussian value because of Eq.~(\ref{etad2}),
i.e. the first coefficient of the expansion of the anomalous 
dimension is zero to $O(1/N)$.
One might only observe (suppressed)
logarithmic corrections to canonical scaling for all $l$.
It is easy to check in perturbation theory that this holds exactly 
for all $N\ge 3$.

Alternatively the exponents $\eta_{2l}$ could have been computed 
from Eqs. (\ref{gamma2ldef}) and (\ref{etal}).

The computation of the universal function $\widehat{g}_4(y)$ is 
particularly involved and is presented in App.~\ref{app1oncalc}. In this 
Section we will only give the values of the coefficients $d_i$ which appear
in the low-momentum expansion (cfr. Eq. (\ref{g2def})).
We found 
\begin{eqnarray} 
d_1&=&-{0.00206468...\over N}+O\left({1\over N^2}\right),\label{s1on}
\\
d_2&=&{0.00007378...\over N}+O\left({1\over N^2}\right),\nonumber \\
d_3&=&-{0.00000424...\over N}+O\left({1\over N^2}\right),
\nonumber 
\end{eqnarray}
etc....

\subsection{$\protect\bbox{g}$-expansion analysis}
\label{sec3sub2c}

The critical exponent $\sigma$ and the scaling
function $\widehat{g}_4(y)$ can also be evaluated in the
$g$-expansion. For this purpose we calculated the 
one-particle irreducible two-point function $\Gamma_{O_4}(k,M_G)$ 
defined in Eq. (\ref{gamma2ldef}).
By a three-loop calculation one finds for the bare correlation function
\begin{eqnarray}
\Gamma_{O_4}(k,M_G) &=& Q_4(k) + g_0^2 {N+2\over 6} J_2(k,M_G) \nonumber \\
&& 
    - g_0^3 {(N+2)(N+8)\over 108} \left[J_{3,1}(k,M_G)+4J_{3,2}(k,M_G)\right] 
        + O(g_0^4)
\end{eqnarray}
where
\begin{eqnarray}
J_2(k,m) &=& \int {d^3 p\over (2\pi)^3}
{Q_4(k-p)A(p,m)\over \left[ (k-p)^2+m^2\right]^2},
\label{JKM2}\\
J_{3,1}(k,m)&=& \int {d^3 p\over (2\pi)^3}
{Q_4(k-p)A(p,m)^2\over \left[ (k-p)^2+m^2\right]^2},
\label{JKM31}\\
J_{3,2}(k,m)&=& \int {d^3 p\over (2\pi)^3}
{A(p,m)A_Q(p,m)\over (k-p)^2+m^2},
\label{JKM32}
\end{eqnarray}
and
\begin{eqnarray}
A(p,m)&=&\int {d^3 q\over (2\pi)^3}
{1\over \left[ q^2+m^2\right] \left[ (q+p)^2+m^2\right]}
= {1\over 4\pi p}{\rm arctan} {p\over 2m},\\
A_Q(p,m) &=& \int {d^3 q\over (2\pi)^3}
{Q_4(q)\over \left[ q^2+m^2\right]^2 \left[ (q+p)^2+m^2\right]}.
\end{eqnarray}
By appropriately renormalizing  $\Gamma_{O_4}(k,M_G)$ at
$k=0$, one derives 
\begin{eqnarray}
&& \overline{Z}_4 = 1 + \bar{g}^2 {N+2\over (N+8)^2} {40 \pi^2\over 3}
Q_4\left( {\partial/\partial k^2}\right) J_2(k,1)|_{k=0}\label{Z4} \\
&&+\bar{g}^3 {N+2\over (N+8)^2} {80 \pi^2\over 9}
Q_4\left( {\partial/\partial k}\right) 
\left[ 3 J_2(k,1) - 4\pi J_{3,1}(k,1) - 16\pi J_{3,2}(k,1)\right]_{k=0} 
+O(\bar{g}^4).\nonumber 
\end{eqnarray}
Defining $Z_{O_4} \equiv \overline{Z}_4/Z_G$ and
using (\ref{zgg}), we obtain 
\begin{equation}
\gamma_{O_4}(\bar{g}) = \beta(\bar{g}) {\partial \ln Z_{O_4}\over \partial
\bar{g}}=\bar{g}^2 {5408\over 25515}{N+2\over (N+8)^2}
\left[ 1 + \bar{g}\times 0.045007+O(\bar{g}^2)\right].
\label{andim}
\end{equation}
The critical exponent $\sigma$ is obtained by evaluating
$\gamma_{O_4}(g)$ at the fixed-point value of the coupling, i.e. 
\begin{equation}
\sigma=\gamma_{O_4}(g^*).
\end{equation}

The scaling function $\widehat{g}_4(y)$ is  obtained from the
zero-momentum renormalized function $\Gamma_{O_4,R}$. Indeed setting 
\begin{equation}
\Gamma_{O_4,R} = Q_4(k)f_4(g,y),
\end{equation} 
we have
\begin{equation}
\widehat{g}_4(y)= \lim_{g\to g^*} f_4(g,y).
\end{equation}
The expansion of $f_4(g,y)$ is 
\begin{eqnarray}
\hskip -5truept
&&f_4(g,y) = 1 + \bar{g}^{2}
{N+2\over (N+8)^2} {40 \pi^2\over 3}
\left[ Q_4\left( {\partial/ \partial k}\right) J_2(k,1) -
Q_4\left( {\partial/ \partial k}\right) J_2(k,1)|_{k=0}\right]
\label{g4gexp} \\
\hskip -5truept
&&+\bar{g}^{3}
{N+2\over (N+8)^2} {80 \pi^2\over 9}
\Bigg\{ Q_4\left( {\partial/ \partial k}\right)
\Big[ 3 J_2(k,1) - 4\pi J_{3,1}(k,1) - 16\pi J_{3,2}(k,1)\Big]
-(k=0)\Bigg\}\nonumber 
+O(\bar{g}^{4}),
\end{eqnarray}

By expanding $f_4(g,y)$ in powers of $y$ around $y=0$, one finds
\begin{equation}
d_i = \bar{g}^{2} {N+2\over (N+8)^2} \bar{d}_i,
\label{digexp}
\end{equation}
and
\begin{eqnarray}
\bar{d}_1 &=&-{380\over 168399}\left[ 1 + \bar{g}\times 0.105400 +
O(\bar{g}^{2})\right] , 
\label{dprimegexp} \\
\bar{d}_2 &=& {3076\over 19702683} \left[ 1 - \bar{g}\times 0.355629 +
O(\bar{g}^{2})\right] , \nonumber \\
\bar{d}_3 &=&-{3112\over 253320210}\left[ 1 - \bar{g}\times 0.696450 +
O(\bar{g}^{2})\right] , \nonumber  
\end{eqnarray}
etc... 

In order to get estimates of $\sigma$ and of the
coefficients $d_i$ from the 
corresponding series, we have employed the procedure outlined
 in Section~\ref{sec2sub2c}. Results for $\sigma$
are reported in Table~\ref{sigma}, and for $d_1$ in Table~\ref{s1}.

\subsection{An $\protect\bbox{\epsilon}$-expansion analysis}
\label{sec3sub2b}

The exponents $\eta_{2l}$ introduced in Sec.~\ref{sec1sub1}
can be evaluated in the $\epsilon$-expansion of the corresponding
$\phi^4$ theory.
A rather simple two-loop calculation gives
\begin{equation}
\eta = {1\over 2} {N+2\over (N+8)^2}\epsilon^2 +O(\epsilon^3)
\label{etaeps}
\end{equation}
and
\begin{equation}
\eta_{2l} = {3\over 2l(2l+1)}{N+2\over (N+8)^2}\epsilon^2 
+O(\epsilon^3).
\label{eta2leps}
\end{equation}
We recall that the exponent $\eta$ is known to $O(\epsilon^4)$~\cite{Zinn}.
For $\sigma$ one then finds
\begin{equation}
\sigma= \eta-\eta_4={7\over 20}{N+2\over (N+8)^2}\epsilon^2 
+O(\epsilon^3).
\label{sigmaeps}
\end{equation}

To compute $\widehat{g}_4(y)$ , we consider 
the renormalized two-point one-particle irreducible function
with an insertion of the operator $O_4(x)$, see (\ref{limiteGamma}).
To order $O(\epsilon^2)$ we find
\begin{equation}
\widehat{g}_4(y) = 1 + \epsilon^2 {N+2\over (N+8)^2}8\pi^4
\left[ Q_4\left( {\partial/ \partial k}\right) J_s(k,1) -
Q_4\left( {\partial/\partial k}\right) J_s(k,1)|_{k=0}\right]
+O(\epsilon^{3}).
\label{g4epsexp}
\end{equation}
The function $J_s(k,m)$ is the finite part of 
the integral
\begin{equation}
J(k,m) = \int {d^d q\over (2\pi)^d}{d^d p\over (2\pi)^d}
{Q_4(k-p)\over \left[ q^2+m^2\right] \left[ (q+p)^2+m^2\right]
\left[ (k-p)^2+m^2\right]^2}.
\label{LKM}
\end{equation}
with the $\overline{MS}$ prescription.
The expansion of $\widehat{g}_4(y)$ in powers of $y$ gives
\begin{equation}
d_i = \epsilon^2 {N+2\over (N+8)^2} \widehat{d}_i + O(\epsilon^3)
\label{diepsexp}
\end{equation}
and
\begin{eqnarray}
\widehat{d}_1 &=&-0.00354500... , \nonumber \\ 
\widehat{d}_2 &=& 0.00011715..., \nonumber \\
\widehat{d}_3 &=&-0.00000599...,\label{dprimeepsexp} 
\end{eqnarray}
etc...

\subsection{A strong-coupling analysis}
\label{sec3sub3}

Anisotropy in the two-point function can be studied at finite $N$
by analyzing the strong-coupling expansion of its lowest
non-spherical moments. 

In order to compute $\sigma$, the correction to the Gaussian
value of $\rho$, we analyze the strong-coupling expansion of the
ratio $q_{4,0}/m_2$, which behaves as
\begin{equation}
{q_{4,0}\over m_2}\sim M_G^\sigma \, \sim (T-T_c)^{\sigma\nu}
\label{b2j}
\end{equation}
for $T\to T_c$.
We recall that in the $1/N$ expansion
$\nu=1+O(1/N)$, and for $N=0,1,2,3$
$\nu\simeq 0.588$, $\nu\simeq 0.630$, $\nu\simeq 0.670$,
$\nu\simeq 0.705$ respectively~\cite{Zinn}.
DPA's and IA's of the available strong-coupling series
of the ratio $q_{4,0}/m_2$ 
 on both cubic and diamond lattice turned out not to be
sufficiently stable to provide satisfactory estimates of $\sigma$
at any finite value of $N$. 

A more satisfactory analysis has been obtained by employing the
so-called critical point renormalization method (CPRM)~\cite{Baker}.
The idea of the CPRM is that from two series $D(x)$ and $E(x)$
which are singular at the same point $x_0$
\begin{eqnarray}
D(x)&=&\sum_i d_i x^i\sim (x_0-x)^{-\delta},\nonumber \\
E(x)&=&\sum_i e_i x^i\sim (x_0-x)^{-\epsilon},
\label{ab}
\end{eqnarray}
one can construct a new series by
\begin{equation}
F(x)=\sum_i {d_i\over e_i}x^i.
\label{cs}
\end{equation}
The function $F(x)$ is singular at $x=1$ and for $x\to 1$ behaves as 
\begin{equation}
F(x)\sim (1-x)^{-\phi}.
\label{css}
\end{equation}
The exponent $\phi$ is related to $\delta$ and $\epsilon$ by
\begin{equation}
\phi = 1+ \delta - \epsilon.
\label{csss}
\end{equation}
Therefore the
analysis of $F(x)$ provides an unbiased estimate of
the difference between the critical exponents of the two functions
$D(x)$ and $E(x)$. Moreover the series $F(x)$ may be analyzed
by employing biased approximants with a singularity at $x_c=1$.

By applying the CPRM to the strong-coupling series of $q_{4,0}$ and $m_2$,
one can extract an unbiased estimate of $\sigma$, by computing the 
exponent $\phi=1-\sigma\nu$ from the resulting series at 
the singularity $x_0=1$. We analyzed this series
by biased IA's, considering those 
indicated in Eq.~(\ref{iaap}). The estimates of $\sigma$
we obtained confirm 
universality between cubic and diamond lattice, 
although the analysis on the diamond
lattice led in general to less stable results.
In Table~\ref{sigma}, for selected values of $N$, 
we report our estimates of $\sigma$,
which are essentially obtained from the analysis on the cubic lattice.
 In order to derive $\sigma$
from $\sigma\nu$, which is the quantity
derived from the strong-coupling analysis,
we have used  the values of $\nu$ available in the literature.  
See e.g. \cite{Bu-Co} for an updated collection of results obtained 
by various numerical and analytic methods.
The errors we report are rough estimates of the uncertainty
obtained by considering the spread of 
all the analyses we performed.
The values of $\sigma$ are very small at all
values of $N$, and at large $N$, say $N\gtrsim 10$,
they are consistent with the corresponding $O(1/N)$ prediction,
cfr. Eq.~(\ref{en19}).
By analyzing separately the strong-coupling series
of $q_{4,0}$ and $m_2$ in the case of $\sigma$, 
and taking the difference
of their exponents, one obtains consistent but less precise
results.

In order to estimate the first non-trivial coefficient $d_1$ of the expansion
of $\widehat{g}_4(y)$, see Eq.~(\ref{g2def}), one may consider 
the quantity $r_1$ defined in Eq. (\ref{definizioner}). However, as 
we did for the analysis of $c_i$ in Sec. \ref{sec2}, 
it is better to consider another quantity $\overline{r}_1$ which is defined 
so that $\overline{r}_1=0$ for $N=\infty$ for all $\beta<\beta_c$.
For the cubic lattice
\begin{equation}
\overline{r}_1 = 2 - {q_{4,1}M_G^2\over 22 q_{4,0}} + {M_G^2\over 22},
\label{r1cub}
\end{equation}
while for the diamond lattice
\begin{equation}
\overline{r}_1 = 2{1 + \case{2}{33}M_G^2 +
\case{1}{528}M_G^4 \over 1+\case{1}{12}M_G^2 }-
{q_{4,1}M_G^2\over 22 q_{4,0}}.  
\label{r1dia}
\end{equation}
In the critical limit ${\overline r}_1\to d_1$.
On the cubic lattice
the available series of $\overline{r}_1$ has
the form $\beta^4\sum_{i=0}^9 a_i\beta^i$,
except for $N=1$ where it has the form 
$\beta^6\sum_{i=0}^7a_i\beta^i$.
These series can be derived from the strong-coupling
expansion of $G(x)$ presented in App.~\ref{appcube}.
On the diamond lattice
the available series of $\overline{r}_1$ 
has the form $\beta^6\sum_{i=0}^{13} a_i\beta^i$, 
except for $N=1$ where it has the form 
$\beta^8\sum_{i=0}^{11}a_i\beta^i$.
The results of the analysis are reported in Table~\ref{s1}. 
Universality between the cubic and diamond lattice is again substantially
verified, although the diamond lattice provides in most cases 
less precise results.
The value of $d_1$ is very small for all
$N$. At large-$N$ the strong-coupling estimate of $d_1$
is in good agreement with the large-$N$ prediction
(\ref{s1on}). The estimates are also in satisfactory agreement 
with the results obtained from the $g$-expansion
and the $\epsilon$-expansion.

We have also obtained estimates of $\sigma_6=\eta-\eta_6$, 
i.e. the anomalous dimension of the irrelevant operator 
\begin{equation}
O_{6}(x)\equiv O_{6,0}(x)=\vec{s}(x)\cdot
Q_6(\partial)\vec{s}(x), 
\label{O6}
\end{equation}
by analyzing the strong-coupling series of
$q_{6,0}$. For the cubic lattice the exponent $\sigma_6$ is
determined from the critical behavior of the ratio $q_{6,0}/m_2$, since
\begin{equation}
{q_{6,0}\over m_2}\sim M_G^{\sigma_6}\sim (T-T_c)^{\sigma_6\nu},
\label{q60m2}
\end{equation}
Notice that Eq. (\ref{q60m2}) is not valid on the diamond lattice, since here
$q_{6,0}$ receives contributions from two operators, from $O_{6}(x)$ and
from the leading irrelevant operator responsible for the 
parity breaking.
Results for various values of $N$ are reported in Table~\ref{sigma6}.
They were obtained by applying the CPRM to the series
of $q_{6,0}$ and $m_2$, and by analyzing the resulting series
by biased IA's. Like $\sigma$, $\sigma_6$ is small
for all values of $N$. At large $N$ our estimates compare
well with 
\begin{equation}
\sigma_6 = {64\over 33 \pi^2 N} + O\left( {1\over N^2}\right). 
\label{sig6}
\end{equation}


Finally we compute $\rho_p$ for the diamond lattice. For a Gaussian
theory $\rho_p=3$ and thus $q_{3,0}\to \hbox{\rm constant}$ for 
$M_G\to 0$.
In general, at finite values of $N$, 
we write $\rho_p=3 + \sigma_p$. The exponent $\sigma_p$ is determined 
from the critical behavior of $\overline{q}_{3,0}$ 
\begin{equation}
\overline{q}_{3,0}\sim M_G^{\sigma_p} .
\label{sigma_p}
\end{equation}
In order to estimate $\sigma_p$, we applied the CPRM to the
series $q_{3,0}$ and $\chi$. 
We found $0\leq \sigma_p\lesssim 0.01$ for all $N\geq 0$.

\subsection{The two-dimensional Ising model}
\label{sec3sub4}

We conclude this section by considering
the two-dimensional Ising model, for which 
we present an argument
showing that the anomalous dimension of the irrelevant
operators breaking rotational invariance is zero.

Let us consider first the square lattice. In this case,
for sufficiently large values of $|x|$ (in units
of the lattice spacing) the asymptotic behavior of $G(x)$ on the
square lattice can be written in the form~\cite{Cheng-Wu}
\begin{equation}
G(x) \approx \, \int {d^2p\over (2\pi)^2} \, e^{ip\cdot x}
{Z(\beta)\over M^2(\beta)+\widehat{p}^2},
\label{eq1is}
\end{equation}
where $\widehat{p}^2 = \sum_\mu 4 \sin^2 (p_\mu/2)$,
\begin{equation}
Z(\beta)=\left[ (1-z^2)^2-4z^2\right]^{1/4} {(1+z^2)^{1/2}\over z},
\label{eq4is}
\end{equation}
and 
\begin{equation}
M^2(\beta)= {(1+z^2)^2\over z(1-z^2)}-4,
\label{eq5is}
\end{equation}
and we have introduced the auxiliary variable 
\begin{equation}
z(\beta)={\rm tanh}\beta.
\label{eq2is}
\end{equation}
This shows that at large distances 
the breaking of rotational invariance is identical to that of the 
massive Gaussian model with nearest-neighbor interactions.
Therefore $\rho=2$ exactly.

This value of $\rho$  is confirmed by a strong-coupling analysis of
the moments $q_{4,m}$ using 
the available 21st order strong-coupling series~\cite{ON-d2-b}.
In particular, on the square lattice we found 
\begin{equation}
\lim_{\beta\rightarrow\beta_c}{q_{4,0}\over m_2} = {1\over 4}
\end{equation}
within an uncertainty of $O(10^{-5})$.

A formula analogous to Eq. (\ref{eq1is}) has been conjectured in 
Ref.~\cite{ON-d2-b} for the Ising model on triangular and honeycomb lattices. 
Thus, also on these lattices, the pattern of breaking of rotation invariance
(and parity in the case of the honeycomb lattice)  
should  be that of the corresponding Gaussian theories,
which have been described in Sec.~\ref{sec3sub1}. 
If the conjecture of Ref.~\cite{ON-d2-b} is correct, 
we have $\rho=4$ for the triangular lattice and $\rho_p=3$ for
the honeycomb lattice.

Again, an analysis of the strong-coupling expansion of $G(x)$
on the triangular and honeycomb lattices supports 
convincingly this conjecture.


\subsection{Summary}
\label{sec3sub5}

For lattice models with $O(N)$ symmetry we studied the problem 
of the recovery of rotational invariance in the critical limit. 
Anisotropic effects vanish as $M_G^\rho$, when $M_G\to 0$.
The universal critical exponent $\rho$, which is related to 
the critical dimension of the leading operator breaking rotational
invariance, turns out to be 2 with very small $N$-dependent corrections 
for the lattices with cubic symmetry.
Notice that this behavior is universal and thus should appear in all 
physical systems which have cubic symmetry. 
The reader should note that $\rho$ is different from the exponent 
$\omega$, which parametrizes the leading correction to
scaling  and which is related to a different 
rotationally-invariant irrelevant operator.
Models defined on lattices with basis, such as the diamond 
lattice show also a breaking of the parity symmetry. We find that
these effects vanish as $M_G^{\rho_p}$, with $\rho_p\approx 3$ 
for all values of $N$.

We have also calculated the 
universal function $\widehat{g}_4(y)$. For $y\lesssim 1$, we find 
$\widehat{g}_4(y) \approx 1$ with very small corrections.

In our study we considered several approaches, 
based on $1/N$, $g$-, $\epsilon$-, and strong-coupling expansions.
All results are in good agreement.

In two dimensions we showed that $\rho=2$ for the square lattice 
for all $N\ge 3$ and $N=1$. We conjecture that this is a general result,
valid for all values of $N$.
Similar arguments apply to the triangular (honeycomb) lattice: 
we conjecture $\rho = 4$ (resp. $\rho_p=3$) for all $N$.

\acknowledgments

We thank Robert Shrock for useful correspondence on the Ising model.
Discussions with Alan Sokal are also gratefully acknowledged.

\appendix

\section{The large-$\protect\bbox{N}$ 
limit on the cubic, f.c.c. and diamond lattice.}
\label{appexlargeN}

In this appendix we present the large-$N$ limit of the two-point function
in ${\rm O}(N)$ $\sigma$ model with nearest-neighbor interaction 
on cubic, f.c.c. and diamond lattices.

\subsection{The cubic lattice}
\label{cubicNi}

The large-$N$ two-point Green's function on the cubic lattice
is~\cite{sqNi}
\begin{equation}
G(\vec{x};\vec{y})=
{1\over \beta}{1\over N_s}\sum_{\vec{k}} 
e^{i\vec{k}\cdot\left(\vec{x}-\vec{y}\right)}
{1\over \widehat{k}^2 +z},
\label{grcubic}
\end{equation}
where
\begin{eqnarray}
&&\widehat{k}^2=\sum_i \widehat{k}_i^2,\nonumber \\
&&\widehat{k}^2_i= 2(1-\cos k_i),
\label{hatk}
\end{eqnarray}
and $N_s$ is the number of sites.
In Eq.~(\ref{grcubic}) $z=M_G^2$, where $M_G$ is the zero-momentum mass scale.
The lowest spherical and non-spherical moments of $G(x)$
are reported in Table~\ref{momd3}.

The relation between $\beta$ and $z$ is determined by the condition
$G(0) = 1$. In the infinite volume limit one has
\begin{equation}
\beta = \int_{-\pi}^{\pi} {d^3k\over (2\pi)^3}
{1\over \widehat{k}^2 +z} .
\label{betacubic}
\end{equation}
An expression in terms of elliptic integrals can be found in 
\cite{Joyce}.
The critical point $\beta_c$ is obtained by setting $z=0$. 
One finds~\cite{Watson}
\begin{equation}
\beta_c={2\over \pi^2} 
\left( 18 + 12\sqrt{2}- 10\sqrt{3} - 7 \sqrt{6}\right)
K\left( 2\sqrt{3} - 2\sqrt{2} - 3 + \sqrt{6}\right)^2= 
0.252731...,
\label{betacsq}
\end{equation}
where $K$ is the complete elliptic integral of the first kind. 
The expression for $\beta_c$ in terms of $\Gamma$-functions 
reported in~\cite{Glasser} (and 
quoted in~\cite{Itzykson-Drouffe}) is unfortunately wrong.

Concerning tha mass-gap estimator introduced in Eq.~(\ref{MM}),
it is easy to check that for $N=\infty$ and for $\beta\leq\beta_c$
\begin{equation}
{M_{\rm c}^2\over M_G^2} = 1,
\end{equation} 
independently of $\beta$.

\subsection{The face-centered cubic lattice}
\label{fccNi}

The f.c.c. lattice is a regular lattice, thus the solution of 
the corresponding Gaussian model does not present difficulties.
Its coordination number is $c=12$ and the volume per site
is $v_s=\case{1}{\sqrt{2}}$ (in unity of $a^3$, where $a$ is the
length of a link).

The sites $\vec{x}$ of a finite periodic f.c.c.
lattice can be represented in cartesian coordinates by
\begin{eqnarray}
&&\vec{x}= l_1\vec{\eta}_1 
+ l_2\vec{\eta}_2+l_3\vec{\eta}_3,\nonumber \\
&& l_i=1,...L_i, \nonumber \\
&& \vec{\eta}_1=\case{1}{\sqrt{2}}\left( 0,1,1\right),\nonumber \\
&& \vec{\eta}_2=\case{1}{\sqrt{2}}\left( 1,0,1\right),\nonumber \\
&& \vec{\eta}_3=\case{1}{\sqrt{2}}\left( 1,1,0\right).
\label{trcoord}
\end{eqnarray}

It is easy to derive the massive Gaussian propagator
and therefore  the
large-$N$ two-point Green's function, which is
\begin{equation}
G(\vec{x};\vec{y})=
{1\over 2 \beta}{1\over N_s}\sum_{\vec{k}} 
e^{i\vec{k}\cdot\left(\vec{x}-\vec{y}\right)}
{1\over \Delta(k) +z},
\label{grfcc}
\end{equation}
where
\begin{equation}
\Delta(k) = 2\left( 3 -
\cos\case{k_1}{\sqrt{2}}\cos\case{k_2}{\sqrt{2}}
-\cos\case{k_2}{\sqrt{2}}\cos\case{k_3}{\sqrt{2}}
-\cos\case{k_3}{\sqrt{2}}\cos\case{k_1}{\sqrt{2}}\right),
\label{defcc}
\end{equation}
and $N_s=L_1L_2L_3$ is the number of sites.
The momenta $\vec{k}$ run over the reciprocal
lattice:
\begin{eqnarray}
&&\vec{k}= {2\pi\over L_1}m_1\vec{\rho}_1 
+ {2\pi\over L_2}m_2\vec{\rho}_2+
{2\pi\over L_3}m_3\vec{\rho}_3,\nonumber \\
&& m_i=1,...L_i, \nonumber \\
&& \vec{\rho}_1=\case{1}{\sqrt{2}}\left( -1,1,1\right),\nonumber \\
&& \vec{\rho}_2=\case{1}{\sqrt{2}}\left( 1,-1,1\right),\nonumber \\
&& \vec{\rho}_3=\case{1}{\sqrt{2}}\left( 1,1,-1\right).
\label{trmom}
\end{eqnarray}
Again $z=M_G^2$.
The lowest moments derived from Eq.~(\ref{grfcc}) are
reported in Table~\ref{momd3}.

The relation between $\beta$ and $z$ is determined by the condition
$G(0) = 1$. In the infinite volume limit one obtains
\begin{equation}
\beta = {1 \over 2}
\int^{\pi}_{-\pi} {d^3 q\over (2\pi)^3}
{1\over 2D(q)+z},
\label{betafcc}
\end{equation}
where
\begin{equation}
D(q) = 
\sin^2\case{q_1}{2}+ \sin^2\case{q_2}{2} + \sin^2\case{q_3}{2}
+ \sin^2\case{q_1-q_2}{2} + \sin^2\case{q_2-q_3}{2} 
+ \sin^2\case{q_3-q_1}{2}.
\label{dqq}
\end{equation}
By setting $z=0$ one finds~\cite{Watson} 
\begin{equation}
\beta_c= {3\Gamma\left( \case{1}{3}\right)\over
2^{20/3} \pi^4}=0.112055....
\end{equation}

It is worth noticing that the function $e_0(k^2)$,
defined in Eq.~(\ref{expansion-tildeJ}), turns out 
to be the same for cubic and f.c.c. lattices, as a consequence
of some trivial symmetries of the angular integration.
Therefore the first few spherical moments are equal as functions
of $M_G^2$ even off-criticality, as shown in Table~\ref{momd3}.

\subsection{The diamond lattice}
\label{diamondNi}

The diamond lattice has coordination number $c=4$. It
is not a regular lattice, because not all the sites
are related by a translation. It consists of two interpenetrating
f.c.c. lattices, and can be regarded as a f.c.c. lattice with a 
two-point basis.  The absence of translation invariance 
causes a few subtleties in the analysis of models
defined on it. One cannot define a 
Fourier transform which diagonalizes the corresponding
Gaussian propagator.
Neverthless, observing that sites at even distance
in the number of lattice links form regular f.c.c. lattices, 
one can define a Fourier-like transformation that partially 
diagonalizes the Gaussian propagator (up to $2\times 2$ matrices).

Setting the lattice spacing (i.e. the length of a link)
$a=1$, the sites $\vec{x}$ of a finite periodic diamond
lattice can be represented in cartesian coordinates by
\begin{eqnarray}
&&\vec{x}= \vec{x}\,'+p\,\vec{\eta}_p
\nonumber \\
&&\vec{x}\,'= l_1\,\vec{\eta}_1 
+ l_2\,\vec{\eta}_2+l_3 \vec{\eta}_3,\nonumber \\
&& l_i=1,...L_i, \;\;\;\;p=0,1,
\label{diacoord}
\end{eqnarray}
where
\begin{eqnarray}
&& \vec{\eta}_p=\case{1}{\sqrt{3}}\left( 1,1,1\right),\nonumber \\
&& \vec{\eta}_1=\case{2}{\sqrt{3}}\left( 0,1,1\right),\nonumber \\
&& \vec{\eta}_2=\case{2}{\sqrt{3}}\left( 1,0,1\right),\nonumber \\
&& \vec{\eta}_3=\case{2}{\sqrt{3}}\left( 1,1,0\right).
\label{diacoord2}
\end{eqnarray}
The total number of sites on the diamond lattice is $N_s=2L_1 L_2 L_3$,
and the volume per site is
$v_s={8\over 3\sqrt{3}}$ (in unit of $a^3$).
The variable $p$ can be interpreted as the parity of the corresponding
lattice site: sites with the same parity are connected by an even
number of links. 

The two sublattices identified by
$\vec{x}_+(l_1,l_2,l_3)\equiv\vec{x}(l_1,l_2,l_3,0)$ 
and $\vec{x}_-(l_1,l_2,l_3)\equiv
\vec{x}(l_1,l_2,l_3,1)$ form two f.c.c. lattices
having $N'_s=N_s/2$ sites.
Each link of the diamond lattice connects sites belonging to different
sublattices.
It is convenient to rewrite a field $\phi(\vec{x})\equiv\phi(l_1,l_2,l_3,p)$
in terms of two new fields $\phi_+(\vec{x}_+)\equiv \phi(\vec{x}_+)$
and $\phi_-(\vec{x}_-)\equiv \phi(\vec{x}_-)$ defined respectively 
on the sublattices
$\vec{x}_+$ and $\vec{x}_-$.
A finite lattice Fourier transform can be consistently defined by
\begin{eqnarray}
&&\phi_\pm(\vec{k})=\sum_{\vec{x}_\pm} e^{i\vec{k}\cdot
\vec{x}_\pm}\,\phi_\pm (\vec{x}_\pm),\nonumber \\
&&\phi_\pm(\vec{x}_\pm)={1\over N'_s}\sum_{\vec{k}} e^{-i\vec{k}\cdot
\vec{x}_\pm}\phi_\pm (\vec{k}),
\label{a3}
\end{eqnarray}
and the set of momenta is
\begin{eqnarray}
&&\vec{k}={2\pi\over L_1}m_1\vec{\rho}_1+{2\pi\over
L_2}m_2\vec{\rho}_2+
{2\pi\over L_3}m_3\vec{\rho}_3,\nonumber \\
&&m_i=1,...L_i,
\label{kk}
\end{eqnarray}
where
\begin{eqnarray}
&& \vec{\rho}_1=\case{\sqrt{3}}{4}\left( -1,1,1\right),\nonumber \\
&& \vec{\rho}_2=\case{\sqrt{3}}{4}\left( 1,-1,1\right),\nonumber \\
&& \vec{\rho}_3=\case{\sqrt{3}}{4}\left( 1,1,-1\right),
\label{a4}
\end{eqnarray}
so that
\begin{equation}
\vec{k}\cdot \vec{x}\,'={2\pi\over L_1}m_1l_1 
+{2\pi\over L_2}m_2l_2+{2\pi\over L_3}m_3l_3.
\label{ooo}
\end{equation}

The large-$N$ limit of the two-point
function is the massive Gaussian propagator 
defined on the same lattice.
By using the Fourier transform (\ref{kk}), straightforward
calculations allow to derive a rather simple expression
of the massive Gaussian propagator, and therefore
of the large-$N$ two-point function:
\begin{eqnarray}
G(\vec{x};\vec{y})=&&G(\vec{x}\,',p_x;\vec{y}\,',p_y)
= \nonumber \\
=&&{3\over 4 \beta}{1\over N'_s}\sum_{\vec{k}} 
e^{i\vec{k}\cdot\left(\vec{x}\,'-\vec{y}\,'\right)}
{1\over \Delta(k)+z\left(1 + {1\over 12}z\right)} 
\left( \matrix{ 1 + {1\over 6}z& e^{-ik_1}H(k)^*\cr
e^{ik_1} H(k)& 1+{1\over 6}z\cr}\right)
\label{greengex}
\end{eqnarray}
where
\begin{equation}
H(k)= \prod_{i=1}^3 \cos\frac{k_i}{\sqrt{3}} - 
i\prod_{i=1}^3 \sin\frac{k_i}{\sqrt{3}},\label{HH}
\end{equation}
and
\begin{eqnarray}
\Delta(k)=&& 3\left( 1 - |H(k)|^2\right)=\nonumber \\
=&&
\frac{3}{4}\left( 3 -
\cos\case{2k_1}{\sqrt{3}}\cos\case{2k_2}{\sqrt{3}}
-\cos\case{2k_2}{\sqrt{3}}\cos\case{2k_3}{\sqrt{3}}
-\cos\case{2k_3}{\sqrt{3}}\cos\case{2k_1}{\sqrt{3}}\right).
\label{deltaex2}
\end{eqnarray}
Notice that $\Delta(k)$ has the same structure of the inverse
propagator of the f.c.c. lattice, cfr. Eq.~(\ref{defcc}).
One can easily verify that $z=M_G^2$, where $M_G$ is the
second-moment mass. Using Eq.~(\ref{greengex}) one can 
derive the expression of the lowest moments of $G(x)$ 
reported in Table~\ref{momd3}.

The relation between $\beta$ and $z$ is determined by the condition
$G(0)=1$. In the infinite volume limit one can write
\begin{equation}
\beta = {3 \over 4}
\int^{\pi}_{-\pi} {d^3q\over (2\pi)^3}
{1+{1\over 6}z\over 
\frac{3}{4}D(q)+z\left(1+{1\over 12}z\right)},
\label{seex}
\end{equation}
where $D(q)$ has been already defined in Eq.~(\ref{dqq}).
Comparing Eqs.~(\ref{seex}) and (\ref{betafcc}) for $z=0$,
one notes that 
$\beta_c^{\rm diamond}=4\beta_c^{\rm f.c.c.}$. Therefore
for the diamond lattice $\beta_c=0.448220...$.

In order to define a mass-gap estimator on the diamond lattice,
one may consider the wall-wall correlation function
defined constructing walls orthogonal to  
$\vec{w}=\case{1}{\sqrt{2}}(-1,1,0)$, which is the  direction
orthogonal to two among the links starting from a site.
We define
\begin{equation}
G_w(t\equiv \vec{x}\cdot\vec{w}) =
\sum_{t={\rm cst}} G(\vec{x})
\label{g2}
\end{equation}
where the sum is performed over all sites with the same
$t\equiv \vec{x}\cdot \vec{w}=\case{2}{\sqrt{3}}(l_1-l_2)$
(the coordinates of the sites $\vec{x}$ are given in Eq.~(\ref{diacoord})).
Using Eq.~(\ref{greengex})
one can easily prove that $G_w(t)$ enjoys the property of exponentiation.
The mass-gap $\mu$ can be
extracted from the long-distance behavior of 
$G_w(t)$. For $t\gg 1$
\begin{equation}
G_w(t)\propto e^{-\mu t}
\label{ldgho}
\end{equation}
In view of a strong-coupling analysis, it is convenient to use
the following quantity
\begin{equation}
M_{\rm d}^2\equiv {4\over 3}
\left( {\rm cosh} \sqrt{\case{3}{2}} \mu - 1\right)
\label{Md}
\end{equation}
which has the property $M_d\to \mu$ for $\mu \to 0$ and has 
a regular strong-coupling series.
In the large-$N$ limit and for $\beta\leq \beta_c$
\begin{equation}
{M_{\rm d}^2\over M_G^2} = 1.
\end{equation}

\section{Perturbative expansion of scaling functions}
\label{appscfun}

In this appendix we present a simple derivation of all
the results that are needed in order to construct
explicitly the $1/N$, and $g$- and $\epsilon$-expansions up to three loops 
presented in Sec.~\ref{sec2}.
Our starting point is the observation that most of the
two- and three-loop calculations needed in the relevant perturbative 
calculations are included, apart from rather trivial
algebraic dependences on $N$, in the one-loop calculation 
of the $1/N$ expansion for the two-point function.
As we shall show, the $1/N$ results can be expanded in $g$ and
$\epsilon$ in order to recover all the corresponding
contributions. Let's therefore start with the evaluation
of the renormalized self-energy to $O(1/N)$ in arbitrary
dimension $d$ and for arbitrary bare coupling $g_0$ in 
the $N$-component $\phi^4$ theory. 

We introduce the dressed composite propagator
(geometric sum of bubble insertions in the $\phi^4$ vertex):
\begin{equation}
\Delta^{-1}(y,g_0) \equiv
\left[ {1\over 2} \int {d^dp\over (2\pi)^d} 
{1\over p^2+m^2}{1\over (p+k)^2+m^2} + {3\over Ng}\right]
m^{4-d}
\equiv \Delta_r^{-1}(y) + {3\over Ng},
\label{x1}
\end{equation}
where $y\equiv k^2/m^2$, and
we have defined the (zero-momentum subtracted)
dimensionless renormalized dressed (inverse) propagator:
\begin{eqnarray}
\Delta_r^{-1}(y) \equiv&&
\Delta^{-1}(y,g_0) - \Delta^{-1}(0,g_0)\nonumber \\&& =
{1\over 2} \int {d^dp\over (2\pi)^d} 
{1\over p^2+m^2}\left[ {1\over (p+k)^2+m^2} - {1\over p^2+m^2}\right],
\label{x2}
\end{eqnarray}
and the four-point (large-$N$) coupling renormalized at zero momentum
\begin{equation}
{3\over Ng} = 
\Delta^{-1}(0,g_0) = {\Gamma(2-\case{d}{2})\over
2(4\pi)^{d/2}} + {3m^{4-d}\over Ng_0}\equiv {\Gamma\left( 2 -
\case{d}{2}\right)\over 2(4\pi)^{d/2}} {N+8\over N\bar{g}},
\label{x3}
\end{equation}
where we have rescaled the coupling for convenience,
generalizing a rather standard three-dimensional prescription.
The integration (\ref{x2}) can be explicitly performed, and 
one obtains
\begin{equation}
\Delta_r^{-1}(y)  = 
{1\over 2} {\Gamma(2-\case{d}{2})\over
(4\pi)^{d/2}} \left[ \left( 1 + {y\over 4}\right)^{d/2-2}
F\left( 2 - {d\over 2}, {1\over 2},{3\over 2}, {y\over y+4}
\right) - 1\right]
\equiv {\Gamma\left(2 - \case{d}{2}\right) 
\over 2(4\pi)^{d/2}} \delta_r(y),
\label{x4}
\end{equation}
which is a regular function of $d$ for all $d\leq 4$.

The renormalized $1/N$ contribution to the self-energy can now be
evaluated by the formal expression
\begin{equation}
\phi_1 (y,g) = 
\sigma(y,g) - \sigma(0,g) - y {\partial\over \partial y}
\sigma (y,g)|_{y=0},
\label{x5}
\end{equation}
\begin{equation}
\sigma(y,g) = m^{2-d} {2(4\pi)^{d/2}\over \Gamma\left( 2 -
\case{d}{2}\right)}
\int {d^d p \over (2\pi)^d}
{\bar{g}\over 1 + \bar{g}\delta_r(p^2/m^2)}
{1\over (p+k)^2 + m^2 },
\label{x6}
\end{equation}
and the subtractions that are symbolically indicated in Eq.~(\ref{x5})
must be done before performing the integration in Eq.~(\ref{x6})
in order to obtain finite quantities in all steps of the derivation.
To this purpose, it is convenient to perform first the angular integration,
by noticing that
\begin{equation}
{2(4\pi)^{d/2}\over \Gamma\left( 2 -
\case{d}{2}\right)}
\int {d^d p \over (2\pi)^d}
{m^{2-d}\over (p+k)^2 + m^2} f(p^2/m^2) = 
2 B\left( \case{d}{2},2-\case{d}{2}\right)^{-1}
\int_0^{\infty} \left( z\right)^{d/2-1} dz f(z) h(z,y)
\label{x7}
\end{equation}
where
\begin{equation}
h(z,y) = {2\over B\left(\case{d-1}{2},\case{1}{2}\right)}
\int_0^\pi d\theta {\left( \sin\theta\right)^{d-1} 
\over z + y + 1 + 2\sqrt{z y}\cos\theta}.
\label{x8}
\end{equation}
The subtraction indicated in Eq.~(\ref{x5})
now simply amounts to replacing in Eq.~(\ref{x6})
\begin{eqnarray}
h(z,y) \longrightarrow &&h(z,y) - h(z,0) - y{\partial \over \partial y}
h(z,y)|_{y=0}\nonumber \\
&&= h(z,y) - {1\over 1+z} + {y\over (1 + z)^2}
- {4zy\over d(1+z)^3}.
\label{x9}
\end{eqnarray}
The relevant results for some integer values of $d$ are:
\begin{eqnarray}
\delta_r(y) &=& {4\over (4+y)} - 1,\nonumber \\
h(z,y) &=& {z + y + 1\over (z + y + 1)^2 - 4zy}
\label{t1}
\end{eqnarray}
for $d=1$;
\begin{eqnarray}
\delta_r(y) &=& {2\over  y \xi}\ln {\xi+1\over \xi-1} -1,
\nonumber \\
h(z,y) &=& \left[ ( z + y + 1)^2 - 4zy \right]^{-1/2}
\label{t2}
\end{eqnarray}
for $d=2$; here $\xi=\sqrt{1 + \case{4}{y}}$;
\begin{eqnarray}
\delta_r(y) &=& {2\over \sqrt{y}}{\rm arctan}
{\sqrt{y}\over 2} - 1,
\nonumber \\
h(z,y) &=& {1\over 4\sqrt{zy}}\ln  
{z + y + 1 + 2\sqrt{zy}\over z + y + 1 - 2\sqrt{zy} }
\label{t3}
\end{eqnarray}
for $d=3$;
\begin{eqnarray}
\lim_{d\rightarrow 4}\Gamma\left(2 - \case{d}{2}\right)
\delta_r(y) &=& -\left( \xi\ln {\xi+1\over \xi-1} -
2\right),
\nonumber \\
\lim_{d\rightarrow 4}h(z,y) 
&=& {1\over 2zy} \left[ z + y + 1 - \sqrt{(z + y + 1 )^2 - 4zy}\right].
\label{t4}
\end{eqnarray}

Eqs.~(\ref{x5}) and (\ref{x6}) are now ready for our purposes.
Let us immediately notice that, since we know the leading large-$N$ fixed
point value of $g$, which happens to coincide with
the $Ng_0\rightarrow \infty$ limit of Eq.~(\ref{x3}),
we may replace
\begin{equation}
\bar{g}\longrightarrow \bar{g}^* = 1,
\label{x10}
\end{equation}
and find the $1/N$ contribution to the scaling
function $\widehat{g}_0(y)$, which in turn is simply the
continuum non-linear $\sigma$ model evaluation 
of the self-energy.
This is the way Eq.~(\ref{g01oN2}) is generated,
by setting $d=3$ in the general expression.
Eq.~(\ref{t4}) shows that, at least in the naive $d\rightarrow 4$
limit, there is no $O(1/N)$-non-Gaussian contribution to the
self-energy
scaling function.

Eq.~(\ref{x6}) is also the starting point for the 
$g$- and $\epsilon$-expansion up to three loops. It     
is indeed straightforward to obtain a representation
of the leading $O(1/N)$ contributions to the self-energy
as a power series in $g$:
\begin{equation}
\phi_1(y,\bar{g}) = 
- \bar{g}^2 \widetilde{\varphi}_2(y) + \bar{g}^3\widetilde{\varphi}_3(y) + 
O(\bar{g}^4),
\label{x11}
\end{equation}
where we have defined the functions
\begin{equation}
\widetilde{\varphi}_n(y) = (-1)^n
{2\over B\left(\case{d}{2},2-\case{d}{2}\right)}
\int_0^\infty  z^{\case{d}{2} -1}dz
[\delta_r(z)]^{n-1} \left[
h(z,y) - {1\over 1+z} + {y\over (1 + z)^2}
- {4zy\over d(1+z)^3}\right],
\label{x12}
\end{equation}
and we exploited the trivial consequence
of the definition Eq.~(\ref{x12}):
$\widetilde{\varphi}_1(y) \equiv 0$.
Restoring the correct dependence on $N$ for arbitrary
(and not only very large) values
of $N$ in front of the functions $\varphi_2$ and
$\varphi_3$ is now simply a combinatorial problem,
whose solution leads to the complete three-loops result
\begin{equation}
f(y,\bar{g}) = 
1 + y + \bar{g}^2 {N+2\over (N+8)^2}\widetilde{\varphi}_2(y)
+ \bar{g}^3 {(N+2)\over (N+8)^2}\widetilde{\varphi}_3(y) +
O(\bar{g}^4)
\label{x13}
\end{equation}
We must keep in mind that the functions $\widetilde{\varphi}_n(y)$
carry a dependence on the dimensionality $d$,
and the scaling function $\widehat{g}_0(y)$
is the value taken by $f(y,\bar{g})$ when evaluated at the fixed point
value $\bar{g}^*$ of the renormalized coupling, where
$\bar{g}^*$ is in turn a function of the dimensionality
and it is obtained by evaluating the zero of the $\beta$-function.
We may now choose two different strategies.
The first simply amounts to fixing $d$ to the
physical value we are interested in and replacing $\bar{g}^*$ with the
numerical value (possibly evaluated by a higher-order expansion
of the $\beta$-function at fixed dimension).
We may however decide to expand the functions $\widetilde{\varphi}_n(y)$
in the parameter $\epsilon\equiv 4-d$ around their value at $d=4$,
perform a similar expansion for the $\beta$-function, then finding
$g^*$ as a series in $\epsilon$~\cite{B-L-Z-2}:
\begin{equation}
g^* = 
\Gamma(d/2)(4\pi)^{d/2}{3\over N+8}\epsilon
\left[ 1 + \epsilon\left( {1\over 2} + 
{3(3N+14)\over (N+8)^2}\right)\right]
+O(\epsilon^3).
\label{x14}
\end{equation}
The functions $\varphi_2(y)$ and $\varphi_3(y)$ we have introduced
in Sec.~\ref{sec2sub2c} are strictly related to
$\widetilde{\varphi}_2(y)$ and $\widetilde{\varphi}_3(y)$
calculated for $d=3$, indeed
\begin{eqnarray}
\varphi_2(y)&=&\widetilde{\varphi}_2(y)|_{d=3}\nonumber \\
\varphi_3(y)&=&\left[\widetilde{\varphi}_3(y)-
2\widetilde{\varphi}_2(y)\right]_{d=3}.
\label{wwphi}
\end{eqnarray}

\section{$\protect\bbox{O(1/N)}$  calculation of 
$\protect\bbox{\widehat{\lowercase{g}}_4(\lowercase{y})}$}
\label{app1oncalc}

In this appendix we present the calculation to order $1/N$ 
of the scaling function $\hat{g}_4(y)$. We will
first study the small-$p$ behaviour of the auxiliary propagator
\begin{equation}
\Delta^{-1}(p)  = {1\over2} 
   \int {d^d q\over (2\pi)^d} {1\over 
   [ \overline{q+p}^2 + M_G^2] [ \overline{q}^2 + M^2_G] } .
\label{Deltam1lattice}
\end{equation}
We are interested in the behaviour for $p\to 0$, $M_G\to 0$ with arbitrary
ratio $p/ M_G$. For $d < 4$, the only case we will 
consider, the leading order is simply given by the continuum expression
\begin{equation}
\Delta^{-1}_0(p)  = {1\over2} 
   \int_{cont} {d^d q\over (2\pi)^d} {1\over 
   \left[ (q+p)^2 + M_G^2\right] ( q^2 + M^2_G) } .
\end{equation}
Here and in the following, when  we will append the subscript ``cont" 
to the integral, we will mean that the integration domain is the 
whole $d$-dimensional space; otherwise the integration should be 
intended over the first Brillouin zone. We want now to compute the first
correction. Using 
\begin{equation}
\overline{q}^2 = q^2 + e_{0,2} (q^2)^2 + e_{4,0} Q_4(q) + 
O(q^6), 
\end{equation}
we could expand (\ref{Deltam1lattice}) in powers of $q$.
This however cannot be the correct answer as the integral one obtains 
in this way is ultraviolet-divergent.
There is however a standard way out. Introduce a fixed cut-off $\Lambda$
and define a sharp-momentum regularized quantity
\begin{eqnarray}
&& I(q,p;M_G,\Lambda)  =  
     {1\over ( (q+p)^2 + M_G^2) ( q^2 + M^2_G)^2 } \times 
\nonumber \\
&& \quad \left[ q^2 + M^2_G - 
   2 (e_{0,2} (q^2)^2 + e_{4,0} Q_4(q) ) + 
   2 e_{0,2} ( q^2 + M^2_G)^2 \theta( (q+p)^2 - \Lambda^2)\right]
\end{eqnarray}
where $\theta(x)$ is the step-function [$\theta(x) = 1$ for $x\ge 0$,
$\theta(x) = 0$ for $x < 0$]. Then we rewrite (\ref{Deltam1lattice}) 
as 
\begin{eqnarray}
\Delta^{-1}(p)  &=& {1\over2} \int_{cont} {d^d q\over (2\pi)^d}
   \left[ 
    {\chi_{latt} (q)\over 
         [ \overline{q+p}^2 + M_G^2] [ \overline{q}^2 + M^2_G] }
   - I(q,p;M_G,\Lambda) \right] \nonumber \\
&& \qquad + {1\over2} \int_{cont} {d^d q\over (2\pi)^d}
       I(q,p;M_G,\Lambda)
\end{eqnarray}
where $\chi_{latt} (q)$ is the characteristic function of the first 
Brillouin zone, i.e. $\chi_{latt} (q) = 1$ if $q\in [-\pi,\pi]^d$,
$\chi_{latt} (q) = 0$ otherwise.
For $p\to 0$ and $M_G \to 0$ (but only for $2\le d < 4$), we can 
simply set $M_G=p=0$ in the first integral thus obtaining a constant
which depends on $\Lambda$ and on the specific 
lattice Hamiltonian. We must now compute the second integral.
A completely straightforward calculation gives
\begin{eqnarray}
{1\over2} \int_{cont} {d^d q\over (2\pi)^d}
       I(q,p;M_G,\Lambda) =&& \Delta^{-1}_0(p) 
 \;- e_{0,2} \left[ 
   \int_{q^2\le \Lambda^2} {d^d q\over (2\pi)^d} {1\over q^2 + M^2_G} 
   - 4 M^2_G \Delta^{-1}_0(p) - M^4_G 
    {\partial \Delta^{-1}_0(p)\over \partial p^2}\right]\nonumber \\
&&- e_{4,0} {Q_4(p)\over (p^2)^2} 
     \left[ A(p^2) + B(p^2) \Delta^{-1}_0(p) \right]
\label{intI}
\end{eqnarray}
where 
\begin{eqnarray}
\hskip -6pt
A(p^2) &=& {(4\pi)^{-d/2} \Gamma(3-d/2)\over 2 (d-1) (d-4)} 
       {p^2 (d+8) + 36 M^2_G \over p^2 + 4 M^2_G}M_G^{d-2}  ,\\
\hskip -6pt
B(p^2) &=& {1\over 8(d-1)} {1\over p^2 + 4 M^2_G} \left [
      (4 - d) (d+2) (p^2)^2 - 8 (d-10) p^2 M^2_G + 144 M^4_G \right]  .
\end{eqnarray}
Summing the two terms we can rewrite the auxiliary propagator as 
\begin{equation}
\Delta^{-1}(p) \approx \Delta^{-1}_0(p) + C(M_G) + 
    e_{0,2} f_0(p^2,M_G) + e_{4,0} Q_4(p) f_4(p^2,M_G), 
\label{auxpropfinale}
\end{equation}
where
\begin{eqnarray}
C(M_G) &=& {1\over2} \int_{cont} {d^d q\over (2\pi)^d} 
    \left[ {\chi_{latt}(q)\over (\overline{q}^2)^2} - 
       {1\over (q^2)^2} + 2 e_{0,2} 
      \left( {1\over q^2} - {1\over q^2 + M_G^2} \right) \right], \\
f_0(p^2,M_G) &=&  - M^2_G \left( 4 \Delta^{-1}_0(p) +
         M^2_G {\partial \Delta^{-1}_0(p)\over \partial p^2}\right), \\
f_4(p^2,M_G) &=& - {1\over (p^2)^2} \left(
              A(p^2) + B(p^2) \Delta^{-1}_0(p) \right).
\end{eqnarray}
The functions $f_0(p^2,M_G)$ and $f_4(p^2,M_G)$ are universal: 
they do not depend on the lattice structure. The lattice dependence is 
contained in the two constants $e_{0,2}$ and $e_{4,0}$ and 
in $C(M_G)$ which depends on the explicit form of the lattice 
hamiltonian.

Let us now consider $\tilde{G}^{-1}(k,M_G)$ to order $1/N$. We will first
compute $Z_4(M_G)\equiv g_4(0,M_G)$ which can be easily obtained from 
\begin{equation}
Z_4(M_G) = {d+2\over 24 d (d-1)} 
     Q\left( {\partial/ \partial k}\right) \tilde{G}^{-1}(k)|_{k=0} .
\end{equation}
Using the expression for $\tilde{G}^{-1}(k)$ to order $1/N$,
cfr. formula (\ref{en1}), we get
\begin{equation}
Z_4(M_G) = e_{4,0} + {1\over N} {d+2\over 24 d(d-1)}
        \int {d^d q\over (2\pi)^d} \Delta(q) \delta z_4(q)
\end{equation}
where
\begin{eqnarray}
\delta z_4(q) &=& 
   - {1\over D(q)^2} \left[ {d-1\over d+2} \sum_\mu w_{\mu\mu\mu\mu} 
          - {3\over d+2} \sum_{\mu\not=\nu} w_{\mu\mu\nu\nu} \right]
\nonumber \\
  && + {8\over D(q)^3} \left[ {d-1\over d+2} \sum_\mu w_{\mu\mu\mu} w_{\mu} 
          - {3\over d+2} \sum_{\mu\not=\nu} w_{\mu\mu\nu} w_{\nu} \right]
\nonumber \\
  && - {36\over D(q)^4} \left[ {d-1\over d+2} \sum_\mu w_{\mu\mu} w_{\mu}^2 
          - {1\over d+2} \sum_{\mu\not=\nu} w_{\mu\mu} w_{\nu}^2 \right]
\nonumber \\
  && + {6\over D(q)^3} \left[ {d-1\over d+2} \sum_\mu w_{\mu\mu}^2
          - {1\over d+2} \sum_{\mu\not=\nu} 
             (w_{\mu\mu} w_{\nu\nu} + 2 w_{\mu\nu}^2) \right]
\nonumber \\
  && + {72\over D(q)^4} {1\over d+2} \sum_{\mu\not=\nu} 
           w_{\mu\nu} w_\mu w_\nu 
\nonumber \\
  && + {24\over D(q)^5} \left[ {d-1\over d+2} \sum_\mu w_\mu^4 -
           {3\over d+2} \sum_{\mu\not=\nu} w_\mu^2 w_\nu^2  \right] \;\; .
\end{eqnarray}
Here $D(q) = \overline{q}^2 + M^2_G$, $w_\mu = \partial_\mu \overline{q}^2$,
$w_{\mu\nu} = \partial_\nu \partial_\mu \overline{q}^2$ and so on.
From this expression one can easily compute the exponent $\eta_4$: 
indeed one must expand $\delta z_4(q)$ and $\Delta(q)$ for $q\to 0$ 
and keep only those terms that behave as $\Delta_0(q) (q^2 + M_G^2)^{-2}$.
We obtain in this way the results (\ref{en8}-\ref{en9}) for $l=2$.

We want now to compute the scaling function $\hat{g}_4(y)$. 
First of all notice that 
\begin{equation} 
g_4(y,M_G) = \, {1\over {\cal N}_4} \int d^d\Omega(\hat{k}) 
      {Q_4(k)\over (k^2)^4} \tilde{G}^{-1} (k,M_G)
\end{equation}
where $d^d\Omega(\hat{k})$ indicates the normalized measure on the 
$(d-1)$-dimensional sphere and we obtained 
\begin{eqnarray}
\int d^d\Omega(\hat{k}) Q_4(k)^2 &=& 
    {24 (d-1)\over (d+2)^2 (d+4) (d+6)} (k^2)^4 \equiv {\cal N}_4
    (k^2)^4,\\
\int d^d\Omega(\hat{k}) Q_6(k)^2 &=& 
    {720 (d-2)(d-1)\over (d+2) (d+4)^2 (d+6)(d+8)^2(d+10)} (k^2)^6 
    \equiv {\cal N}_6
    (k^2)^6.
\label{nnn}
\end{eqnarray}
(We shall not need Eq.~(\ref{nnn}), but we quote it for further reference.)
Using then Eq.~(\ref{en1}) we get
\begin{equation} 
g_4(y,M_G) = \, e_{4_,0} + {1\over N} {1\over {\cal N}_4}
       \int d^d\Omega(\hat{k}) 
      {Q_4(k)\over (k^2)^4} \int {d^d q\over (2\pi)^d}
      {\Delta(q)\over \overline{(q+k)}^2 + M^2_G},
\end{equation}
so that
\begin{eqnarray} 
\hat{g}_4(y) &=& 1 + {1\over N} {1\over e_{4,0} {\cal N}_4} 
         \lim_{M^2_G\to 0} 
        \int d^d\Omega(\hat{k}) {Q_4(k)\over (k^2)^4} \times \nonumber \\
&& \quad \left\{ \int {d^d q\over (2\pi)^d}
  \left[ {\Delta(q)\over \overline{(q+k)}^2 + M^2_G} -
   {d+2\over 24 d(d-1)} Q_4(k) \Delta(q) \delta z_4(q) \right]\right\}.
\end{eqnarray}
Now because of the subtraction, we can take the limit $M_G\to 0$
by expanding the integrand in powers of $q$ and using 
(\ref{auxpropfinale}). We now integrate over the angular variables.
In particular we need to compute the following two integrals:
\begin{eqnarray}
I_1 (k^2,q^2,M^2_G) &=& {1\over {\cal N}_4} 
       \int d^d\Omega(\hat{k}) \int d^d\Omega(\hat{q})
         {Q_4(k) Q_4(q)\over (q+k)^2 + M^2_G}, \\
I_2 (k^2,q^2,M^2_G) &=& {1\over {\cal N}_4}
     \int d^d\Omega(\hat{k}) \int d^d\Omega(\hat{q})
      {Q_4(k) Q_4(q+k)\over ((q+k)^2 + M^2_G)^2}. 
\end{eqnarray}
Using the techniques of Ref.~\cite{CMPS} we get 
\begin{eqnarray}
&& I_1 (k^2,q^2,M^2_G)  = 
   (k^2 q^2)^{3/2} F_{d,4}(z), \\
&& I_2 (k^2,q^2,M^2_G)  =  -
    {(k^2)^3 \over 2 q^2} 
     \left[
     {F'}_{d,0}(z) + 4 \left( {q^2\over k^2}\right)^{1/2} {F'}_{d,1}(z) +
     {6 q^2\over k^2} {F'}_{d,2}(z) + \right. \nonumber \\
&& \qquad    \left. 
   4 \left( {q^2\over k^2}\right)^{3/2} {F'}_{d,3}(z) +
     \left( {q^2\over k^2}\right)^{2} {F'}_{d,4}(z) \right],
\end{eqnarray}
where we have defined:
\begin{eqnarray}
z &=& {q^2 + k^2 + M^2_G \over 2 \sqrt{q^2 k^2}} \;\; ,\\
F_{d,l}(z) &=& {2^{(1-d)/2} l! (d-2)!\over \Gamma((d-1)/2) (d+l-3)!} \,
     (-1)^l e^{-(d-3)\pi i/2}
     (z^2 - 1)^{(d-3)/4} Q_{l+(d-3)/2}^{(d-3)/2} (z) \;\; .
\end{eqnarray}
Here $Q_\mu^\nu(z)$ is the associated Legendre function of the 
second kind (see Ref.~\cite{Gradshteyn}, Sec. 8.7 and 8.8). 
Notice that for $l=0$ the functions $F_{d,0}(z)$ are related to 
$h(z,y)$ defined in (\ref{x8}). Indeed
\begin{equation}
 F_{d,0}(z) = \sqrt{q^2 k^2} \, 
h\left({q^2\over M^2_G},{k^2\over M^2_G}\right).
\end{equation}
In two dimensions it is easy to see that 
\begin{equation}
F_{2,l}(z) = {1\over 2 \sqrt{z^2-1}} \left( \sqrt{z^2-1} - z\right)^l .
\end{equation}
For $d=3$ we have 
\begin{equation}
F_{3,l}(z) = {(-1)^l\over 4} Q_l (z),
\end{equation}
and for $d=4$
\begin{equation}
F_{4,l}(z) = - {1\over l+1} \left( \sqrt{z^2 -1} - z\right)^{l+1} .
\end{equation}
Putting all together, we get finally 
\begin{equation}
\widehat{g}_4(y) = 1 - {1\over N} {1\over y^4} 
   \int {d^d q\over (2\pi)^d} \left[
  \Delta_0(q)^2 f_4(q^2,1) I_1(y,q^2,1) +  
  \Delta_0(q) I_2(y,q^2,1) \, -\hbox{subtr}\right]
\end{equation}
where ``subtr" indicates the integrand computed for $y\to 0$. As expected 
the final result is universal: any dependence on the lattice hamiltonian
has disappeared.

\section{Strong-coupling expansion of $\protect\bbox{G(\lowercase{x})}$  
on the cubic lattice}
\label{appcube}

Presenting $l$-th order strong-coupling results 
for the two-point Green's function would naively imply writing down
as many coefficients as the number of lattice sites that can be
reached by a $l$-step random walk starting from the origin (up to 
discrete lattice symmetries). 
It is interesting to notice the relationship existing between the
number $n_l$ of lattice points (not related by a lattice symmetry)
that lie at a given lattice distance $l$ from the origin and the
number of independent lattice-symmetric functions
$Q_{2m}(k)(k^2)^{l-m}$. One can easily get convinced that, on a hypercubic
lattice, the number of functions $Q_{2m}^{(p)}(k)(k^2)^{l-m}$
is the same as the number of monomials of total degree $l$ in
the $d$ variables $k_i^2$ that are not related by a lattice symmetry
(that is, the number of independent, homogeneous lattice-symmetric
degree-$l$ polynomials in the $k_i^2$). This number in turn is equal
to that of the partitions of $l$ into $d$ ordered non-negative
integers, and this is nothing but the number of independent 
lattice points at a lattice distance $l$ (where ordering insures
independence by elimination of copies obtained by permutation).
As a corollary, the relationship $p_l=n_l-n_{l-1}$
holds for arbitrary $d$ on hypercubic lattices. 
A generating function for $n_l$, for a given value of $d$,
is
\begin{equation}
\sum_{l=0}^\infty n_lt^l=\prod_{n=1}^d {1\over 1-t^m},
\end{equation}
implying the asymptotic behavior 
\begin{equation}
n_l\longrightarrow {l^{d-1}\over d!(d-1)!}.
\end{equation}  
In the case of three-dimensional hypercubic lattices, one
can show that $p_l=\lfloor l/6\rfloor +1$ 
with the exception of $l=6k+1$ in which
case $p_l=k$, while $n_l$ is the integer nearest to $(l+3)^2/12$ and
the sum $\sum_{i={\rm even}}^l n_i$ is the 
integer nearest to $(l+4)^3/72$.
This would mean roughly $(l+4)^3/72$
coefficients for the $l$-th order of the strong-coupling expansion on
the cubic lattice. This number can be sensibly reduced (asymptotically
by a factor 27 on the cubic lattice), without losing any physical
information, by noticing that the inverse two-point function, when
represented in coordinate space, involves only points that can be
reached by a $\lfloor l/3\rfloor$-step random walk. 
This fact can be traced to the 
one-particle irreducible nature of the inverse correlation. As a
matter of fact, instead of the 93 coefficients needed to represent the
15-th order contributions to $G(x)$, only 8 coefficients are enough for
the corresponding contribution to the inverse function $G^{-1}(x)$,
which we construct by the following procedure (a similar representation
was used for the Ising model in a magnetic field in~\cite{Ta-Fi}).

We introduce the equivalence classes of lattice sites under symmetry
transformations and choose a representative $y$ for each class:
whenever $x\sim y$ then $G(x)=G(y)$.
We define the ``form factor'' of the equivalence class
\begin{equation}
Q(y) = \sum_{x\sim y} e^{ipx},
\label{apc1}
\end{equation}
and represent the Fourier transform of $G(x)$ according to
\begin{equation}
\widetilde{G}(p) = \sum_y Q(y) G(y).
\label{apc2}
\end{equation}
The momentum-space inverse Green's function is defined by
\begin{equation}
\widetilde{G}^{-1}(p) \widetilde{G}(p) = 1.
\label{apc3}
\end{equation}
Therefore its inverse Fourier transform enjoys the symmetries of
$G(x)$ and satisfies the relationships
\begin{equation}
\widetilde{G}^{-1}(p) = \sum_x e^{ipx} G^{-1}(x)=\sum_y Q(y)G^{-1}(y).
\label{apc4}
\end{equation}
In practice we exploit the property
\begin{equation}
Q(v)Q(y) = \sum_z n(z;v,y) Q(z)
\label{apc5}
\end{equation}
where 
\begin{equation}
n(z;v,y) = \sum_{u\sim v,x\sim y} \delta_{z,u+x}
\label{apc6}
\end{equation}
are integer numbers, and reduce the problem of evaluating
$G^{-1}(y)$ to that of solving the linear system of equations
\begin{equation}
\sum_v G^{-1}(v)M(v,z) = \delta_{z,0}
\label{apc7}
\end{equation}
where 
\begin{equation}
M(v,z) = \sum_y G(y) n(z;v,y).
\label{apc8}
\end{equation}
When expanding in powers of $\beta$, the system takes a triangular
structure and , as expected, it admits a solution whose non-trivial
terms are only those corresponding to the equivalence classes of sites
that can be reached by $l/3$ random steps. 

Solutions for $G^{-1}(x)$
can be found for arbitrary $N$. In Table~\ref{gxSC} 
we only exhibit $G^{-1}(x)$ for $N=0,1,2,3,4$ and 16.
We choose a representative of the equivalence class by the
prescription $x_1\geq x_2\geq x_3\geq 0$. We may notice as a general
feature that in the class represented by $x_1>1$, $x_2=x_3=0$ the first 
non-trivial contribution is of order $3x_1+2$ ($3x_1+4$ when $N=1$).
When $N=0,1$ a number of seemingly non-trivial coefficients turn out
to be zero.

\section{Strong-coupling series of $\protect\bbox{\chi}$  and
$\protect\bbox{\lowercase{m}_2}$  
on the diamond lattice}
\label{appdia}

On the diamond  lattice we have calculated  the strong-coupling
expansion of $G(x)$ up to 21st order.  In the character-like
approach, the possibility
of reaching larger orders than on the cubic lattice
is related to the smaller coordination
number. However longer series do not necessarily mean that more
precise results can be obtained from their analysis. This
is essentially related to the approach to the asymptotic regime of
the correspoding strong coupling expansion, 
which is expected to occur later on lattices 
with smaller coordination number. The
21th order series on the diamond lattice provide 
estimates of the exponents $\gamma$ and $\nu$
which are, as we shall see for $N=1,2,3$,
 substantially consistent with the results obtained by analyzing
series on cubic-like lattices (see for example Ref.~\cite{Bu-Co}
where series to $O(\beta^{21})$ for the
cubic and b.c.c. lattice have been presented  and analyzed),
but less precise.

In this appendix we report the 21st order strong-coupling series
of $\chi$ and $m_2$ calculated on the diamond lattice, for 
$N=1,2,3$. 27th order strong-coupling series for $N=0$, 
i.e. for the self-avoiding 
walk model, can be found in Ref.~\cite{Guttmann_di}.

\subsection{$\protect\bbox{N=1}$}
\label{neq1}

\begin{eqnarray}
\chi = &&
  1 + 4\,\beta  + 12\,{{\beta }^2} + {{104\,{{\beta }^3}}\over 3} + 
   100\,{{\beta }^4} + {{4328\,{{\beta }^5}}\over {15}} + 
   {{12128\,{{\beta }^6}}\over {15}} + {{711328\,{{\beta }^7}}\over {315}} + 
   {{132452\,{{\beta }^8}}\over {21}} + 
\nonumber \\ && 
   {{49894088\,{{\beta }^9}}\over {2835}} + 
   {{230044448\,{{\beta }^{10}}}\over {4725}} + 
   {{20986492048\,{{\beta }^{11}}}\over {155925}} + 
   {{11593048528\,{{\beta }^{12}}}\over {31185}} + 
\nonumber \\ && 
   {{6239638466896\,{{\beta }^{13}}}\over {6081075}} + 
   {{40044715794736\,{{\beta }^{14}}}\over {14189175}} + 
   {{381115667726672\,{{\beta }^{15}}}\over {49116375}} + 
   {{907261838473556\,{{\beta }^{16}}}\over {42567525}} + 
\nonumber \\ && 
   {{635228192216156408\,{{\beta }^{17}}}\over {10854718875}} + 
   {{5223277546855685888\,{{\beta }^{18}}}\over {32564156625}} + 
   {{815815904018756584288\,{{\beta }^{19}}}\over {1856156927625}} + 
\nonumber \\ && 
   {{744572898253973823856\,{{\beta }^{20}}}\over {618718975875}} + 
   {{642020997051581736673936\,{{\beta }^{21}}}\over {194896477400625}}
+ O(\beta^{22}).
\label{chin1dia}
\end{eqnarray}

\begin{eqnarray}
m_2=&&
  4\,\beta  + 32\,{{\beta }^2} + {{488\,{{\beta }^3}}\over 3} + 
   {{2048\,{{\beta }^4}}\over 3} + {{38888\,{{\beta }^5}}\over {15}} + 
   {{417664\,{{\beta }^6}}\over {45}} + 
   {{10027936\,{{\beta }^7}}\over {315}} + 
   {{33306368\,{{\beta }^8}}\over {315}} + 
\nonumber \\ && 
   {{971601608\,{{\beta }^9}}\over {2835}} + 
   {{15453950464\,{{\beta }^{10}}}\over {14175}} + 
   {{532482065296\,{{\beta }^{11}}}\over {155925}} + 
   {{4939730085376\,{{\beta }^{12}}}\over {467775}} + 
\nonumber \\ && 
   {{196443743845456\,{{\beta }^{13}}}\over {6081075}} + 
   {{4168605624019328\,{{\beta }^{14}}}\over {42567525}} + 
   {{188065240470724112\,{{\beta }^{15}}}\over {638512875}} + 
\nonumber \\ && 
   {{561744708980235008\,{{\beta }^{16}}}\over {638512875}} + 
   {{28352355075085440248\,{{\beta }^{17}}}\over {10854718875}} + 
\nonumber \\ && 
   {{57966531061027107328\,{{\beta }^{18}}}\over {7514805375}} + 
   {{42081281751167641189216\,{{\beta }^{19}}}\over {1856156927625}} + 
\nonumber \\ && 
   {{615759333006052918694656\,{{\beta }^{20}}}\over {9280784638125}} + 
   {{37696556941296724618984336\,{{\beta }^{21}}}\over {194896477400625}}
+ O(\beta^{22}).
\label{m2n1dia}
\end{eqnarray}

We have analyzed the series of $\chi$ by using the $[m/l/k]$ 
first order IA's with
\begin{eqnarray}
&&m+l+k+2=21,\nonumber \\
&&\lfloor (n-2)/3 \rfloor -2\leq m,l,k \leq \lceil (n-2)/3\rceil +2. 
\label{iaap2}
\end{eqnarray}
We have obtained $\beta_c=0.3697(1)$ and
$\gamma=1.238(14)$. An estimate of $\gamma$ can be also
obtained by applying CPRM to the series $\chi^2$ and $\chi$, as
explained in Sec.~\ref{sec3sub3}. By employing biased IA's (those
indicated in (\ref{iaap2}),
one finds $\gamma=1.253(4)$. 
By applying CPRM to the series $m_2$ and $\chi$, and using
biased IA's, one finds $\nu=0.645(4)$. These values of
$\gamma$ and $\nu$ are slithly larger
than the available estimates obtained by other tecniques
(field-theoretical approaches give $\gamma \simeq 1.240$ and
$\nu\simeq 0.630$), or strong
coupling expansion on other lattices, but we would not consider
them inconsistent. One should not forget 
that the error displayed does not take into account possible
systematic errors
(due, for example, to confluent singularities), 
but just the spread of the results of the
various IA's indicated in (\ref{iaap2}).

\subsection{$\protect\bbox{N=2}$}
\label{neq2}

\begin{eqnarray}
\chi = &&
  1 + 4\,\beta  + 12\,{{\beta }^2} + 34\,{{\beta }^3} + 96\,{{\beta }^4} + 
   {{814\,{{\beta }^5}}\over 3} + 743\,{{\beta }^6} + 
   {{24145\,{{\beta }^7}}\over {12}} + {{10925\,{{\beta }^8}}\over 2} + 
\nonumber \\ && 
   {{889703\,{{\beta }^9}}\over {60}} + 
   {{2387483\,{{\beta }^{10}}}\over {60}} + 
   {{22968773\,{{\beta }^{11}}}\over {216}} + 
   {{25617551\,{{\beta }^{12}}}\over {90}} + 
\nonumber \\ && 
   {{11516036093\,{{\beta }^{13}}}\over {15120}} + 
   {{40849680041\,{{\beta }^{14}}}\over {20160}} + 
   {{520550507027\,{{\beta }^{15}}}\over {96768}} + 
\nonumber \\ && 
   {{3457894675397\,{{\beta }^{16}}}\over {241920}} + 
   {{495995794312009\,{{\beta }^{17}}}\over {13063680}} + 
\nonumber \\ && 
   {{2188572410969059\,{{\beta }^{18}}}\over {21772800}} + 
   {{173608313274399461\,{{\beta }^{19}}}\over {653184000}} + 
\nonumber \\ && 
   {{76543471229019871\,{{\beta }^{20}}}\over {108864000}} + 
   {{5344313242348050991\,{{\beta }^{21}}}\over {2874009600}}
+ O(\beta^{22}).
\label{chin2dia}
\end{eqnarray}

\begin{eqnarray}
m_2=&&
  4\,\beta  + 32\,{{\beta }^2} + 162\,{{\beta }^3} + 672\,{{\beta }^4} + 
   {{7534\,{{\beta }^5}}\over 3} + {{26488\,{{\beta }^6}}\over 3} + 
   {{356305\,{{\beta }^7}}\over {12}} + {{289444\,{{\beta }^8}}\over 3} + 
\nonumber \\ && 
   {{18326503\,{{\beta }^9}}\over {60}} + 
   {{42659326\,{{\beta }^{10}}}\over {45}} + 
   {{3125910649\,{{\beta }^{11}}}\over {1080}} + 
   {{1176454982\,{{\beta }^{12}}}\over {135}} + 
\nonumber \\ && 
   {{78423473449\,{{\beta }^{13}}}\over {3024}} + 
   {{577822206313\,{{\beta }^{14}}}\over {7560}} + 
   {{108069034519903\,{{\beta }^{15}}}\over {483840}} + 
\nonumber \\ && 
   {{58770348791597\,{{\beta }^{16}}}\over {90720}} + 
   {{24384512261505001\,{{\beta }^{17}}}\over {13063680}} + 
\nonumber \\ && 
   {{43660988509648999\,{{\beta }^{18}}}\over {8164800}} + 
   {{9954936929950180901\,{{\beta }^{19}}}\over {653184000}} + 
\nonumber \\ && 
   {{1764942584095467281\,{{\beta }^{20}}}\over {40824000}} + 
   {{1754883256361403082267\,{{\beta }^{21}}}\over {14370048000}}
+ O(\beta^{22}).
\label{m2n2dia}
\end{eqnarray}

By performing an IA analysis of the series of $\chi$, 
one finds $\beta_c=0.3845(2)$ and
$\gamma=1.33(2)$. By applying CPRM to the series $\chi^2$ and $\chi$, 
and employing biased IA's, 
one finds $\gamma=1.34(1)$. 
By applying CPRM to the series $m_2$ and $\chi$, and using
biased IA's, one finds $\nu=0.689(8)$. 
These results are substantially consistent
with the available estimates of $\gamma$ obtained on other lattices
and by other approaches (see e.g. Refs.~\cite{Bu-Co} and \cite{Zinn}).

\subsection{$\protect\bbox{N=3}$}
\label{neq3}

\begin{eqnarray}
\chi=&&  1 + 4\,\beta  + 12\,{{\beta }^2} + {{168\,{{\beta }^3}}\over 5} + 
   {{468\,{{\beta }^4}}\over 5} + {{9144\,{{\beta }^5}}\over {35}} +
   {{123456\,{{\beta }^6}}\over {175}} + 
   {{65568\,{{\beta }^7}}\over {35}} + 
   {{873708\,{{\beta }^8}}\over {175}} + 
\nonumber \\ && 
   {{128270568\,{{\beta }^9}}\over {9625}} + 
   {{11818853472\,{{\beta }^{10}}}\over {336875}} +
   {{2007117038928\,{{\beta }^{11}}}\over {21896875}} + 
   {{5262987995856\,{{\beta }^{12}}}\over {21896875}} + 
\nonumber \\ &&  
   {{1973906542032\,{{\beta }^{13}}}\over {3128125}} + 
   {{25696714370736\,{{\beta }^{14}}}\over {15640625}} + 
   {{277395071474138256\,{{\beta }^{15}}}\over {65143203125}} + 
\nonumber \\ && 
   {{5048136975344060076\,{{\beta }^{16}}}\over {456002421875}} + 
   {{1747312876419771883176\,{{\beta }^{17}}}\over {60648322109375}} + 
\nonumber \\ && 
   {{35523883350405253078656\,{{\beta }^{18}}}\over {476522530859375}} + 
   {{3207088211587054727672352\,{{\beta }^{19}}}\over
   {16678288580078125}} +
\nonumber \\ && 
   {{8294186293466843988864336\,{{\beta }^{20}}}\over {16678288580078125}} + 
   {{285612671193686662161552\,{{\beta }^{21}}}\over
   {221862716796875}} + O(\beta^{22}).
\label{chin3dia}
\end{eqnarray}

\begin{eqnarray}
m_2=&&
  4\,\beta  + 32\,{{\beta }^2} + {{808\,{{\beta }^3}}\over 5} + 
   {{3328\,{{\beta }^4}}\over 5} + {{17240\,{{\beta }^5}}\over 7} + 
   {{1498496\,{{\beta }^6}}\over {175}} + 
   {{4978592\,{{\beta }^7}}\over {175}} + 
   {{15959296\,{{\beta }^8}}\over {175}} + 
\nonumber \\ &&  
   {{391158744\,{{\beta }^9}}\over {1375}} + 
   {{292871549952\,{{\beta }^{10}}}\over {336875}} + 
   {{8170771755824\,{{\beta }^{11}}}\over {3128125}} + 
   {{169326765636096\,{{\beta }^{12}}}\over {21896875}} + 
\nonumber \\ &&  
   {{495146153921968\,{{\beta }^{13}}}\over {21896875}} + 
   {{7166586778308992\,{{\beta }^{14}}}\over {109484375}} + 
   {{1747788082514945008\,{{\beta }^{15}}}\over {9306171875}} + 
\nonumber \\ && 
   {{243755148694999429888\,{{\beta }^{16}}}\over {456002421875}} + 
   {{91627122308762345759912\,{{\beta }^{17}}}\over {60648322109375}} + 
\nonumber \\ && 
   {{2022510678813614989101568\,{{\beta }^{18}}}\over {476522530859375}} + 
   {{197787508407138584345236512\,{{\beta }^{19}}}\over 
     {16678288580078125}} + 
\nonumber \\ && 
{{21994072978677629556242688\,
       {{\beta }^{20}}}\over {667131543203125}} +  
   {{34998691725014346631615751056\,{{\beta }^{21}}}\over 
     {383600637341796875}} + O(\beta^{22}).
\label{m2n3dia}
\end{eqnarray}

By performing an IA analysis of the series of $\chi$, 
one finds $\beta_c=0.3951(2)$ and
$\gamma=1.42(2)$. 
We mention that singularities approximately as far to the origin
as $\beta_c$ have been detected by our analysis,
indeed we found two singularities at $\beta\simeq \pm i 0.39$.
By applying CPRM to the series $\chi^2$ and $\chi$, 
and employing biased IA's, 
we obtained  $\gamma=1.42(1)$. 
By applying CPRM to the series $m_2$ and $\chi$, and using
biased IA's, one finds $\nu=0.726(4)$. 
These results are slightly
larger (and less precise)
than the values obtained on other lattices (see
e.g. Ref.~\cite{Bu-Co}, or by other
techniques (see e.g. Ref.~\cite{Zinn}),
but substantially consistent.



\begin{table}
\caption{Numerical values of
the coefficients $h_i^{(n,j)}$ 
defined in Eq.~(\ref{kkk}).
\label{hcoeff}}
\begin{tabular}{ccr@{}lr@{}lr@{}lr@{}l}
\multicolumn{1}{c}{$i$}&
\multicolumn{1}{c}{$h_i^{(2)}$}&
\multicolumn{2}{c}{$h_i^{(3)}$}&
\multicolumn{2}{c}{$h_i^{(4,1)}$}&
\multicolumn{2}{c}{$h_i^{(4,2)}$}&
\multicolumn{2}{c}{$h_i^{(4,3)}$}\\
\tableline \hline
2 & $-\case{2}{405}$ & 0&.00949125 & 0&.000765804 & 
$-$0&.0134856 & $-$0&.0345992 \\
3 & $\case{4}{15309}$ & $-$0&.000612784 & $-$0&.00000341189&
0&.00102554 & 0&.00253490 \\
4 & $-\case{1}{59049}$ & 0&.0000450060 & $-$0&.00000233206 &
$-$0&.0000841861 & $-$0&.00020327 \\
5 & $\case{4}{3247695}$ & $-$0&.0000035762 & 0&.00000035769 &
0&.0000072651 & 0&.00002390 \\
\end{tabular}
\end{table}

\begin{table}
\squeezetable
\caption{
For several values of $N$
and for the cubic and diamond lattice, we report the values of $\beta_c$ 
we used in our strong-coupling calculations. 
We also report the fixed-point value $\bar{g}^*$  of the rescaled
zero-momentum
four-point renormalized coupling, as obtained by  
field-theoretical methods.
\label{betac}}
\begin{tabular}{cr@{}lr@{}lr@{}l}
\multicolumn{1}{c}{$N$}&
\multicolumn{2}{c}{cubic}&
\multicolumn{2}{c}{diamond}&
\multicolumn{2}{c}{$\bar{g}^*$}\\
\tableline \hline
0&$\beta_c=$&0.213492(1)\cite{n0MC2} &
$\beta_c=$&0.34737(1)~\cite{Guttmann_di} & 1&.421(8)\cite{Zinn} \\
1&$\beta_c=$&0.2216544(3)\cite{Blote}& 
$\beta_c=$&0.3697(1) & 1&.416(5)\cite{Zinn} \\
2 &$\beta_c=$&0.22710(1)\cite{Hasenbush}& 
$\beta_c=$&0.3845(2) & 1&.406(4)\cite{Zinn} \\
3&$\beta_c=$&0.231012(12)\cite{Chen-Holm}&
$\beta_c=$&0.3951(2) & 1&.391(4)\cite{Zinn} \\
4&$\beta_c=$&0.23398(2)\cite{Bu-Co}& 
$\beta_c=$&0.4027(2) & 1&.369\cite{Antonenko} \\
8&$\beta_c=$&0.24084(3)\cite{Bu-Co}& 
$\beta_c=$&0.4200(2) & 1&.303\cite{Antonenko} \\
16 &$\beta_c=$&0.24587(6) & 
$\beta_c=$&0.4327(2) & 1&.207\cite{Antonenko}\\
32 &$\beta_c=$&0.2491(1) & $\beta_c=$&0.4401(1) & 1&.122\cite{Antonenko}\\
$\infty $ &$\beta_c=$&0.252731... & $\beta_c=$&0.448220... & 1 &
\end{tabular}
\end{table}

\begin{table}
\squeezetable
\caption{
Estimates of $c_2$, $c_3$,  and $S_M-1$ for selected values of $N$ 
from various analyses of the strong-coupling
series on the cubic lattice (see the text).
An asterisk indicates that most of the approximants are
defective, or, in the
cases where numbers are not shown, that all approximants are defective, so
that no estimate can be extracted.
\label{ancub}}
\begin{tabular}{ccr@{}lr@{}lr@{}lr@{}l}
\multicolumn{1}{r}{}&
\multicolumn{1}{r}{$N$}&
\multicolumn{2}{c}{PS}&
\multicolumn{2}{c}{PA}&
\multicolumn{2}{c}{DPA}&
\multicolumn{2}{c}{IA}\\
\tableline \hline
$10^4 c_2$ &  0   & $-$0&.6(1.8) &$^*-$2&(2) & *& & $-$1&.0(3) \\ 
      &  1   & $-$3&.0(4) & $-$3&.2(5) & $-$3&.0(6) & $-$2&.9(1) \\  
      &  2   & $-$3&.9(3) & $-$4&.1(7) & $-$4&.0(5)& $-$3&.9(1)\\ 
      &  3   & $-$4&.1(1) & $-$4&.5(1.2) & $-$4&(3) & $^*-$3&.7(4)\\ 
      &  4   & $-$4&.06 & $-$4&.4(6) & $-$4&.3(2)& $-$4&.09(3) \\
      &  8   & $-$3&.4(1) & $-$3&.6(6) & $-$3&.59(9) & $-$3&.5(1) \\ 
      &  16  & $-$2&.26   & $-$2&.5(3)     & $-$2&.4(1) & $-$2&.43(7) \\ 
      &  32  & $-$1&.28 & $-$1&.5(2) &$-$1&.46(9) &$-$1&.45(4) \\\hline 

$10^4 c_3$ &  0   &0&.257 & 0&.12(2) & *&       & 0&.125(3) \\ 
      &  1   &0&.170 & 0&.11(4) & 0&.107(3) & 0&.10(2)  \\  
      &  2   & 0&.205 & 0&.11(2) & 0&.11(1) & 0&.11(2) \\ 
      &  3   &0&.212 & 0&.11(2)&0&.12(1) & *& \\ 
      &  4   & 0&.133  & 0&.12(1) & 0&.12(1)  & *& \\
      &  8   & $-$0&.286 & 0&.07(4) & 0&.10(1) & *& \\
      &  16  & $-$0&.582 & 0&.04(3) & 0&.070(5) & *& \\
      &  32  & $-$0&.559 & 0&.02(2) & 0&.041(3) & $^*$0&.03(2) \\\hline

$10^4(S_M-1)$& 0     &$-$2&.03 & 0&(2) & *& &*& \\  
       & 1   &$-$1&.76  & $-$2&(1) & $-$3&(1) & *&  \\  
       & 2   &$-$2&.72  & $-$4&(1) & $^*-$2&(4) & $^*-$3&.3(6)  \\ 
       & 3   &$-$3&.34  & $-$4&(1) & $-$4&(3) & $-$3&.9(3) \\ 
       & 4   & $-$3&.65 & $-$4&.4(9) & $-$4&(3) & $-$3&.9(4) \\ 
       & 8   & $-$3&.63 & $-$3&.9(5) & $-$4&(2) & $-$3&.9(5) \\ 
       & 16  & $-$2&.72 & $-$2&.8(3)&$-$3&(1) & $-$2&.8(2) \\ 
       & 32  & $-$1&.71 & $^*-$1&.8(3)&$-$2&.0(7)  & *& 
\end{tabular}
\end{table}

\begin{table}
\squeezetable
\caption{
Estimates of $c_2$, $c_3$ and $S_M-1$
from various analyses of the strong-coupling series
on the diamond lattice.
An asterisk indicates that most of the approximants considered are
defective, or, in the
cases where numbers are not shown, that all approximants are defective, so
that no estimate can be extracted.
\label{andia}}
\begin{tabular}{ccr@{}lr@{}lr@{}l}
\multicolumn{1}{r}{}&
\multicolumn{1}{r}{$N$}&
\multicolumn{2}{c}{PA}&
\multicolumn{2}{c}{DPA}&
\multicolumn{2}{c}{IA}\\
\tableline \hline
$10^4 c_2$ & 0   & $-$2&(1) & *& &  $-$0&.6(3)\\ 
           & 1   & $-$3&.0(3)  & $-$2&.9(5) & $-$3&.0(2) \\
           & 2   & $-$4&.1(4)  & $-$4&.1(5) & $-$4&.36(4) \\ 
           & 3   & $-$4&.6(3)  & $-$4&.4(3) & $^*-$4&.7(1) \\ 
           & 4   & $-$4&.7(2)  & $^*-$4&.7(3) & $-$4&.8(2) \\ 
           & 8   & $-$4&.0(2)  & $^*-$3&.9(2)    & $-$3&.97(3)\\ 
           & 16  & $-$2&.66(7) &  $^*-$2&.6(1)     & $-$2&.65(2)\\ 
           & 32  & $-$1&.52(3) & $^*-$1&.51(7) & $-$1&.49(5)\\\hline 

$10^4 c_3$ &  0   & 0&.10(3) & *& &  0&.099(6) \\ 
           &  1   & 0&.12(2)   & 0&.08(2) & 0&.11(2) \\ 
           &  2   & 0&.12(4)   & 0&.08(1) & 0&.12(2) \\ 
           &  3   & 0&.12(8)   & 0&.09(1) & 0&.13(2) \\ 
           &  4   & 0&.1(2) & 0&.09(1) & 0&.12(2) \\ 
           &  8   & 0&.2(4)   & $^*$0&.08(2)& 0&.1(1) \\
           &  16  & 0&.1(2)  & 0&.06(2)  & 0&.08(7) \\
           &  32  & 0&.0(1) & 0&.04(1) & *& \\ \hline

$10^4(S_M-1)$& 0   &    0&(1)  & *& &  *& \\ 
             & 1   & $-$2&.3(4)  & $-$2&.2(3) & $-$2&.3(4) \\ 
             & 2   & $-$3&.6(4)  & $-$3&.4(2) &  $-$3&.5(2) \\ 
             & 3   & $-$4&.0(5) & $-$3&.9(3) & $^*-$4&(1) \\
             & 4   & $-$4&.3(7) & $-$4&.1(2) & $^*-$5&(3) \\ 
             & 8   & $-$4&.1(7) &  $-$3&.6(3)& $-$4&.0(3) \\ 
             & 16  & $-$3&.0(4) & $-$2&.4(2)  & $-$2&.8(2)\\ 
             & 32  & $-$1&.9(3) & $-$1&.5(2) &$-$1&.7(2)  

\end{tabular}
\end{table}

\begin{table}
\squeezetable
\caption{
Estimates of the coefficients $c_2$ and $c_3$, and of
the mass-ratio $S_M$, from the 
analysis of the strong-coupling
series of $\bar{u}_2$ and $\bar{u}_3$, 
$M_{\rm c}^2/M_G^2$ (on the cubic lattice) and 
$M_{\rm d}^2/M_G^2$ (on the diamond lattice). 
We report also
results from the $1/N$-expansion, 
from the $g$-expansion
and the $\epsilon$-expansion.
In the latter cases we give two numbers corresponding to the
two  choices: resumming $R(x)$ or $R(x)/x^2$.
For $N=1$ we also give the estimates from an ``improved"
resummation of the $\epsilon$-expansion which takes into account
the exactly known results in two dimensions.
\label{summary}}
\begin{tabular}{ccr@{}lr@{}lr@{}l}
\multicolumn{1}{c}{$N$}&
\multicolumn{1}{c}{}&
\multicolumn{2}{c}{$10^4 c_2$}&
\multicolumn{2}{c}{$10^4 c_3$}&
\multicolumn{2}{c}{$10^4(S_M-1)$}\\
\tableline \hline
0 & cubic   & $-$1&(1) & 0&.12(1)& 0&(2)  \\

  & diamond & $-$1&(1) & 0&.10(1)& 0&(1)\\

  & $g$-expansion & $-$3&.29,$\;$$-$3.63  
& 0&.108,$\;$0.102   & $-$2&.95,$\;$$-$3.50 \\

& $\epsilon$-expansion & $-$2&.48,$\;$$-$4.26  
& 0&.065,$\;$0.114   & $-$2&.55,$\;$$-$4.38 \\\hline

1 & cubic   & $-$3&.0(2) & 0&.10(1) & $-$2&.5(1.0)  \\

  & diamond & $-$3&.0(2) & 0&.10(2) & $-$2&.3(4) \\

  & $g$-expansion & $-$3&.92,$\;$$-$4.27 
& 0&.126,$\;$0.120 & $-$3&.50,$\;$$-$4.12 \\

& $\epsilon$-expansion & $-$3&.06,$\;$$-$4.99 
& 0&.080,$\;$0.134 & $-$3&.14,$\;$$-$5.13 \\

& impr-$\epsilon$-expansion& 
     $-$2&.80,$\;$$-$3.64 & 0&.060,$\;$0.089 & $-$2&.86,$\;$$-$3.73 \\\hline


2 & cubic   & $-$3&.9(2) & 0&.11(1) & $-$3&.5(1.0) \\

  & diamond & $-$4&.1(4) & 0&.10(2) & $-$3&.5(3) \\

  & $g$-expansion & $-$4&.22,$\;$$-$4.54 
& 0&.133,$\;$0.128 &  $-$3&.85,$\;$$-$4.40 \\

  & $\epsilon$-expansion & $-$3&.39,$\;$$-$5.29 
& 0&.089,$\;$0.142 & $-$3&.48,$\;$$-$5.44 \\\hline

3 & cubic   & $-$4&.1(1) & 0&.11(2) & $-$4&.1(4) \\

  & diamond & $-$4&.5(3) & 0&.11(3) & $-$4&.0(4) \\

  & $g$-expansion & $-$4&.29,$\;$$-$4.58 & 0&.134,$\;$0.128 
&  $-$3&.96,$\;$$-$4.45 \\

  & $\epsilon$-expansion & $-$3&.56,$\;$$-$4.55 
& 0&.094,$\;$0.144 &  $-$3&.66,$\;$$-$5.50 \\\hline


4 & cubic   & $-$4&.1(2) & 0&.12(1) & $-$4&(1) \\

  & diamond & $-$4&.7(2) & 0&.10(2) & $-$4&.2(4) \\

  & $g$-expansion & $-$4&.21,$\;$$-$4.46 & 0&.130,$\;$0.125 
& $-$3&.92,$\;$$-$4.34  \\

  & $\epsilon$-expansion & $-$3&.64,$\;$$-$5.28 
& 0&.096,$\;$0.143 & $-$3&.74,$\;$$-$5.43  \\

  & $1/N$-expansion & $-$11&.12  & 0&.336 & $-$11&.48 \\\hline

8 & cubic   & $-$3&.5(1)& 0&.09(2) & $-$3&.8(5) \\

  & diamond & $-$4&.0(1) & 0&.05(5) & $-$3&.8(4) \\

  & $g$-expansion& $-$3&.60,$\;$$-$3.72 & 0&.108,$\;$0.103 
& $-$3&.44,$\;$$-$3.68 \\

  & $\epsilon$-expansion & $-$3&.48,$\;$$-$4.55 
& 0&.093,$\;$0.124 &  $-$3&.58,$\;$$-$4.68 \\

  & $1/N$-expansion  & $-$5&.56  & 0&.118 & $-$5&.74 \\\hline

16& cubic   & $-$2&.4(1) & 0&.06(1) & $-$2&.8(2) \\

  & diamond & $-$2&.65(5)& 0&.05(3) & $-$2&.7(3) \\

  & $g$-expansion & $-$2&.46,$\;$$-$2.49 & 0&.072,$\;$0.069 
&$-$2&.43,$\;$$-$2.52  \\

  & $\epsilon$-expansion & $-$2&.73,$\;$$-$3.19 & 0&.074,$\;$0.088 
&$-$2&.81,$\;$$-$3.28  \\

  & $1/N$-expansion    & $-$2&.78 & 0&.084 & $-$2&.87 \\\hline

32& cubic   & $-$1&.45(5) & 0&.04(1) & $-$1&.8(3)  \\

  & diamond & $-$1&.50(5) & 0&.04(1)  & $-$1&.7(3) \\

  & $g$-expansion & $-$1&.427,$\;$$-$1.429 & 0&.041,$\;$0.040 
& $-$1&.45,$\;$$-$1.48  \\

  & $\epsilon$-expansion & $-$1&.73,$\;$$-$1.84 & 0&.047,$\;$0.052 
& $-$1&.78,$\;$$-$1.90  \\

  & $1/N$-expansion & $-$1&.39 & 0&.042 & $-$1&.43 \\\hline

$\infty$ &  & 0 &                           
  & 0  &                & 0&
\end{tabular}
\end{table}

\begin{table}
\squeezetable
\caption{Three-dimensional ${\rm O}(N)$ $\sigma$ model
with nearest-neighbor interactions:
lowest moments of $G(x)$ at $N=\infty$ on the cubic
and diamond lattice.
\label{momd3}}
\begin{tabular}{cccc}
\multicolumn{1}{c}{moments}&
\multicolumn{1}{c}{cubic}&
\multicolumn{1}{c}{f.c.c.}&
\multicolumn{1}{c}{diamond}\\
\tableline \hline
$\chi$ & $\case{1}{\beta z}$ & $\case{1}{2\beta z}$ & $\case{3}{2\beta z}$ \\

$\overline{m}_2$ & $\case{6}{z}$ & $\case{6}{z}$ & $\case{6}{z}$ \\

$M_G^2$ & $z$ & $z$ & $z$ \\

$\overline{q}_{3,0}$ & 0 & 0 & $\case{1}{6\sqrt{3}}
\left(1+\case{z}{12}\right)^{-1}$  \\

$\overline{m}_4$ & $\case{120}{z^2}\left(1+\case{z}{20}\right)$ & 
$\case{120}{z^2}\left(1+\case{z}{20}\right)$ &
$\case{120}{z^2}\left(1+\case{z}{20}\right)$   \\

$\overline{q}_{4,0}$ & $\case{12}{5z}$ & 
$-\case{3}{5z}$ & $-\case{8}{5z}$ \\

$\overline{q}_{3,1}$ & 0 & 0 & $\sqrt{3}
\left(1+\case{z}{18}+\case{z^2}{216}\right)
\left(1+\case{z}{12}\right)^{-2}$  \\

$\overline{m}_6$ & $\case{5040}{z^3}
\left(1+\case{z}{10}+\case{z^2}{840}\right)$ &  
$\case{5040}{z^3}\left(1+\case{z}{10}+\case{z^2}{840}\right)$ &
$\case{5040}{z^3}
\left(1+\case{11z}{60}+\case{8z^2}{945}+\case{z^3}{10080}\right)$  \\ 

$\overline{q}_{4,1}$ & $\case{528}{5z^2}\left(1+\case{z}{44}\right)$& 
$-\case{132}{5z^2}\left(1+\case{z}{44}\right)$ &
$-\case{352}{5 z^2}\left(1+\case{z}{33}+\case{z^2}{528}\right)
\left(1+\case{z}{12}\right)^{-1}$  \\

$\overline{q}_{6,0}$ & 
$\case{12}{77z}$&
$-\case{39}{154z}$ & 
$-\case{416}{231z}\left(1-\case{z}{78}\right)
\left(1+\case{z}{12}\right)^{-1}$  \\

$\overline{m}_8$ &
$\case{362880}{z^4}\left(1+\case{3z}{20}+\case{11z^2}{2160}
+\case{z^3}{60480}\right)$&  
$\case{362880}{z^4}\left(1+\case{3z}{20}+\case{37z^2}{7560}
+\case{z^3}{60480}\right)$&  
$\case{362880}{z^4}\left(1+\case{3z}{20}+\case{389z^2}{136080}+O(z^3)\right)$\\

$\overline{q}_{4,2}$ &
$\case{41184}{5z^3}\left(1+\case{19z}{286}+\case{z^2}{3432}\right)$&  
$-\case{10296}{5z^3}\left(1+\case{9z}{143}+\case{z^2}{3432}\right)$&  
$-\case{27456}{5z^3}\left(1+\case{z}{858}+\case{139z^2}{30888}+O(z^3)\right)$\\

$\overline{q}_{6,1}$ &
$\case{240}{11z^2}\left(1+\case{z}{140}\right)$&  
$-\case{1110}{77z^2}\left(1+\case{13z}{740}\right)$&  
$-\case{23680}{231z^2}\left(1-\case{129z}{1480}+\case{5z^2}{444}+O(z^3)\right)$
\end{tabular}
\end{table}

\begin{table}
\squeezetable
\caption{Two-dimensional ${\rm O}(N)$ $\sigma$ model
with nearest-neighbor interactions:
lowest moments of $G(x)$ at $N=\infty$ 
on the square, triangular, and honeycomb lattice.
\label{momd2}}
\begin{tabular}{cccc}
\multicolumn{1}{c}{moments}&
\multicolumn{1}{c}{square}&
\multicolumn{1}{c}{triangular}&
\multicolumn{1}{c}{honeycomb}\\
\tableline \hline
$\chi$
&$\case{1}{\beta z}$ & $\case{2}{3\beta z}$ & $\case{4}{3\beta z}$ \\

$\overline{m}_2$&$\case{4}{z}$ & $\case{4}{z}$ & $\case{4}{z}$ \\
$M_G^2$ & $z$ & $z$ &$z$ \\

$\overline{t}_{3,0}$ & 0 & 0 & $\case{1}{2}\left(1+\case{z}{8}\right)^{-1}$ \\

$\overline{m}_4$& $\case{64}{z^2}\left(1+\case{z}{16}\right)$ 
& $\case{64}{z^2}\left(1+\case{z}{16}\right)$ & 
$\case{64}{z^2}\left(1+\case{z}{16}\right)$ \\

$\overline{q}_{4,0}$ & $\case{1}{z}$ & 0 & 0 \\

$\overline{m}_6$& $\case{2304}{z^3}\left(1+\case{z}{8}+
\case{z^2}{576}\right)$ & 
$\case{2304}{z^3}\left(1+\case{z}{8}+\case{z^2}{576}\right)$ &
$\case{2304}{z^3}(1+\case{z}{4}+\case{z^2}{64}+
\case{z^3}{4608})(1+\case{z}{8})^{-1}$ \\

$\overline{q}_{4,1}$ & $\case{40}{z^2}
\left( 1 + \case{z}{40}\right)$ & 0 & 0 \\

$\overline{t}_{6,0}$ & 0 & $-\case{4}{z}$ & $\case{36}{z}
\left(1-\case{z}{72}\right)\left(1+\case{z}{8}\right)^{-1}$ \\
\end{tabular}
\end{table}

\begin{table}
\squeezetable
\caption{
For various values of $N$, we report estimates of
$\sigma$ obtained by our strong-coupling analysis,
from the $1/N$-expansion, from the 
resummation of the $g$-expansion (see Section~\ref{sec2sub2b})
(in this case we give two numbers corresponding to the
two  choices: resumming $R(x)$ or $R(x)/x^2$),
and from the $O(\epsilon^2)$ term of the $\epsilon$-expansion.
In order to derive $\sigma$ from  $\sigma\nu$, which
is what is computed in the strong-coupling analysis, we used 
the following values of $\nu$:
$\nu\simeq 0.59$ for $N=0$;  $\nu\simeq 0.63$ for $N=1$;  
$\nu\simeq 0.67$ for $N=2$;  $\nu\simeq 0.71$ for $N=3$;  
$\nu\simeq 0.74$ for $N=4$;  $\nu\simeq 0.83$ for $N=8$;  
$\nu\simeq 0.91$ for $N=16$; 
$\nu\simeq 0.96$ for $N=32$.
The errors we report for the strong-coupling estimates
take into account all the analyses we performed. 
\label{sigma}}
\begin{tabular}{cr@{}lr@{}lr@{}lr@{}l}
\multicolumn{1}{c}{$N$}&
\multicolumn{2}{c}{s.c.-expansion}&
\multicolumn{2}{c}{$1/N$-expansion}&
\multicolumn{2}{c}{$g$-expansion}&
\multicolumn{2}{c}{$\epsilon$-expansion}\\
\tableline \hline
0 & 0&.00(1) & &    &  0&.0119,$\;$0.0141 & 0&.0109 \\
1 & 0&.01(1) & &    &  0&.0143,$\;$0.0166 & 0&.0130 \\
2 & 0&.02(1) & &    &  0&.0156,$\;$0.0177 & 0&.0140 \\
3 & 0&.03(2) & 0&.0515 & 0&.0160,$\;$0.0179 & 0&.0145 \\
4 & 0&.03(2) & 0&.0386 & 0&.0158,$\;$0.0174 & 0&.0147 \\
8 & 0&.02(1) & 0&.0193 & 0&.0139,$\;$0.0148 & 0&.0137 \\
16 & 0&.009(3)& 0&.0096 & 0&.0098,$\;$0.0109 & 0&.0109 \\
32 & 0&.004(2)& 0&.0048 & 0&.0058,$\;$0.0059 & 0&.0074 
\end{tabular}
\end{table}

\begin{table}
\squeezetable
\caption{
Estimates of
$\sigma_6$, obtained by applying the CPRM to the series
$q_{6,0}$ and $m_2$ on the cubic lattice.
The errors reported in the Table 
take into account all the analyses we  performed. 
\label{sigma6}}
\begin{tabular}{cr@{}lr@{}lr@{}l}
\multicolumn{1}{c}{$N$}&
\multicolumn{2}{c}{s.c.-expansion}&
\multicolumn{2}{c}{$1/N$-expansion}&
\multicolumn{2}{c}{$\epsilon$-expansion}\\
\tableline \hline
0 & 0&.01(1) && & 0&.0134  \\
1 & 0&.03(2) && & 0&.0159 \\
2 & 0&.04(2) && & 0&.0171 \\
3 & 0&.04(2) && & 0&.0177 \\
4 & 0&.036(10) & 0&.0491 & 0&.0178\\
8 & 0&.024(4) & 0&.0245 & 0&.0167\\
16 & 0&.013(2) &0&.0123 & 0&.0134\\
32 & 0&.0065(8) &0&.0061 & 0&.0091 \\
\end{tabular}
\end{table}

\begin{table}
\squeezetable
\caption{
Estimates of $d_1$ 
from the analysis of the strong-coupling
series on the cubic and diamond lattice. The last column is
our final estimate.
An asterisk indicates that most of the approximants we considered are
defective, or, in the
cases where numbers are not shown, that all the approximants are defective, so
that no estimate can be extracted. 
We also report results obtained by resumming the available terms of the
$g$-expansion, from the $1/N$ calculation.
\label{s1}}
\begin{tabular}{ccr@{}lr@{}lr@{}lr@{}l}
\multicolumn{1}{r}{$N$}&
\multicolumn{1}{c}{}&
\multicolumn{2}{c}{$10^4$PA}&
\multicolumn{2}{c}{$10^4$DPA}&
\multicolumn{2}{c}{$10^4$ IA}&
\multicolumn{2}{c}{$10^4 d_1$}\\
\tableline \hline
0 & cubic  & 1&.3(7)   & *&      &$^*$0&.4(9)   & 1&(1) \\  
  & diamond&$-$0&.9(5)   & *&      &$^*-$1&.2(2)  &$-$1&.0(5)  \\
  & $g$-expansion &  &  &  &   &   &       &$-$1&.31,$\;$$-$1.60  \\ \hline  
     
1 & cubic  &$-$1&.6(8)   & *&      &$^*-$1&.7(1)  &$-$1&.7(5) \\  
  & diamond&$-$3&(1)     &$-$3&.1(7) &$^*-$3&.1(3)  &$-$3&(1) \\
  & $g$-expansion &  &  &  &   &   &       &$-$1&.59,$\;$$-$1.89 \\   \hline  

2 & cubic  &$-$2&.2(3)   & *&      & $-$2&.3(1)  &$-$2&.3(2) \\
  & diamond&$-$3&(1)     &$-$2&.8(3) &$^*-$3&.7(9)  &$-$3&(1) \\
  & $g$-expansion &  &  &  &   &   &       &$-$1&.72,$\;$$-$2.01 \\   \hline  

3 & cubic  &$-$2&.4(3)   & *&      & $-$2&.5(1)  &$-$2&.5(2)\\
  & diamond&$-$2&.2(9)   &$-$2&.6(3) & $-$2&.6(7)  &$-$2&.6(3)\\
  & $g$-expansion &  &  &  &   &   &       &$-$1&.77,$\;$$-$2.03\\   \hline  

4 & cubic  &$-$2&.4(3)   & *&      & $-$2&.5(2)  &$-$2&.5(2)\\
  & diamond&$-$5&(3)     &$-$2&.3(3) &$^*-$4&(3)    &$-$2&.5(5)\\
  & $g$-expansion &  &  &  &   &   &       &$-$1&.76,$\;$$-$1.99\\   
  & $1/N$-expansion &  &  &  &   &   &       &$-$5&.12\\\hline       

8 & cubic  &$^*$-2&.0(4)  & *&      & $-$2&.1(2)  &$-$2&.1(2) \\
  & diamond&$-$5&(3)     &$-$2&(2)   &$^*-$4&(1)    &$-$3&(2) \\  
  & $g$-expansion &  &  &  &   &   &       &$-$1&.55,$\;$$-$1.68\\   
  & $1/N$-expansion &  &  &  &   &   &       &$-$2&.56 \\  \hline     

16& cubic  &$-$1&.3(3)   & *&      & $-$1&.4(2)  &$-$1&.4(2)\\  
  & diamond&$-$3&(3)     &$-$1&.2(8) &  *&       &$-$1&.2(8)\\
  & $g$-expansion &  &  &  &   &   &       &$-$1&.10,$\;$$-$1.15\\   
  & $1/N$-expansion &  &  &  &   &   &       &$-$1&.28\\  \hline     

32& cubic  &$-$0&.7(3)   & *&      & $-$0&.7(3)  &$-$0&.7(2) \\  
  & diamond&$-$1&(1)     &$-$0&.4(2) &  *&       &$-$0&.5(3)\\
  & $g$-expansion &  &  &  &   &   &       &$-$0&.66,$\;$$-$0.67 \\   
  & $1/N$-expansion &  &  &  &   &   &       &$-$0&.64\\ \hline

$\infty$ & & & & & & & & 0 & 
\end{tabular}
\end{table}

\begin{table}
\squeezetable
\caption{Coefficients of the strong-coupling expansion
of $G^{-1}(x)$ on the cubic lattice.
The representative of each equivalence class is chosen by 
$x_1\geq x_2\geq x_3\geq 0$. $l$ indicates the order.
\label{gxSC}}
\renewcommand\arraystretch{1.3}

\begin{tabular}{rrrr|cccccc}
\multicolumn{1}{c}{$x_1$} & \multicolumn{1}{c}{$x_2$}
 & \multicolumn{1}{c}{$x_3$} & \multicolumn{1}{c|}{$l$}
 & $N=0$ & $N=1$ & $N=2$ & $N=3$ & $N=4$ & $N=16$ \\
\tableline \hline

0 & 0 & 0 & 0 & $1$ & $1$ & $1$ & $1$ & $1$ & $1$ \\
\hline
1 & 0 & 0 & 1 & $-1$ & $-1$ & $-1$ & $-1$ & $-1$ & $-1$ \\
\hline
0 & 0 & 0 & 2 & $6$ & $6$ & $6$ & $6$ & $6$ & $6$ \\
\hline
1 & 0 & 0 & 3 & $-1$ & $-{2\over 3}$ & 
$-{1\over 2}$ & $-{2\over 5}$ & $-{1\over 3}$ & $-{1\over 9}$ \\
\hline
0 & 0 & 0 & 4 & $30$ & $26$ & $24$ & ${{114}\over 5}$ 
& $22$ & ${{58}\over 3}$ \\
\hline
1 & 0 & 0 & 5 & $-13$ & $-{{122}\over {15}}$ & $-{{35}\over 6}$ 
& $-{{158}\over {35}}$ & $-{{11}\over 3}$ & $-{{49}\over {45}}$ \\
\hline
0 & 0 & 0 & 6 & $366$ & ${{4204}\over {15}}$ & ${{479}\over 2}$ 
& ${{37788}\over {175}}$ & ${{602}\over 3}$ & ${{20674}\over {135}}$ \\
1 & 1 & 0 & 6 & $2$ & $0$ & $-{1\over 2}$ & $-{{16}\over {25}}$ 
& $-{2\over 3}$ & $-{{10}\over {27}}$ \\
\hline
1 & 0 & 0 & 7 & $-197$ & $-{{33604}\over {315}}$ & $-{{1123}\over {16}}$ 
& $-{{44812}\over {875}}$ & $-{{1793}\over {45}}$ 
& $-{{132961}\over {13365}}$ \\
\hline
0 & 0 & 0 & 8 & $5022$ & ${{348266}\over {105}}$ & ${{10265}\over 4}$ 
& ${{1882494}\over {875}}$ & ${{9454}\over 5}$ & ${{5012914}\over {4455}}$ \\
1 & 1 & 0 & 8 & $24$ & $-16$ & $-{{65}\over 3}$ & $-{{18576}\over {875}}$ 
& $-{{176}\over 9}$ & $-{{1912}\over {243}}$ \\
2 & 0 & 0 & 8 & $4$ & $0$ & $-1$ & $-{{32}\over {25}}$ & $-{4\over 3}$ 
& $-{{20}\over {27}}$ \\
\hline
1 & 0 & 0 & 9 & $-2889$ & $-{{3805202}\over {2835}}$ 
& $-{{191503}\over {240}}$ & $-{{5197194}\over {9625}}$ 
& $-{{53443}\over {135}}$ & $-{{2940271}\over {40095}}$ \\
1 & 1 & 1 & 9 & $6$ & $-8$ & $-{{57}\over 4}$ & $-{{408}\over {25}}$ 
& $-{{50}\over 3}$ & $-{{778}\over {81}}$ \\
2 & 1 & 0 & 9 & $-1$ & $0$ & $-{1\over 4}$ & $-{{56}\over {125}}$ 
& $-{5\over 9}$ & $-{{115}\over {243}}$ \\
\hline
0 & 0 & 0 & 10 & $76062$ & ${{211434604}\over {4725}}$ 
& ${{3833513}\over {120}}$ & ${{8484905796}\over {336875}}$ & $21078$ 
& ${{1969903274}\over {200475}}$ \\
1 & 1 & 0 & 10 & $258$ & $-{{952}\over 3}$ & $-{{24851}\over {72}}$ 
& $-{{1878984}\over {6125}}$ & $-{{35786}\over {135}}$ 
& $-{{10005158}\over {120285}}$ \\
2 & 0 & 0 & 10 & $116$ & $-24$ & $-{{583}\over {12}}$ 
& $-{{45272}\over {875}}$ & $-{{1340}\over {27}}$ & $-{{49060}\over {2187}}$ \\
\hline
1 & 0 & 0 & 11 & $-45357$ & $-{{2874597004}\over {155925}}$ 
& $-{{42875903}\over {4320}}$ & $-{{949030894596}\over {153278125}}$ 
& $-{{12000923}\over {2835}}$ & $-{{34414485049}\over {70366725}}$ \\
1 & 1 & 1 & 11 & $72$ & $-240$ & $-{{2635}\over 8}$ 
& $-{{7370064}\over {21875}}$ & $-{{8636}\over {27}}$ & $-{{915928}\over {6561}}$ \\
2 & 1 & 0 & 11 & $-15$ & $-24$ & $-{{133}\over 3}$ 
& $-{{1142304}\over {21875}}$ & $-{{4369}\over {81}}$ 
& $-{{605717}\over {19683}}$ \\
3 & 0 & 0 & 11 & $0$ & $0$ & $-{1\over 2}$ & $-{{96}\over {125}}$ 
& $-{8\over 9}$ & $-{{160}\over {243}}$ \\
\hline
0 & 0 & 0 & 12 & $1230462$ & ${{101355262012}\over {155925}}$ 
& ${{77441167}\over {180}}$ & ${{49050932335452}\over {153278125}}$ 
& ${{26831414}\over {105}}$ & ${{2184687177202}\over {23455575}}$ \\
1 & 1 & 0 & 12 & $2460$ & $-{{254288}\over {45}}$ & $-{{470591}\over {90}}$ 
& $-{{49887963408}\over {11790625}}$ & $-{{1379716}\over {405}}$ 
& $-{{3960090404}\over {5412825}}$ \\
2 & 0 & 0 & 12 & $1944$ & $-896$ & $-{{43417}\over {36}}$ 
& $-{{35397888}\over {30625}}$ & $-{{423976}\over {405}}$ 
& $-{{427970056}\over {1082565}}$ \\
2 & 1 & 1 & 12 & $0$ & $-24$ & $-{{393}\over 8}$ & $-{{39096}\over {625}}$ 
& $-{{616}\over 9}$ & $-{{35456}\over {729}}$ \\
2 & 2 & 0 & 12 & $0$ & $0$ & $-{7\over 8}$ & $-{{1104}\over {625}}$ 
& $-{{64}\over {27}}$ & $-{{5600}\over {2187}}$ \\
3 & 1 & 0 & 12 & $0$ & $0$ & $-{1\over 8}$ & $-{{192}\over {625}}$ 
& $-{4\over 9}$ & $-{{400}\over {729}}$ \\
\hline
1 & 0 & 0 & 13 & $-745189$ & $-{{1629512844964}\over {6081075}}$ 
& $-{{2652479497}\over {20160}}$ & $-{{11642408503972}\over {153278125}}$ 
& $-{{45795581}\over {945}}$ & $-{{470143113389}\over {164189025}}$ \\
1 & 1 & 1 & 13 & $-678$ & $-{{28728}\over 5}$ & $-{{614743}\over {96}}$ 
& $-{{903833736}\over {153125}}$ & $-{{704326}\over {135}}$ 
& $-{{600662842}\over {360855}}$ \\
2 & 1 & 0 & 13 & $-476$ & $-1120$ & $-{{437317}\over {288}}$ 
& $-{{241988968}\over {153125}}$ & $-{{616348}\over {405}}$ 
& $-{{765424036}\over {1082565}}$ \\
3 & 0 & 0 & 13 & $0$ & $-24$ & $-{{1337}\over {24}}$ 
& $-{{1500984}\over {21875}}$ & $-{{5864}\over {81}}$ 
& $-{{868672}\over {19683}}$ \\
\hline
0 & 0 & 0 & 14 & $20787102$ & ${{140597688722408}\over {14189175}}$ 
& ${{245838046393}\over {40320}}$ & ${{298587843693288}\over {69671875}}$ 
& ${{46330187578}\over {14175}}$ 
& ${{15116615635191578}\over {16254713475}}$ \\
1 & 1 & 0 & 14 & $17378$ & $-{{18013984}\over {189}}$ 
& $-{{1330017637}\over {17280}}$ & $-{{43519901193056}\over {766390625}}$ 
& $-{{1801726502}\over {42525}}$ & $-{{35538833424418}\over {6966305775}}$ \\
2 & 0 & 0 & 14 & $29088$ & $-20744$ & $-{{8258723}\over {360}}$ 
& $-{{1192015283352}\over {58953125}}$ & $-{{21049024}\over {1215}}$ 
& $-{{253843671008}\over {48715425}}$ \\
2 & 1 & 1 & 14 & $0$ & $-1456$ & $-{{54377}\over {24}}$ 
& $-{{278755152}\over {109375}}$ & $-{{628336}\over {243}}$ 
& $-{{253847360}\over {177147}}$ \\
2 & 2 & 0 & 14 & $32$ & $-144$ & $-{{2425}\over 8}$ 
& $-{{42568048}\over {109375}}$ & $-{{103640}\over {243}}$ 
& $-{{52536352}\over {177147}}$ \\
3 & 1 & 0 & 14 & $-2$ & $-32$ & $-{{206}\over 3}$ 
& $-{{1427216}\over {15625}}$ & $-{{8294}\over {81}}$ 
& $-{{4477778}\over {59049}}$ \\
4 & 0 & 0 & 14 & $0$ & $0$ & $-{1\over 4}$ & $-{{288}\over {625}}$ 
& $-{{16}\over {27}}$ & $-{{1280}\over {2187}}$ \\
\hline
1 & 0 & 0 & 15 & $-12672757$ & $-{{2597638257068408}\over {638512875}}$ 
& $-{{3549389785799}\over {1935360}}$ 
& $-{{64161784918165784}\over {65143203125}}$ 
& $-{{25033873577}\over {42525}}$ 
& $-{{305112195055173211}\over {24138249510375}}$ \\
1 & 1 & 1 & 15 & $-48624$ & $-{{7317712}\over {63}}$ 
& $-{{317601607}\over {2880}}$ & $-{{555516678768}\over {6015625}}$ 
& $-{{92268604}\over {1215}}$ & $-{{2363546376656}\over {146146275}}$ \\
2 & 1 & 0 & 15 & $-16428$ & $-{{154436}\over 5}$ 
& $-{{605816971}\over {17280}}$ & $-{{1403163936156}\over {42109375}}$ 
& $-{{110231638}\over {3645}}$ & $-{{4917927730652}\over {438438825}}$ \\
2 & 2 & 1 & 15 & $30$ & $-232$ & $-{{8199}\over {16}}$ 
& $-{{2141976}\over {3125}}$ & $-{{20954}\over {27}}$ 
& $-{{4066162}\over {6561}}$ \\
3 & 0 & 0 & 15 & $-400$ & $-1752$ & $-{{192203}\over {72}}$ 
& $-{{11182524776}\over {3828125}}$ & $-{{10592632}\over {3645}}$ 
& $-{{133926221584}\over {87687765}}$ \\
3 & 1 & 1 & 15 & $-10$ & $-48$ & $-{{1681}\over {16}}$ 
& $-{{90832}\over {625}}$ & $-{{506}\over 3}$ & $-{{953338}\over {6561}}$ \\
3 & 2 & 0 & 15 & $-1$ & $0$ & $-{{13}\over 8}$ & $-{{11936}\over {3125}}$ 
& $-{{151}\over {27}}$ & $-{{50545}\over {6561}}$ \\
4 & 1 & 0 & 15 & $0$ & $0$ & $-{1\over {16}}$ & $-{{648}\over {3125}}$ 
& $-{{28}\over {81}}$ & $-{{11840}\over {19683}}$ \\
\end{tabular}

\end{table}

\end{document}